\documentclass[pre,twocolumn,preprintnumbers,amsmath,amssymb]{revtex4}

\usepackage{graphicx}
\usepackage{dcolumn}
\usepackage{bm}
\usepackage{parskip}
\usepackage{color}
\usepackage{subfigure}
\usepackage{amssymb}
\usepackage{amsmath}
\usepackage{amssymb}
\usepackage{graphicx}

\def\be{\begin{eqnarray}}
\def\ee{\end{eqnarray}}
\def\be{\begin{equation}}
\def\ee{\end{equation}}
\begin{document}

\title{
Multiparameter universality and intrinsic diversity of critical phenomena \\in weakly anisotropic systems}

\author{Volker Dohm}

\affiliation{Institute for Theoretical Physics, RWTH Aachen University, 52056 Aachen, Germany}

\date {July 4, 2023}

\begin{abstract}
Recently a unified hypothesis of multiparameter universality for the critical behavior of bulk and confined anisotropic systems has been formulated
[V. Dohm, Phys. Rev. E {\bf 97}, 062128 (2018)]. We prove the validity of this hypothesis
in $d\geq 2$ dimensions on the basis of the principle of two-scale-factor universality
for isotropic systems. We introduce an angular-dependent correlation vector and a generalized shear transformation that transforms weakly anisotropic systems to isotropic systems.
As examples we consider the $O(n)$-symmetric $\varphi^4$ model, Gaussian model, and $n$-vector model.
By means of the inverse of the shear transformation we determine the general structure of the bulk order-parameter correlation function, of the singular bulk part of the critical free energy, and of critical bulk amplitude relations
of anisotropic systems at and away from $T_c$. It is shown that weakly anisotropic systems exhibit a high degree of intrinsic diversity due to $d(d+1)/2-1$ independent parameters that cannot be determined by thermodynamic measurements.
Exact results are derived for the $d=2$ Ising universality class and for the spherical  and Gaussian universality classes in $d\geq 2$ dimensions. For the $d=3$ Ising universality class we identify the universal scaling function of the isotropic bulk correlation function from the nonuniversal result of the functional renormalization group.
A proof is presented for the validity of multiparameter universality of the  exact critical free energy and critical Casimir amplitude in a finite rectangular geometry
of weakly anisotropic systems with periodic boundary conditions in the Ising universality class.
This confirms the validity of recent predictions of self-similar structures of finite-size effects in the $(d=2,n=1)$  universality class
at $T=T_c$ derived from conformal field theory
[V. Dohm and S. Wessel, Phys. Rev. Lett. {\bf 126}, 060601 (2021)].
This also substantiates the previous notion of an effective shear transformation for anisotropic two-dimensional Ising models. Our theory paves the way for a quantitative theory of nonuniversal critical Casimir forces in anisotropic superconductors for which experiments have been proposed by G.A. Williams, Phys. Rev. Lett. {\bf 92},197003 (2004).
\end{abstract}
\maketitle

\renewcommand{\thesection}{\Roman{section}}
\renewcommand{\theequation}{1.\arabic{equation}}
\setcounter{equation}{0}
\section{Introduction}
Spatial anisotropy is a fundamental property that is omnipresent in condensed matter physics where it is the origin of a wide variety of nonuniversal effects. Substantial evidence has emerged over the last two decades \cite{cd2004,dohm2005,dohm-arXiv,selke2005,selke2009,dohm2006,chen-zhang,dohm2008,
dohmphysik2009,DG,dohm2011,kastening-dohm,dohm2018,dohm2019,DW2021,DWKS2021} that part of this nonuniversal diversity persists  near ordinary critical points and in the near-critical Goldstone regime  of so-called weakly anisotropic systems. This includes magnetic materials, superconductors, alloys, solids with structural phase transitions, liquid crystals, compressible solids, and ultracold gases in anisotropic optical lattices. Recently an unexpected complex form of self-similarity of anisotropy effects in finite weakly anisotropic systems with periodic boundary conditions has been found \cite{DW2021} near the instability where weak anisotropy breaks down.

Ordinary bulk critical phenomena can be divided into universality classes characterized
by the dimension $d$
and the symmetry properties of the ordered state \cite{priv,pelissetto}.
As an example we consider $O(n)$-symmetric systems with short-range interactions and
with an $n$-component order parameter.
Within each $(d,n)$ universality class, all systems have the same universal quantities
(critical exponents, amplitude ratios, and scaling functions) which includes isotropic and weakly anisotropic systems since spatial anisotropy is only a marginal perturbation in the renormalization-group sense \cite{pri,priv,Aharony1976,weg-1,bruce,zia}. The principle of two-scale-factor universality
(or hyperuniversality) \cite{stau,priv,aharony1975,hohenberg1976,pri,weg-1,gerber,privmanbuch}
predicts that, once the universal quantities of a universality class are known,
the asymptotic critical behavior of any particular system of this universality class is known completely provided that only two nonuniversal amplitudes are given.
This principle was stated to be valid for all systems in a universality class
\cite{priv,henkelprivmanbuch,privman1988}. Furthermore it was asserted \cite{cardy1987,cardy1983,cardybuch,Indekeu,binder-wang,nightingaleprivmanbuch,barber1984,zia} that asymptotic isotropy can be restored in weakly anisotropic systems by a suitable anisotropic scale transformation and that universality can be restored \cite{Indekeu}, reintroduced \cite{nightingaleprivmanbuch}, or repaired \cite{priv} in some cases. However
the question was left unanswered how the various nonuniversal effects due to non-cubic anisotropy in bulk \cite{CoyWu,bruce,cardybuch,Aharony1976,aharony1980,Vaidya1976,WuCoy,Perk1,Perk2,Perk3,Perk4} and confined \cite{Izmailian, HH2019,cardyfinite,cardy1987,kim1987,Yurishchev,Indekeu,night1983,nightingaleprivmanbuch,NO1999,barber1984,binder-wang} systems could be reconciled with the principle of two-scale-factor universality. So far the issue of weak anisotropy has not been addressed in the approach based on the nonperturbative functional renormalization-group (FRG) \cite{wetterich2002,hassel2007,sinner2008,AdamR,metzner2021,baloq,dupuis2008}
where an extended universality was claimed to be valid for the isotropic bulk order-parameter correlation function in the  nonasymptotic critical region with only two nonuniversal parameters \cite{hassel2007,sinner2008}.

A systematic study of the effect of non-cubic spatial anisotropy on the critical behavior in bulk and confined systems was begun \cite{cd2004} by introducing a nondiagonal anisotropy matrix ${\bf A}$ into the $O(n)$-symmetric $d$-dimensional $\varphi^4$ field theory. An anisotropy-induced nonuniversality was discovered for the bulk correlation function as well as for the  critical Casimir amplitude \cite{DDPhysRep,bloete,krech,toldin2013,diehl2014} and the critical Binder cumulant ratio \cite{priv} which were the hallmarks of finite-size universality. The analytic results \cite{cd2004} were based on the exact large-$n$ limit and on the $\varphi^4$ theory in the minimal subtraction scheme in three dimensions \cite{dohm1985,schl}. These results demonstrated that restoring isotropy relates the critical behavior of anisotropic systems to that of isotropic systems but does not restore two-scale-factor universality. The violation of two-scale-factor universality in weakly anisotropic bulk and confined systems was subsequently confirmed and substantiated \cite{selke2005,dohm2006,DG,dohm2008,dohm2018,dohm2019,chen-zhang,selke2009,kastening-dohm,dohm-arXiv}, most recently by exact analytic results for the critical free energy and Casimir amplitude in two dimensions \cite{DW2021} based on conformal field theory \cite{franc1997}, as well as by Monte Carlo simulations \cite{DWKS2021}.

The main analytical and conceptional progress was achieved within the anisotropic $\varphi^4$ theory in three \cite{dohm2018} and two \cite{dohm2019,DW2021} dimensions. A central role is played by a shear transformation between anisotropic and isotropic systems which is formulated in terms of eigenvectors and eigenvalues of the anisotropy matrix ${\bf A}$. It was shown that weakly anisotropic $\varphi^4$ models do not have a unique single bulk correlation length but rather an angular-dependent correlation length \cite{dohm2019} with $d$ independent nonuniversal amplitudes in the $d$ principal directions which are determined by $d(d-1)/2$ nonuniversal angles. As a consequence, two-scale-factor universality is absent in the anisotropic $\varphi^4$ theory and is replaced by  multiparameter universality which allows for up to $d(d+1)/2+1$ independent nonuniversal parameters. This implies a revised notion of a universality class: it must be divided into subclasses \cite{dohm2008,dohmphysik2009} of isotropic and weakly anisotropic systems with different scaling forms for the  bulk correlation function and for the free energy of confined systems within the same universality class. These anisotropic scaling forms are governed by a reduced anisotropy matrix ${\bf \bar A}$ which has a universal structure in terms of principal correlation lengths and principal directions.

It was hypothesized \cite{dohm2018} that this revised picture of a universality class holds not only for the subclass of anisotropic $\varphi^4$ models but is valid quite generally for all other weakly anisotropic systems beyond the $\varphi^4$ theory. So far no general proof has been given for this hypothesis.

It is the aim of this paper to present a general proof for the validity of
multiparameter universality of the bulk critical behavior of weakly
anisotropic $d$-dimensional systems at and away from $T_c$ for $d\geq 2$.
Our strategy is to perform a generalized shear transformation from the anisotropic to an isotropic system, then to invoke two-scale-factor universality for the isotropic system and to derive the anisotropic structures by inverting the shear transformation. It is shown that
among the $d(d+1)/2+1$ nonuniversal parameters of weakly anisotropic systems there are only two parameters that can be determined by thermodynamic measurements whereas the $d(d+1)/2-1$ independent parameters contained in  ${\bf \bar A}$ cause a high degree of intrinsic diversity arising from
the nonuniversal angular dependence of the critical correlations. In most cases their principal directions depend in a  generically unknown way on the anisotropic interactions \cite{dohm2019,night1983}. This implies that in most cases, e.g., in anisotropic $n$-vector models and for real systems, it is unknown {\it a priori} how to perform the anisotropic scale transformation invoked in the literature \cite{cardy1987,cardy1983,cardybuch,Indekeu,binder-wang,nightingaleprivmanbuch,barber1984,zia},
with the exception of anisotropic $\varphi^4$ models with known eigenvectors of the anisotropy matrix ${\bf A}$.

Special attention is payed to the bulk order-parameter correlation function whose universal scaling function $\Psi_\pm$ in the asymptotic critical region is the same for both isotropic and weakly anisotropic systems in the same universality class. This function is known exactly for the $d=2$ Ising universality class \cite{WuCoy,dohm2019}. For the $d=3$ Ising universality class we determine the universal Fourier transform $\hat \Psi_\pm$ from an approximate nonasymptotic scaling form
that has been derived within the FRG approach \cite{hassel2007,sinner2008}.  We refute the claim \cite{hassel2007,sinner2008} that an extended universality is valid in the nonasymptotic region.

We also reanalyze the finite-size effects on the critical free energy and Casimir amplitude at $T_c$ of anisotropic two-dimensional systems
on a rectangle with periodic boundary conditions  \cite{DW2021,DWKS2021}. We prove the validity of multiparameter universality for these quantities of the $d=2$ Ising universality class including the feature of self-similarity on the basis of conformal field theory \cite{franc1997} without recourse to the anisotropic $\varphi^4$ theory.

The outline of this paper is as follows.
In Sec. II we give basic definitions within the anisotropic $\varphi^4$  and $n$-vector
models. In Sec. III we recall important facts with regard to two-scale-factor universality
of isotropic systems. Secs. IV and V are the central sections of this paper. In Sec. IV we introduce the notion of principal correlation vectors and a generalized shear transformation that depends on a free parameter and that is applicable to all weakly anisotropic systems, in contrast to the special shear transformation for the $\varphi^4$ model introduced earlier \cite{cd2004}. In particular we demonstrate the universality of the structure of the reduced $d \times d$ anisotropy matrix ${\bf \bar A}$ as a function of the nonuniversal angles of the principal axes and of the nonuniversal ratios of the principal correlation lengths.
In Sec. V we prove that multiparameter universality is generally
valid for the bulk order-parameter correlation function in the scaling
limit of weakly anisotropic systems in $d\geq 2$ dimensions.
As applications we consider in Sec. V the anisotropic correlation functions
of the Gaussian universality class for $d\geq 2$  and  of the spherical
universality class for $d>2$. The latter universality class includes the exactly
solvable large-$n$ limit of the  anisotropic $\varphi^4$  and $n$-vector models
as well as the exactly solvable spherical and the mean spherical models
\cite{berlin-kac,joyce,brankov,Baxter1982,stanley1968}. We also reanalyze the nonuniversal scaling form of the isotropic bulk correlation function derived within the $d=3$ FRG approach \cite{hassel2007,sinner2008} and invoke an exact sum rule that leads to an identification of the universal part of this scaling form in the asymptotic region.
Our proof of multiparameter universality is extended in Sec. VI to the bulk part
of the free energy and to critical bulk amplitude relations of anisotropic systems for $d\geq 2$. As a major application we discuss in Sec. VII the anisotropic bulk correlation function of the $(d=2,n=1)$ universality class which includes both anisotropic Ising and scalar $\varphi^4$ models. The exactly solvable triangular-lattice Ising model \cite{dohm2019,Vaidya1976,stephenson} is discussed as an example.
In Sec. VIII we present the shear transformation of the two-dimensional
angular-dependent correlation vector and derive a general transformation
formula that is applicable to both bulk and confined systems. An application is given in Sec. IX where we take up the analysis of finite-size effects on the critical free energy and Casimir amplitude of the anisotropic two-dimensional Ising model in \cite{DW2021}. We show that our generalized shear transformation substantiates and specifies the notion of the "effective shear transformation" mentioned in \cite{DW2021}.
We also discuss the critical Casimir force in anisotropic superconducting films.
Sec. X contains a summary.
\renewcommand{\thesection}{\Roman{section}}
\renewcommand{\theequation}{2.\arabic{equation}}
\setcounter{equation}{0}
\section{Basic definitions}
We consider isotropic and weakly anisotropic systems
near their critical points. In order to discuss the issue of multiparameter universality it is
appropriate to analyze and compare different types of anisotropic models that belong to the
same universality classes. As examples we consider (i) the $O(n)$-symmetric anisotropic
"soft-spin" $\varphi^4$ lattice model and (ii) the $O(n)$-symmetric anisotropic $n$-vector
model which is a fixed-length spin model.
We assume short-range interactions in a range of couplings where the systems have an
isotropic or  weakly anisotropic critical point.
Although the models (i) and (ii) have the same critical exponents and therefore belong to the
same  $(d,n)$ universality classes the analytic description of their anisotropy
properties require different approaches. The reason is that the principal directions
of the anisotropic $\varphi^4$ model are well defined at the outset through the anisotropy matrix {\bf A} whereas these axes
are generically unknown for the $n$-vector model. We also consider the $O(n)$-symmetric
Gaussian model which is defined only for $T\geq T_c$.
\subsection{${\bf \varphi^4}$ model}
The $\varphi^4$ lattice Hamiltonian divided by $k_B T$ reads \cite{dohm2006,dohm2008}
\begin{eqnarray}
\label{2a} H  &=&   v \Bigg[\sum_{i=1}^N \left(\frac{r_0}{2}
\varphi_i^2 + u_0 (\varphi_i^2)^2 - {\bf h}\cdot\varphi_i \right)
\nonumber\\&+& \sum_{i, j=1}^N \frac{K_{i,j}} {2} (\varphi_i -
\varphi_j)^2 \Bigg], \;
\end{eqnarray}
with $u_0>0$ and an ordering field ${\bf h}= h\;{\bf e}_h$ with a unit vector ${\bf e}_h$. The variables $\varphi_i \equiv \varphi ({\bf x}_i)$ are $n$-component vectors
on $N$ lattice points ${\bf x}_i \equiv(x_{i1}, x_{i2},\ldots, x_{id})$ of a
$d$-dimensional Bravais lattice of volume $V = Nv$
where $v$ is the volume of the primitive cell.
We assume periodic boundary conditions.
The components $\varphi_i^{(\mu)} \; , \mu = 1, 2, \ldots, n$ of $\varphi_i$
vary in the continuous range $- \infty \leq
\varphi_i^{(\mu)} \leq \infty$.
For an appropriate class of pair interactions  $K_{i,j}$
the model undergoes a phase transition in the bulk limit at a finite bulk critical
temperature $T_c$ for $n=1, d>1$ and  $n\geq2,d>2$. For $n=2,d=2$ this is a
Kosterlitz-Thouless transition.
No finite $T_c$ exists for $n>2$ in $d\leq 2$ dimensions.
For $n=1,2,3, \infty$ the $\varphi^4$ model
belongs to the Ising, $XY$, Heisenberg and spherical universality classes
\cite{priv,pelissetto}, respectively. The distance from bulk criticality is described
by the variable
\be
\label{rnullc}
r_0(T)-r_{0c}= a_0 t, \;\;\;t = (T - T_c) / T_c
\ee
with $a_0>0$  where $T_c$ is the bulk critical temperature.
The critical value $r_0(T_c)=r_{0c}= r_{0c}(v,u_0,K_{i,j}, d,n)$ is determined by the
divergence of the bulk susceptibility at $h=0$ \cite{dohm2008} and depends on the lattice structure,
on $v$, on $u_0$, and on all couplings $K_{i,j}$.
If $u_0=0$ the model (\ref{2a}) with $r_{0c}=0$ and $r_0=a_0t\geq 0$ belongs to the
Gaussian universality class
with a critical point at $r_0=0,h=0$ in $d>0$ dimensions.
The large-distance anisotropy is described by a dimensionless symmetric anisotropy matrix ${\bf A}(\{K_{i,j}\})$ \cite{cd2004} whose matrix elements
\begin{equation}
\label{2i} A_{\alpha \beta}(\{K_{i,j}\}) =  N^{-1} \sum^{ N}_{i,
j = 1} (x_{i \alpha} - x_{j \alpha}) (x_{i \beta} - x_{j \beta})\;K_{i,j}
\end{equation}
are determined by the second moments of the microscopic pair interactions $K_{i,j}$ \cite{dohm2006}. 
A characteristic feature of weakly anisotropy systems is that they have the same critical exponents as the isotropic system \cite{cd2004}. This requires
\be
\label{condition}
\det {\bf A}(\{K_{i,j}\})>0
\ee
as a necessary (but not yet sufficient \cite{dohm2008,dohm2018}) condition. The condition for isotropy in the large-distance regime is
\be
\label{isocondition}
{\bf A}(\{K_{i,j}\})= c^{\rm iso}_0{\bf 1}
\ee
with $c^{\rm iso}_0>0$ where ${\bf 1} $ is the unity matrix.
Eq. (\ref{2i})
and the criteria (\ref{condition}) and (\ref{isocondition}) are valid for general $d$ and $n$
\cite{dohm2008,dohm2018} including the large-$n$ limit and the Gaussian model.

The dimensionless partition function $Z$, the total free energy  ${\cal F}_{\rm tot}$ (divided by $k_BT$), and the total free-energy density $f$ of the $\varphi^4$ model are defined by
\begin{eqnarray}
\label{partphi}
Z&=& \Big[\prod_{i = 1}^{ N} \frac{\int
d^n {\varphi}_i}{v^{n (2-d) / (2d)}} \Big] \exp \left(- H \right),\\
\label{Ftotphi}
{\cal F}_{\rm tot}&=&-\ln Z,
\\
\label{free density}
f &=&{\cal F}_{\rm{tot}}/V,
\end{eqnarray}
and the bulk parts above $(+)$ and below $(-)$ $T_c$ by
\begin{eqnarray}
\label{Fdensphi}
f_{b,\pm}&=&\lim_{V\to \infty} {\cal F}_{\rm tot}/V,\\
\label{Fbulkphi}
{\cal F}_{b,\pm}&=& V f_{b,\pm}.
\end{eqnarray}
Near $T_c$ we use the decompositions
\begin{eqnarray}
{\cal F}_{\rm tot}&=&{\cal F}_s+{\cal F}_{ns},\\
f&=&f_s+f_{ns}, \\
f_{b,\pm}&=&f_{b,s,\pm}+f_{b,ns,\pm}
\end{eqnarray}
into singular and nonsingular parts.
The singular bulk part $f_{b,s,\pm}(t,h)$ has the scaling form of Eq. (1.1) of \cite{dohm2008} with the two nonuniversal constants $A_1$ and $A_2$, and it vanishes at $t=0,h=0$.
We shall discuss the leading singular bulk part at $h=0$
\begin{eqnarray}
\label{singbulkF}
{\cal F}_{b,s,\pm}&=& V f_{b,s,\pm}
\end{eqnarray}
of ${\cal F}_{\rm tot}$ for $d\geq2$ and, for $d=2$, the finite critical free energy
\be
\label{criticalF}
{\cal F}_c=\lim_{h\to 0, T\to T_c}{\cal F}_s
\ee
which is identical with the singular finite-size part of ${\cal F}_{\rm tot}$ at $T_c$, $h=0$ studied in \cite{DW2021}.

 The bulk order-parameter correlation function is
\be
\label{corrfctphi}
 G_\pm({\bf x_i}-{\bf x_j},t,h)
= \lim_{V \to \infty}\big[<{\varphi}_i \cdot {\varphi}_j> - {\cal M}^2\big].
\ee
where ${\cal M}^2 =  \lim_{|{\bf x_i-x_j}| \to \infty}< \varphi_i \cdot \varphi_j>$ is the square of the bulk order parameter.
The static version of the dissipation-fluctuation theorem \cite{halperin,hohenberg} yields the exact sum rule \cite{dohm2008}
\begin{eqnarray}
\label{3-n-lattice}
v \sum_{\bf
x}\; G_\pm ({\bf x}, t,h)=\chi_\pm (t,h)=\partial^2 f_{b,\pm} (t, h) / \partial h^2\;\;\;
\end{eqnarray}
where $\chi_\pm (t,h)$ is the bulk susceptibility.
Here we have assumed general $n\geq 1$ for $T\geq T_c$  and  $n=1$ for $T<T_c$.
For $n>1,T<T_c$ it is necessary to distinguish between longitudinal and transverse correlation functions \cite{priv}.

For the description of the asymptotic critical behavior on a long-distance scale it
suffices to study the continuum version of this model in terms of the vector
field $\varphi({\bf x})$. The continuum Hamiltonian reads \cite{cd2004,dohm2008}
\begin{eqnarray}
\label{contin} H^{\rm field} &=& \int_V d^d x \Bigg[\frac{r_0}{2}
\varphi^2 + \sum_{\alpha,
\beta=1}^d \frac{A_{\alpha \beta}}{2} \frac{\partial \varphi}
{\partial x_\alpha} \frac{\partial \varphi} {\partial x_\beta}
 \nonumber\\ &+& u_0 (\varphi^2)^2 - h \varphi \Bigg]
\end{eqnarray}
with a finite anisotropic cutoff in momentum space where now the sum rule has the form
\begin{eqnarray}
\label{sumrulexcontin}
\int d^d{\bf x}\;\;G_\pm (|{\bf x}|,t,h)=\chi_\pm(t,h)=\partial^2 f^{\rm field}_{b,\pm} (t, h) / \partial h^2.\;\;\;\;\;
\end{eqnarray}
The sum rules (\ref{3-n-lattice}) and (\ref{sumrulexcontin}) have nothing to do with the existence of a critical point and therefore are exactly valid in the entire range of $t$ and $h$, not only in the asymptotic critical region.
In particular the sum rule (\ref{sumrulexcontin}) remains exactly valid for any choice of a finite cutoff and for both isotropic and anisotropic $\varphi^4$ models. The susceptibility
plays an important role for the structure of the Fourier transform $ \hat G_\pm({\bf k}, t,h) $ of the correlation function. It can be uniquely divided into a thermodynamic part $\chi_\pm(t)$ and a correlation part $\hat D({\bf k}, t,h)$ in the Fisher-Aharony scaling form \cite{Fisher-Aharony-1974,tarko,hohenberg1976},
\begin{eqnarray}
\label{separationk}
\hat G_\pm({\bf k}, t,h)&=&\chi_\pm(t,h)\;\hat D_\pm({\bf k}, t,h)
\end{eqnarray}
with the normalization
\begin{eqnarray}
\label{normalizationk}
\hat D_\pm({\bf 0}, t,h)=1.
\end{eqnarray}
Here $\chi_\pm(t,h)$ is determined entirely by the bulk free-energy density defined at ${\bf k}={\bf 0}$ whereas $\hat D({\bf k}, t,h)$ requires ${\bf k}$-dependent investigations. Experimentally this means that $\chi_\pm(t,h)$ is determined by thermodynamic measurements  whereas $\hat D({\bf k}, t,h)$ requires spatially resolved scattering experiments. This physical distinction is of fundamental importance in the asymptotic critical region where $\hat D_\pm({\bf k}, t,h)$ becomes a universal function whereas $\chi_\pm(t,h)$ still contains a nonuniversal thermodynamic amplitude that is independent of the correlation-length amplitude. Thus near criticality the relevance of (\ref{separationk}) is that it provides a unique decomposition into a nonuniversal and a universal part \cite{hohenberg1976,tarko,priv,dohm2019}.
Right at criticality with $t=0,h=0$, $\chi_\pm(0,0)$  does not exist and two alternative decompositions have been introduced \cite{pri,dohm2019} that remain applicable to the asymptotic critical region including the critical point $t=0,h=0$, as discussed in Sec. III.C. These general considerations are valid for both isotropic and weakly anisotropic systems. They are of relevance for a discussion in Sec. V. E of the claims with regard to the universality properties of the correlation function of the isotropic $\varphi^4$ model derived in the framework of the FRG \cite{hassel2007,sinner2008}
where $\hat G_\pm({\bf k}, t,0)$ was divided into two nonuniversal parts, as we shall see.

In the anisotropic case, the principal correlation lengths
and the principal axes of $G_\pm$  are determined by the eigenvalues
and the eigenvectors
of $ {\bf A}$ \cite{cd2004,dohm2006}, thus they are well-defined functions of the couplings
and the lattice structure, as shown explicitly for $d=2$ in \cite{dohm2019}.
This differs fundamentally from the $n$-vector model.
\subsection{Fixed-length spin model}
We consider an anisotropic $O(n)$-symmetric fixed-length spin model defined on the
same lattice with the same boundary conditions as the $\varphi^4$ model.
It is called  $n$-vector model and has the Hamiltonian \cite{stanley1968,priv}
\begin{eqnarray}
\label{Hspin}
H^{\rm sp} = - \sum_{i,j} E_{i,j} {\bf S}_i \cdot {\bf S}_j
- \sum_{i} {\bf h}^{\rm sp} \cdot {\bf S}_i
\end{eqnarray}
with the statistical weight $\exp (- H^{\rm sp}/(k_B T))$. The classical dimensionless spin variables ${\bf S}_i$ are $n$-component vectors with a fixed length  ${\bf S}_i^2=1$. They have components $S_i^{(\mu)}$, $\mu= 1,2,...,n$
which are continuous variables for $n>1$. For $n=2,3$ the model is called $XY$ and Heisenberg model, respectively. For $n\to \infty$ the model belongs to the spherical universality class. For $n=1$ (\ref{Hspin}) is the anisotropic Ising model with the discrete variables $S_i \equiv\sigma_i=\pm 1$.

The  dimensionless partition function $Z^{\rm sp}$, the total free energy  ${\cal F}_{\rm tot}^{sp}$ (divided by $k_BT$), and the bulk correlation function are defined by
\begin{eqnarray}
\label{Zspin}
&&Z^{\rm sp}= \Big[\prod_{i = 1}^{N} \int
d^n {\bf S}_i \Big] \exp \left(- \beta H^{\rm sp} \right),\\
\label{Ftotspin}
&&{\cal F}^{\rm sp}_{\text{tot}}=-\ln Z^{\rm sp},\\
\label{corrfctsp}
&&G^{\rm sp}_\pm({\bf x_i}-{\bf x_j},t,{\bf h}^{\rm sp})
=\lim_{V\to \infty} \big[<{\bf S}_i \cdot {\bf S}_j> - ({\cal M}^{\rm sp})^2\big]\;\;\;\;\;\;\;\;\;\;
\end{eqnarray}
with $\beta= 1/(k_B T)$ and the constraint ${\bf S}_i^2=1$.
The definitions of $f^{\rm sp}$, ${\cal F}^{\rm sp}_{b,\pm}$,
${\cal F}^{\rm sp}_c$, $\chi^{\rm sp}_\pm$, etc.
are analogous to (\ref{free density})-(\ref{3-n-lattice}). The counterpart of the sum rule (\ref{3-n-lattice}) in the $n$-vector model is formultated for the example of the Ising model in Eq. (2.43) of \cite{WuCoy}.
If the variables $S_i^{(\mu)}$ have an unrestricted range, $- \infty \leq S_i^{(\mu)} \leq \infty$, the model (\ref{Hspin}) with $T \geq T_c$ belongs to the Gaussian universality class.
The general considerations with regard to the structure of the correlation functions $G_\pm$ and $\hat G_\pm$ remain valid also for $G^{\rm sp}_\pm$ and $\hat G^{\rm sp}_\pm$.

In view of the generalized shear transformation which is a pure coordinate transformation
to be introduced in Sec. IV we make the following general remark on the fixed-length
spin model (\ref{Hspin}). The definition of the Hamiltonian $H^{\rm sp}$ and the partition
function $Z^{\rm sp}$ in terms of the dimensionless variables ${\bf S}_i$ implies that $Z^{\rm sp}$
and the total free energy (\ref{Ftotspin}) do not depend on the details of the position of the
lattice points but only on the value and the topology of the pair interactions $E_{i,j}$.
This means that, for given $n$, given boundary conditions, and given number of lattice points,  two $n$-vector models $H^{\rm sp}$
 with the same coupling constants $E_{i,j}$ on two different lattices with the same topology of the pair interactions have the same partition function and free energy. Thus, if a transformation is performed at fixed boundary conditions that involves only a smooth change of the coordinates of the lattice points without changing the value and topology of the pair interactions $E_{i,j}$ and without changing the amplitudes of the spin variables ${\bf S}_i$, the partition function and total free energy are not affected and remain invariant under such a transformation. This implies that also the decomposition into singular and nonsingular parts
\be
\label{freesingnonsing}
{\cal F}^{\rm sp}_{\rm tot}={\cal F}^{\rm sp}_s + {\cal F}^{\rm sp}_{ns}
\ee
remains unchanged and that ${\cal F}^{\rm sp}_s$ remains invariant. In particular the critical free energy
\be
{\cal F}^{\rm sp}_c=\lim_{{\bf h}^{\rm sp} \to {\bf 0}, T\to T_c}{\cal F}^{\rm sp}_s
\ee
of the finite system remains invariant.
Furthermore such a pure coordinate transformation changes only the spatial argument of the correlation function $G_\pm^{\rm sp}$ but leaves its amplitude invariant.

No general approach to an analytic construction of the principal axes and correlation lengths
has been developed so far for the model (\ref{Hspin}).
Furthermore, no analytic condition for criticality is known for general couplings $E_{i,j}$, and
general criteria for weak anisotropy and isotropy analogous to the general conditions (\ref{condition}) and (\ref{isocondition})
for $\varphi^4$ models are as yet unknown for the $n$-vector model.
Apart from anisotropic $d=2$ Ising models, very little is known with regard to the anisotropy
properties of $n$-vector models for $d>2$ as compared to those of $\varphi^4$ models where
quantitative renormalized perturbation calculations can be performed within the minimal subtraction
scheme at fixed dimension $d$ \cite{schl,dohm1985,dohm2008,dohm2018}. Nevertheless it is possible to derive general
structural properties of the anisotropic critical behavior of the model (\ref{Hspin}) for
general $d$ and $n$ in the scaling limit, as will be
shown below.

In the remainder of this paper we consider only the case of vanishing external field.
\renewcommand{\thesection}{\Roman{section}}
\renewcommand{\theequation}{3.\arabic{equation}}
\setcounter{equation}{0}
\section{Isotropic case: Two-scale-factor universality}
For any given Bravais lattice of the  models (\ref{2a})  and (\ref{Hspin}) the couplings $K_{i,j}$ and $E_{i,j}$ can be chosen such that the bulk system has isotropic correlations in the large-distance scaling regime near $T_c$.
Since the bulk critical behavior of isotropic systems plays a fundamental role for
the development of
the theory of weakly anisotropic systems we recall and reformulate important definitions and facts in an appropriate way such that they can be extended to weakly anisotropic systems. A detailed exposition of the universal isotropic reformulation is also necessary for a comparison with the nonuniversal  formulation for the correlation function of the FRG \cite{hassel2007,sinner2008}  discussed in Sec. V.E.

\subsection{Isotropic bulk correlation function}
In the isotropic case, the  correlation function  (\ref{corrfctphi}) has the established Privman-Fisher scaling form in the asymptotic critical scaling region above $(+)$ and below $(-)$ $T_c$  below $d=4$ dimensions  \cite{pri}
\begin{eqnarray}
\label{3c} G^{\rm iso}_\pm (|{\bf x}|, t) &=& \frac{D^{\rm iso}_1}{ | {\bf x} |^{ d - 2 +\eta}}
\;\Phi_\pm \Big(\frac{|{\bf x}|}{ \xi^{\rm iso}_{\pm}(t)}\Big) \; ,\;\;\;\;\;\;
\\
\label{isocorr}
\xi^{\rm iso}_\pm(t)&=&\xi^{\rm iso}_{0\pm}\;|t|^{-\nu}
\end{eqnarray}
with the universal critical exponents $\eta$ and $\nu$ and the universal scaling function $\Phi_\pm$.
For an alternative form see (\ref{3calt}). The constant $D^{\rm iso}_1$ is the same above, at, and below $T_c$. The scaling region is defined by  $|{\bf x}| \gg  v^{1/d}$ and $ \xi^{\rm iso}_\pm \gg  v^{1/d}$ at fixed finite ratio $|{\bf x}|/ \xi^{\rm iso}_{\pm}$.
The function $G^{\rm iso}_\pm$ is finite and continuous at $T_c$ with $\Phi_+(0)=\Phi_-(0)$.
The correlation lengths $\xi^{\rm iso}_\pm$
may be chosen to be either the "true" (exponential) \cite{pri,pelissetto,dohm2019} or the second-moment \cite{dohm2008} correlation lengths. The ratios $Q^+_\xi$  and $Q^-_\xi$ of these correlation lengths above and below $T_c$ defined in Table 2 of  \cite{pelissetto} are universal.
In this paper we employ the exponential correlation length defined by the exponential decay $\sim \exp (-|{\bf x}|/\xi^{\rm iso}_{\pm})$ for large but finite $|{\bf x}|/\xi^{\rm iso}_{ \pm}$ within the scaling region.
We recall, however, that the
universal scaling form (\ref{3c}) is not valid for $|{\bf x}| / \xi^{\rm iso}_{\pm}\gg 1$ where two-scale-factor universality and scaling in the sense of (\ref{3c}) is violated
even arbitrarily close to $T_c$
\cite{cd2000-1,cd2000-2,dan-2,dohm2008,dohm2011}, see the shaded region in Fig. 2 of \cite{dohm2008}.

Among the three nonuniversal amplitudes $D^{\rm iso}_1$,  $\xi^{\rm iso}_{0+}$, and $ \xi^{\rm iso}_{0-}$ there are only two independent amplitudes since the amplitudes $\xi^{\rm iso}_{0\pm}$ are universally related by
\be
\label{ratioxi}
\xi^{\rm iso}_{0 +}/\xi^{\rm iso}_{0 -}=X_\xi= {\rm universal}.
\ee
In \cite{pelissetto} $X_\xi$ is denoted by $U_{\xi_{\rm gap}}$ for the exponential correlation length. It corresponds to
$1/X_-(0)$ in \cite{dohm2008,dohm2018} (denoted by $U_{\xi}$ in \cite{pelissetto}) where the second-moment correlation length is employed. We shall also consider the asymptotic susceptibility above and below $T_c$
\begin{eqnarray}
\label{chi}
\chi^{\rm iso}_\pm(t)=\Gamma^{\rm iso}_\pm |t|^{-\gamma},
\end{eqnarray}
with $\gamma= (2-\eta)\nu$ and the bulk order parameter
\be
{\cal M}^{\rm iso}(t)=B^{\rm iso}|t|^\beta
 \ee
below $T_c$ with the nonuniversal amplitudes $\Gamma^{\rm iso}_\pm $ and $B^{\rm iso}$ with the universal ratio \cite{pelissetto}
\begin{eqnarray}
\label{chiratio}
\Gamma^{\rm iso}_+/\Gamma^{\rm iso}_- = U_2\;=\;{\rm universal}.
\end{eqnarray}
 The amplitude  $B^{\rm iso}$ can be expressed in terms of $\Gamma^{\rm iso}_+$ and $\xi^{\rm iso}_{0 +}$  through the universal relation
\cite{priv,pelissetto,dohm2019}
\begin{eqnarray}
\label{amplrelisoxxy}
&&(B^{\rm iso})^2 (\Gamma_+^{\rm iso})^{-1}(\xi^{\rm iso}_{0+})^d= Q_c= \text {universal},
\end{eqnarray}
for general $n$ with the universal constant $Q_c$.

For $n>1, T<T_c$ the transverse bulk correlation function of  isotropic systems has the algebraic large-distance behavior for $d>2$ \cite{priv}
\begin{eqnarray}
\label{trans correl large}
G^{\rm iso}_{\rm T} (|{\bf x}|, t) &=& {\cal C}_{\rm T}  [{\cal M}^{\rm iso}(t)]^2 \big[\xi^{\rm iso}_{\rm T}(t)/|{\bf x}|\big]^{d-2},\\
\label{constT}
{\cal C}_{\rm T}& = &\Gamma(d/2)/[2\pi^{d/2}(d-2)].
\end{eqnarray}
The bulk transverse correlation length near $T_c$ is
\be
\xi^{\rm iso}_{\rm T}(t) = \xi^{\rm iso}_{0\rm T} |t|^{- \nu}
\ee
with the  nonuniversal amplitude $\xi^{\rm iso}_{0 \rm T}$ and the universal amplitude ratio
\begin{equation}
\label{3exT}
\xi^{\rm iso}_{0 +}/\xi^{\rm iso}_{0\rm T}   = X_{{\rm T}\xi}= {\rm universal}.
\end{equation}
\subsection{Singular part of the bulk free-energy density}
The following relations apply to isotropic systems of both the $(d,n)$ universality class (including $n$-vector and $\varphi^4$ models) and the Gaussian universality class for $t>0$ (with $\nu=1/2$ for general $d$ and $n$). The singular part of the bulk free-energy density has the asymptotic form \cite{priv,pelissetto}
\begin{eqnarray}
\label{3a}
f^{\text {iso}}_{b,s,\pm} (t) =\left\{
\begin{array}{r@{\quad \quad}l}
                         \; A^{\text {iso}}_\pm |t|^{d \nu}\quad          & \mbox{for} \;2<d<4\;, \\
                         \; \frac{1}{2}A^{\rm iso}_\pm |t|^{2\nu}\ln |t| & \mbox{for} \;d=2. \;
                \end{array} \right.
\end{eqnarray}
The nonuniversal amplitudes $A^{\rm iso}_\pm $ have the universal ratio
\be
\label{ratiofiso}
f^{\rm iso}_{b,s+} (t)/f^{\rm iso}_{b,s-} (t) = A^{\rm iso}_+/ A^{\rm iso}_-= \;\;\rm {universal}.
\ee
Due to two-scale-factor universality \cite{pri,priv,weg-1,stau}
these amplitudes
are universally related to the correlation-length amplitude $ \xi_{0+}^{\rm iso}$  through
\begin{eqnarray}
\label{3fy}
 \left(\xi_{0+}^{\rm iso}\right)^d A_+^{\rm iso} = \; Q_1= \;\;\rm {universal}, \;d\geq 2
\end{eqnarray}
with  the universal constant $Q_1$, with
 $Q_1= 1/(2\pi)$ for the $d=2$ Ising universality class (called $(R^+_\xi)^2$ in Eq. (6.31) of \cite{priv}) for the case that the true correlation length is used.
The validity of (\ref{3fy}) has been established by the renormalization-group theory
\cite{weg-1,hohenberg1976,aharony1975,gerber}.
Thus we obtain in $2<d<4$ dimensions
\begin{eqnarray}
\label{3azneu}
 f^{\text {iso}}_{b,s,+} (t)&=& Q_1\left(\xi_{0+}^{\rm iso}\right)^{-d}t^{d \nu},\\
f^{\text {iso}}_{b,s,-} (t) &=&\frac{A_-^{\rm iso}}{A_+^{\rm iso}}
 Q_1\left(\xi_{0+}^{\rm iso}\right)^{-d}|t|^{d \nu},
 \end{eqnarray}
whereas for the ($d=2,n=1$) universality class $f^{\text {iso}}_{b,s,\pm}$ depends on $|t|$ rather than $t$
\begin{eqnarray}
\label{minus3bbzneu} f^{\rm iso}_{b,s,\pm} (t) &=& \frac{1}{2}Q_1\left(\xi_{0+}^{\rm iso}\right)^{-2} \;t^2\ln |t| \;, \; d=2, \; \;\;\;\;\;\;\;\;\;
\end{eqnarray}
because of
\be
\label{aplusminus}
A^{\rm iso}_+/ A^{\rm iso}_-=1, \;\;\nu=1 , \; \;d= 2
\ee
for this universality class \cite{priv,pelissetto}.
We note that the amplitude $\xi_{0+}^{\rm iso}$ defined above $T_c$ appears in the above relations for both $t>0$ and $t<0$.

For the purpose of a structural analysis of critical bulk amplitude relations and a later extension to weakly anisotropic systems it is advantageous to express the relations (\ref{3azneu})-(\ref{minus3bbzneu}) for the bulk free-energy density $f^{\text {iso}}_{b,s,\pm}$ in terms of the singular bulk part ${\cal F}^{\rm iso}_{b,s,\pm}$ of the free energy in a arbitrary finite volume $V^{\rm iso}$ according to the definition (\ref{singbulkF}),
\begin{eqnarray}
\label{calFiso}
{\cal F}^{\rm iso}_{b,s,\pm}(t,V^{\rm iso}) &=& V^{\rm iso}f^{\rm iso}_{b,s,\pm} (t).
\end{eqnarray}
This leads in a natural way to the scaling variable
\begin{eqnarray}
\label{scalinvariablezz}
\widetilde x&=&t[V^{\rm iso}/(\xi_{0+}^{\rm iso})^{d}]^{1/(d\nu)}
\end{eqnarray}
appearing in the scaling form obtained from (\ref{3azneu})- (\ref{calFiso}) in $2<d<4$ dimensions
\begin{eqnarray}
\label{FISOzz}
{ \cal F}^{\rm iso}_{b,s,+}(t,V^{\rm iso})& =&Q_1\; \widetilde x^{d\nu}, \,\,\;\;\;\;\;
\\
{ \cal F}^{\rm iso}_{b,s,-}(t,V^{\rm iso})& =&\frac{A^{\text {iso}}_-}{ A^{\rm iso}_+}Q_1\; |\widetilde x|^{d\nu} , \,\, \;\;\;\;\;\;\;\;\;
\end{eqnarray}
and for the ($d=2,n=1$) universality class
\begin{eqnarray}
\label{Fisominuszz}
{ \cal F}^{\rm iso}_{b,s,\pm}(t,V^{\rm iso})& =&\frac{1}{2 }Q_1\;|\widetilde x|^2 \ln |t|, \,\,\;\;\;
\end{eqnarray}
with a nonscaling logarithmic factor. Unlike $f^{\text {iso}}_{b,s,\pm}$, both ${ \cal F}^{\rm iso}_{b,s,\pm}$ and $\widetilde x$ satisfy important invariance properties with respect to the shear transformations between isotropic and anisotropic systems as we shall see in Eqs. (\ref{FISO})-(\ref{Fisominus}).

The singular part of the Gaussian bulk free-energy density per component divided by $k_BT$ has the asymptotic behavior above $T_c$ \cite{kastening-dohm}
\begin{eqnarray}
\label{Gauss3a}
f^{\rm G,iso}_{b,s,+} (t)  =\left\{
\begin{array}{r@{\quad \quad}l}
                         \; A^{\rm G,iso}_+ t^{d/2}\quad          & \mbox{for} \;2<d<4\;, \\
                         \; \frac{1}{2}A^{\rm G, iso}_+ t\ln t & \mbox{for} \;d=2. \;
                \end{array} \right.
\end{eqnarray}
Due to two-scale-factor universality
the amplitude $A^{\rm G, iso}_+$
is universally related to the correlation-length amplitude $ \xi_{0+}^{\rm G,iso}$  through
\begin{eqnarray}
\label{Gaussxxx3fy}
 \left(\xi_{0+}^{\rm G,iso}\right)^d A_+^{\rm G, iso}
 &=& Q^{\rm G}_1 \;\;= {\rm universal}
\end{eqnarray}
where the  universal constant is \cite{kastening-dohm}
\be
\label{Q1Gaussd}
Q^{\rm G}_1=  -\frac{\Gamma(-d/2)}{2(4\pi)^{d/2}}\;\;, 2<d<4,
\ee
with $Q^{\rm G}_1= -1/(12 \pi)$ for $d=3$,
and \cite{kastening-dohm}
\be
\label{Q1Gauss2}
Q^{\rm G}_1=  -1/(4 \pi), \;\;  d=2,
\ee
thus we obtain for $t>0$
\begin{eqnarray}
\label{Gauss3azneu}
f^{\rm G,iso}_{b,s,+} (t)  =\left\{
\begin{array}{r@{\quad \quad}l}
                         \;  Q^{\rm G}_1\left(\xi_{0+}^{\rm G,iso}\right)^{-d}t^{d/2}\quad          & \mbox{for} \;d>2, \\
                         \; \frac{1}{2} Q^{\rm G}_1\left(\xi_{0+}^{\rm G, iso}\right)^{-2}\;t\ln t & \mbox{for} \;d=2. \;
                \end{array} \right.
\end{eqnarray}
For the singular bulk part
\begin{eqnarray}
\label{Gausssingbulk}
{\cal F}^{\rm G,iso}_{b,s,+}(t,V^{\rm iso}) &=& V^{\rm iso}f^{\rm G,iso}_{b,s,+} (t)
\end{eqnarray}
of the free energy of the isotropic Gaussian model in a finite volume $ V^{\rm iso}$ this leads to the scaling form
\begin{eqnarray}
\label{GaussFISO}
{ \cal F}^{\rm G, iso}_{b,s,+}(t,V^{\rm iso})  =\left\{
\begin{array}{r@{\quad \quad}l}
                         \; Q^{\rm G}_1\; (\widetilde x^{\rm G})^{d/2}\quad          & \mbox{for} \;d>2, \\
                         \; \frac{1}{2} Q^{\rm G}_1\; \widetilde x^{\rm G} \ln t & \mbox{for} \;d=2, \;
                \end{array} \right.
\end{eqnarray}
with a nonscaling logarithmic factor for $d=2$ and with the Gaussian scaling variable
\begin{eqnarray}
\label{Gaussscalinvariable}
\widetilde x^{\rm G}&=&t[V^{\rm iso}/(\xi_{0+}^{\rm G, iso})^{d}]^{2/d}\;\;\; {\rm for}\;\; d\geq 2.
\end{eqnarray}
\subsection{Sum rule and thermodynamic amplitudes of isotropic systems}
Two-scale-factor universality implies that the scaling form of the isotropic order-parameter correlation function is fully determined once two independent nonuniversal parameters are given. In the correlation function $G^{\rm iso}_\pm ({\bf x}, t)$, (\ref{3c}),  these parameters were chosen to be the correlation-length amplitude $\xi^{\rm iso}_{0 +}$ (or $\xi^{\rm iso}_{0 -}$) and the amplitude $D^{\rm iso}_1$. However, in view of the sum rule (\ref{sumrulexcontin})
\begin{eqnarray}
\label{sumrulex}
\chi^{\rm iso}_\pm(t)=\int d^d{\bf x}\;\;G^{\rm iso}_\pm (|{\bf x}|, t)
\end{eqnarray}
it is advantageous to express $D^{\rm iso}_1$ in terms of the amplitude $\Gamma_\pm^{\rm iso}$ of the directly obervable isotropic susceptibility (\ref{chi}).
From (\ref{3c}),(\ref{chi}), and  (\ref{sumrulex}) we obtain two relations
\begin{eqnarray}
\label{b3x}
D_1^{\rm iso} \;
&=&\Gamma^{\rm iso}_+(\xi^{\rm iso}_{0 +})^{-2+\eta}\; \;\widetilde \Phi_+ ^{-1} \;\;\;{\rm for \;general}\; n,\\
\label{a3x}
D_1^{\rm iso} \;
&=&\Gamma^{\rm iso}_-(\xi^{\rm iso}_{0 -})^{-2+\eta}\; \;\widetilde \Phi_- ^{-1}\;\;\;{\rm for}\; n=1,\;\;\;\;\;
\end{eqnarray}
with the two universal constants
\begin{eqnarray}
\label{a2}
\widetilde \Phi_\pm & =& 2 \pi^{d/2} \Gamma (d/2)^{-1}
\int_0^\infty ds s^{1 - \eta} \Phi_\pm (s).
\end{eqnarray}
Thus the susceptibility can be expressed as
\begin{eqnarray}
\label{suscept}
\chi^{\rm iso}_\pm(t)=D_1^{\rm iso}[\xi^{\rm iso}_\pm(t)]^{2-\eta}\widetilde \Phi_\pm .
\end{eqnarray}
This shows that $\chi^{\rm iso}_\pm(t)$ and $\xi^{\rm iso}_\pm(t)$ are not universally related.
Equations (\ref{3c}) and (\ref{b3x}) yield the alternative representation of the correlation function above and below $T_c$ for general $n$ in terms of the observable amplitudes $\Gamma^{\rm iso}_+$ and $\xi_{0+}^{\rm iso} $,
\begin{equation}
\label{3calt} G^{\rm iso}_\pm (|{\bf x}|, t) = \frac{\Gamma^{\rm iso}_+(\xi_{0+}^{\rm iso})^{-2+\eta}}{ | {\bf x} |^{d-2+\eta}} \;\Psi_\pm \Big(\frac{|{\bf x}|}{ \xi^{\rm iso}_{\pm}(t)} \Big) \; ,
\end{equation}
with the universal scaling function $\Psi_\pm$ above and below $T_c$,
\begin{eqnarray}
\label{3d}
&& \Psi_+ (y_+) =  \frac{\Phi_+ (y_+)}{\widetilde \Phi_+},\;\; \;\Psi_- (y_-) =  \frac{\Phi_-(y_-)}{\widetilde \Phi_+}.\;\;\;\;\;
\end{eqnarray}
The value of $\Psi_\pm $ at $T_c$ is a universal constant \cite{dohm2008}
\begin{eqnarray}
\label{tildeQ}
\Psi_+ (0) = \Psi_- (0) =  \widetilde Q_3=  {\rm universal},
\end{eqnarray}
compare Eqs. (3.15) and (A16) of \cite{dohm2008} and Eq.  (5.32) of \cite{dohm2018}.
The sum rule (\ref{sumrulex}) implies the normalization
\be
\label{normpsi}
2 \pi^{d/2} \big[\Gamma(d/2)\big]^{-1}\int_0^\infty dy y^{1 - \eta} \Psi_\pm (y)= 1
\ee
which provides a unique separation of the universal and nonuniversal parts of the correlation function.
The scaling function $\Psi_\pm$ depends on the universality class. Its exact analytic form is known for the $d=2$ Ising universality class \cite{WuCoy,Vaidya1976,dohm2019} and is presented in Secs. V.C and V.D. for the spherical universality class corresponding to the large-$n$ limit and for the Gaussian universality class. 
It is remarkable that, although no finite bulk correlation length and bulk susceptibility exist at $T_c$, the isotropic bulk correlation function at $T_c$
\begin{eqnarray}
\label{3caltTc}
G^{\rm iso}_\pm (|{\bf x}|, 0) &=& \widetilde Q_3\;\frac{\Gamma^{\rm iso}_+(\xi_{0+}^{\rm iso})^{-2+\eta}}{ | {\bf x} |^{d-2+\eta}}
\end{eqnarray}
is expressed in terms of the two independent  bulk amplitude $\xi_{0+}^{\rm iso}$ and $\Gamma^{\rm iso}_+$ defined {\it above} $T_c$ for general $n$.
(According to (\ref{a3x}),  $G^{\rm iso} ({\bf x}, 0)$ can also be expressed in terms of the bulk amplitudes  $\xi_{0-}^{\rm iso}$ and $\Gamma^{\rm iso}_-$ {\it below} $T_c$ but this is applicable only for $n=1$.) The representation (\ref{3caltTc}) demonstrates that, together with $\Gamma^{\rm iso}_+$, the amplitude  $\xi_{0+}^{\rm iso} $
constitutes an important reference length for the spatial decay of the correlation function at $T_c$ which is not apparent in the form (\ref{3c}) with the single amplitude $D^{\rm iso} _1$. The physical significance of  $\xi_{0\pm}^{\rm iso}$ right at $T_c$ is relevant for the applicability of our generalized shear transformation at $T_c$.

We shall also need the Fourier transform
\begin{eqnarray}
\label{3da} \hat G^{\rm iso}_\pm (|{\bf k}|, t)&=&\int d^d {\bf x}\;e^{- i {\bf k} \cdot {\bf x} }G^{\rm iso}_\pm (|{\bf x}|, t)
\\
\label{FourierG}
 &=& \frac{\Gamma^{\rm iso}_+}{ \big(| {\bf k} |\;\xi^{\rm iso}_{0+}\big)^{2-\eta}} \;\hat \Psi_\pm \Big( | {\bf k} |\xi^{\rm iso}_{\pm}(t) \Big)\;
\end{eqnarray}
with the universal scaling function
\begin{eqnarray}
\label{3db}
&&\hat \Psi_\pm (y_\pm) = \frac{2 \pi^{(d-1)/2}}
{\Gamma \big((d-1)/2\big)}
\int\limits_0^\infty ds \; s^{1 - \eta} \nonumber\\
&& \times \int\limits_{-1}^1 d (\cos \vartheta) (\sin
\vartheta)^{d-3}\; e^{- i s \cos \vartheta} \Psi_\pm (s/y_\pm). \;\;\;
\end{eqnarray}
The existence of the  scaling function $\hat \Psi_+ \Big( | {\bf k} |\xi^{\rm iso}_+(t) \Big)$ at infinite cutoff  has been shown in $4-\varepsilon$ dimensions up to $O(\varepsilon^2)$ for general $n$  \cite{Fisher-Aharony-1974}.
Because of the sum rule (\ref{sumrulex}) we have for $t\neq 0$
\begin{eqnarray}
\label{susceptx}
\hat G_\pm^{\rm iso} ( 0, t)= \chi^{\rm iso}_\pm(t)=\Gamma^{\rm iso}_\pm |t|^{-\gamma}.
\end{eqnarray}
The scaling function $\hat \Psi_\pm$ has a universal value at $T_c$ corresponding to $y_\pm=\infty$ \cite{tarko,priv},
\be
\label{Qdrei}
\hat \Psi_+ (\infty)=\hat \Psi_- (\infty)=Q_3=  {\rm universal}.
\ee
This implies the asymptotic  universal scaling form at $T_c$
\begin{eqnarray}
\label{atTcxisoiso} \hat G_\pm^{\rm iso} (|{\bf k}|, 0) = Q_3\frac{\Gamma^{\rm iso}_+}{\big(|{\bf k}|\;\xi^{\rm iso}_{0+}\big)^{2-\eta}}\;\;
\end{eqnarray}
 where again two independent bulk amplitudes $\Gamma^{\rm iso}_+$ and $\xi^{\rm iso}_{0+}$ appear.
The universal constants $Q_3$ and $\widetilde Q_3$  are related by a Fourier transformation which yields  \cite{dohm2008,dohm2019} (see App. A)
\begin{eqnarray}
\label{tildeQ}
 \widetilde Q_3= \frac{2^{d-2+\eta}\Gamma[(d-2+\eta)/2]}{(4\pi)^{d/2})\Gamma[(2-\eta)/2]}\;Q_3.\;\;\;\;
\end{eqnarray}
Both $Q_3$ and $\widetilde Q_3$ are known exactly for the $d=2$ Ising universality class \cite{dohm2019,tarko} and approximately for $d=3,n=1$ \cite{priv,tarko,brezin1974}. In Sec. V. E we shall show how to derive $\hat \Psi_\pm $ and $Q_3$ from an approximate result of the FRG \cite{hassel2007,sinner2008}.

Owing to the sum rule the nonuniversal overall amplitudes of $\hat G_\pm^{\rm iso}$ can be expressed completely in terms of $\chi^{\rm iso}_\pm(t)$ if $t\neq 0$. Using (\ref{FourierG}), (\ref{susceptx}), (\ref{ratioxi}), (\ref{chi})and (\ref{chiratio})) we obtain the representation for $t\neq 0$
\begin{eqnarray}
\label{overallplusminus}
\hat G_\pm^{\rm iso}( | {\bf k} |,t)&=& \Gamma^{\rm iso}_\pm|t|^{-\gamma}\;\hat D_\pm\Big( | {\bf k} |\xi^{\rm iso}_\pm(t) \Big),
 \end{eqnarray}
where the universal single-parameter scaling functions $\hat D_\pm$ are universally related to the universal functions $\hat \Psi_\pm $ as
\begin{eqnarray}
\label{Dplus}
\hat D_+(y_+)&=& y_+^{-2+\eta} \hat\Psi_+(y_+),\\
\label{Dminus}
\hat D_-(y_-)&=&U_2 \;(X_\xi)^{-2+\eta}\;y_-^{-2+\eta} \hat\Psi_-(y_-),
 \end{eqnarray}
with the universal constants  $U_2$ and $X_\xi$. These scaling functions satisfy the normalization condition
\begin{eqnarray}
\label{Dplusnull}
\hat D_\pm(0)&=&1
 \end{eqnarray}
in agreement with the scaling form (\ref{separationk} and (\ref{normalizationk}).
This condition is equivalent to our normalization condition (\ref{normpsi}). The same scaling form was derived within the framework of two-scale-factor universality by Hohenberg et al. \cite{hohenberg1976} where $\hat D_\pm$ was denoted by $\widetilde Z$. Thus, in the asymptotic region where $\hat D_\pm$ is calculated at infinite cutoff, $\hat G_\pm^{\rm iso}( | {\bf k} |,t)$ depends on two independent nonuniversal amplitudes $\Gamma_+^{\rm iso}$ and $\xi^{\rm iso}_{0+}$. Since (\ref{overallplusminus}) does not exist at $t=0$
our representations (\ref{3calt}) and (\ref{FourierG}) have the advantage
that they are applicable to the asymptotic region including the critical point $t=0$. In particular our scaling functions $\Psi_\pm$ and $\hat \Psi_\pm$ immediately capture the universal constants $\widetilde Q_3=\Psi_\pm(0)$ and $ Q_3=\hat\Psi_\pm(\infty)$, in contrast to the scaling functions $\hat D_\pm$ and its inverse Fourier transformed counterpart $D_\pm$ in real space.

The advantage of the representations (\ref{3calt}) and (\ref{normpsi}) over (\ref{3c}) is that, unlike the amplitude  $D_1^{\rm iso}$, both the amplitude $\Gamma^{\rm iso}_+$ of the bulk susceptibility $\chi^{\rm iso}_+$ above $T_c$ and $\xi_{0+}^{\rm iso}$ can be determined directly from independent thermodynamic measurements
 as follows. The bulk susceptibility
\begin{eqnarray}
\label{sush}
\chi^{\rm iso}_+(t)&=&-\lim_{h\to 0}\partial^2f^{\rm iso}_{b,+}(t,h)/\partial h^2\\
&=&\lim_{h\to 0}\partial m^{\rm iso}(t,h)/\partial h
\end{eqnarray}
can be determined via a measurement of the order parameter $m^{\rm iso}(t,h)$ for small $h\to 0$. Furthermore, owing to two-scale-factor universality, the bulk correlation length
\begin{eqnarray}
\label{sush}
\xi_{0+}^{\rm iso}&=&(Q_1/A_+^{\rm iso})^{1/d}
\end{eqnarray}
is determined by the amplitude $A_+^{\rm iso}$ of the free energy density according to (\ref{3fy}).
The latter amplitude can be measured via the singular part of the  bulk specific heat per unit volume (divided by $k_B$) above $T_c$
\begin{eqnarray}
\label{singbulkheatiso}
C^{\rm iso}_{b,s,+}(t)&=&-\partial^2f^{\rm iso}_{b,s,+}(t)/\partial t^2
\end{eqnarray}
of isotropic systems where $f^{\rm iso}_{b,s,+}(t)$ is given by (\ref{3a}).
This yields the dependence on $\xi^{\rm iso}_{0+}$
\begin{eqnarray}
\label{specheat}
C^{\rm iso}_{b,s+}(t) & =&\frac{\big(R^+_\xi\big)^d}{\alpha\; ( \xi^{\rm iso}_{0+})^d} \;t^{-\alpha},\;\;\;d>2,\\
\label{specheatdzwei}
C^{\rm iso}_{b,s+}(t) & =&-\;\frac{Q_1}{ \big( \xi^{\rm iso}_{0+}\big)^2} \;\ln t,\;\;d=2,\;\;\;\;
\end{eqnarray}
with the universal constant
\begin{eqnarray}
\label{Rplus}
\frac{(R^+_\xi)^d}{\alpha(1-\alpha)(2-\alpha)} = -\;Q_1,\;\;\; d>2.\;\;
\end{eqnarray}
Thus the important consequence of two-scale-factor universality for isotropic bulk systems is that all nonuniversal parameters of the isotropic bulk correlation function can be completely determined by the thermodynamic amplitudes of the bulk susceptibility and the bulk specific heat without involvement of measurements of the spatial dependence of the correlation function via scattering experiments. We shall show in Sec. V that weak anisotropy destroys this important feature due to the intrinsic nonuniversality of the correlation function.
 \subsection{Critical free energy in finite isotropic systems}
In Sec. IX we shall also consider the singular part of the free energy at $T_c$ of a finite anisotropic system on the basis of its structure in the isotropic case. For finite isotropic systems, e.g.,
in a $d$-dimensional parallelepiped of volume $V^{\rm iso}$ with given aspect ratios and
given angles between the confining surfaces, the hypothesis of two-scale-factor universality for finite systems \cite{priv,pri} predicts that, for given boundary conditions, the singular part ${\cal F}^{\rm iso}_s$ of
the free energy ${\cal F}_{\rm tot}^{\rm iso}$  has a universal
value at $T=T_c$ with the finite critical amplitude
\begin{eqnarray}
\label{freesingcrit}
{\cal F}^{\rm iso}_c = \lim_{T \to T_c} {\cal F}_s^{\rm  iso} = \text {universal}
\end{eqnarray}
which depends on $d,n,$ and is a universal function of the aspect ratios and angles.

The universal structure of (\ref{3c}), (\ref{amplrelisoxxy}), (\ref{3calt}), (\ref{FourierG}), and (\ref{specheat})-(\ref{freesingcrit}) as well as the scaling relations
(\ref{scalinvariablezz})-(\ref{Fisominuszz}) and (\ref{GaussFISO}) are  a consequence of the principle of two-scale-factor
universality for isotropic systems.
In the subsequent sections we address the question how weak anisotropy affects the structure of these results.
We shall show that multiparameter universality with up to $d(d+1)/2 +1$ independent nonuniversal parameters replaces two-scale-factor universality not only within the anisotropic $\varphi^4$ theory and the anisotropic Gaussian model \cite{dohm2008,dohm2018,dohm2019} but quite generally in all weakly anisotropic systems.
\renewcommand{\thesection}{\Roman{section}}
\renewcommand{\theequation}{4.\arabic{equation}}
\setcounter{equation}{0}
\section{Shear transformations of anisotropic systems}
Although the analysis of this section is formulated primarily for $O(n)$-symmetric systems with general $n\geq 1, T\geq T_c$ and $n=1, T<T_c$ all definitions and results can be extended to systems with $n>1, T<T_c$ where transverse correlation lengths come into play \cite{dohm2018}.

Our strategy
is to perform a shear transformation from the anisotropic to an isotropic system, then to invoke two-scale-factor universality for the isotropic system with the known structures of the correlation function, of the free energy, and of universal amplitude relations, and finally to derive the anisotropic structures by inverting the shear transformation. However, the choice of the shear transformation is not unique. 
\subsection{Special shear transformation of the $\varphi^4$ theory}
For the subsequent development of the anisotropic theory it is indispensable to complement and
reformulate the special shear transformation that is most suitable for the weakly
anisotropic $\varphi^4$ model \cite{cd2004, dohm2006,dohm2008,dohm2018}.
In the absence of an external field this transformation is defined by
\begin{eqnarray}
\label{shearalt}
{\bf x'}&=& {\mbox {\boldmath$\lambda$}}^{-1/2} {\bf U}{\bf x}, \\
\label{shearphi}
\varphi'({\bf x'})&=&(\det {\bf A})^{1/4}\varphi({\bf x}),\\
\label{shearu}
u'_0&=&(\det{\bf A}) ^{-1/2}u_0,
\end{eqnarray}
where ${\bf A}={\bf A}(\{K_{i,j}\})$ is given by (\ref{2i}). This $T$-independent transformation is
defined at fixed couplings $K_{i,j}$ for arbitrary temperatures above, at, and
below $T_c$ and leaves the distance $r_0-r_{0c}$ from criticality invariant \cite{dohm2008}.
It consists of a rotation of the lattice points ${\bf x}$ provided by the orthogonal matrix
${\bf U}$ and a subsequent spatial rescaling by the diagonal rescaling matrix
${\mbox {\boldmath$\lambda$}}$ such that the $O(\bf k^2)$ part of the Fourier
transform  $\delta \hat K ({\bf k})$ of the anisotropic interaction
of the Hamiltonian (\ref{2a}) \cite{dohm2006,dohm2008} is brought into the isotropic form
\begin{eqnarray}
\label{Hamisokprime}
\delta \hat K ({\bf k})&=&\delta \hat K ({\bf U}^{-1}{\mbox {\boldmath$\lambda$}}^{-1/2}{\bf k'})\equiv\delta \hat K' ({\bf k'})={\bf k'}\cdot{\bf k'}, \;\;\;\;\\
\label{sheark}
{\bf k'}&=& {\mbox {\boldmath$\lambda$}}^{1/2} {\bf U}{\bf k}.
\end{eqnarray}
Note, however, that
$O(\bf k^4)$ parts remain anisotropic in general and give
rise to anisotropic corrections outside the isotropic scaling regime of the transformed system (see Fig. 2 of \cite{dohm2008}).
The matrix elements of
\be
\label{Uvone}
{\bf U}={\bf U}\big(\{{\bf e}^{(\alpha)}\}\big)
\ee
are $U_{\alpha \beta} = e_{\beta}^{(\alpha)}$
which are determined by the $d$ eigenvectors ${\bf e}^{(\alpha)}$ of the matrix ${\bf A}$
whose Cartesian components are denoted by $e_{\beta}^{(\alpha)}$. We call
these vectors principal unit vectors. They satisfy the eigenvalue equation
\begin{eqnarray}
\label{eigenequ}
{\bf A}
(\{K_{i,j}\})
\;{\bf e}^{(\alpha)}&=&{\mbox {$\lambda$}_{ \alpha}}\;{\bf e}^{(\alpha)},\;\alpha = 1, 2, ..., d,\;\;
\end{eqnarray}
with ${\bf e}^{(\alpha)}\cdot{\bf e}^{(\beta)}=\delta_{\alpha\beta}$. The matrix ${\bf U}$ diagonalizes the matrix ${\bf A}$ according to
\be
\label{lambdaUAU}
{\mbox {\boldmath$\lambda$}}(\{K_{i,j}\})={\bf U}{\bf A}(\{K_{i,j}\}){\bf U}^{-1}
\ee
whose diagonal elements are the eigenvalues $\lambda_\alpha>0$ of ${\bf A}$. Thus we have
\be
\label{detlambda}
\det {\bf A}=\det {\mbox {\boldmath$\lambda$}} =\prod^d_{\alpha=1}{\mbox {$\lambda$}_{ \alpha}}.
\ee
The eigenvalue equation (\ref{eigenequ}) determines the eigenvalues $\lambda_\alpha$ and the principal unit vectors ${\bf e}^{(\alpha)}$ as a function of the couplings $K_{i,j}$.
The latter determine the directions of the principal axes of the large-distance correlations of the anisotropic system above, at, and below $T_c$ \cite{cd2004,dohm2008,dohm2018}.
For the formulation of bulk and finite-size
properties of anisotropic systems it will be necessary to define the reduced anisotropy matrix
 \begin{eqnarray}
\label{Aquerspecial}
  {\bf \bar A}(\{K_{i,j}\})&=&{\bf A}/(\det {\bf A})^{1/d}.
 \end{eqnarray}
An important alternative definition is
 \begin{eqnarray}
\label{AquerUlambdaU}
  {\bf \bar A}&=&{\bf U}^{-1}{\bar{\mbox {\boldmath$ \lambda$}}}{\bf U}
 \end{eqnarray}
where ${\bar{\mbox {\boldmath$ \lambda$}}}$ is the
reduced rescaling matrix
\begin{eqnarray}
\label{lambdaquerspecial}
 {\bar{\mbox {\boldmath$ \lambda$}}}= {{\mbox {\boldmath$\lambda$}}}  /(\det {\mbox {\boldmath$\lambda$}})^{1/d}
 \end{eqnarray}
as follows from (\ref{lambdaUAU}). According to (\ref{AquerUlambdaU}) and (\ref{lambdaquerspecial}), ${\bf \bar A}$ is known if the {\it ratios} of the eigenvalues $\lambda_\alpha$ of ${\bf  A}$ and the principal unit vectors ${\bf e}^{(\alpha)}$ are given. This makes possible to determine ${\bf  \bar A}$ even if the matrix ${\bf  A}$ is not known. It is the definition (\ref{AquerUlambdaU}) rather than (\ref{Aquerspecial}) that is used in deriving the shear transformation of the bulk correlation functions in Sec. V [compare (\ref{inverseAquer}) and (\ref{inverseAquersp})].

The anisotropic continuum Hamiltonian (\ref{contin}) is invariant under the special shear transformation (\ref{shearalt})-(\ref{shearu}) and is transformed to the Hamiltonian $H_{\text {field}}'$ which has  the standard isotropic form
\begin{eqnarray}
\label{2zxx}
H_{\text {field}}&=& H_{\text {field}}'\nonumber\\
\label{isotrH}
&=& \int_{V'} d^d x'
\Big[\frac{r_0} {2} \varphi'({\bf x}')^2 +  \frac{1} {2} (\nabla'
\varphi')^2 +
 u'_0 (\varphi'^2)^2\Big],\;\;\;\;\;\;\;\;\;\;\;
\end{eqnarray}
with the transformed volume
\begin{eqnarray}
\label{volumeprime}
 V' &=&\int_{V'} d^d x'=\det {\mbox {\boldmath$\lambda$}}^{-1/2}\;\int_{V} d^d x \\
& =&(\det {\mbox {\boldmath$\lambda$}})^{-1/2}\;V
 =(\det{\bf A})^{-1/2}\;V
\end{eqnarray}
and with a transformed cutoff which in general is still anisotropic.
The anisotropic bulk correlation function (\ref{corrfctphi}), its Fourier transform,
and the bulk order parameter ${\cal M}=<\varphi>$ are transformed as
\begin{eqnarray}
\label{isocorrf}
G_\pm'({\bf x'},t)&=& (\det {\bf A})^{1/2} G_\pm({\bf x},t),\;\;\;\;\;
\\
\label{transshearFourier}
\hat G_\pm'({\bf k'},t)&=& \hat G_\pm({\bf k},t),\\
\label{orderpar}
{\cal M'}(t)&=&(\det {\bf A})^{1/4} {\cal M}(t).
\end{eqnarray}
The bulk susceptibility remains invariant as follows
from the sum rules for the bulk correlation functions
\begin{eqnarray}
\label{relationchiphi}
&&\chi_\pm'(t)=\int d^d{\bf x'}\;\;G_\pm' ({\bf x'}, t)\\
&&= \int d^d{\bf x}\;\;G_\pm ({\bf x}, t)=\chi_\pm(t).\;\;\;\;\;\;
\end{eqnarray}
The partition function and the total free energy are not invariant but are
transformed as \cite{dohm2008}
\begin{eqnarray}
\label{ZZstrich}
Z'&=&(\det {\bf A})^{nN/(2d)}\;Z,\\
\label{Ftotstrich} {\cal F}'_{\rm tot}&=&{\cal F}_{\rm tot}\; - \;
[nN/(2d)] \ln (\det {\bf A}) \;
\end{eqnarray}
where the last term is a nonsingular bulk contribution
(which is absent in our generalized shear transformation
of the $n$-vector model to be defined below.) Thus the singular
part ${\cal F}_s$
of the total free energy is invariant under the shear transformation,
\be
\label{totalsingshearphi}
{\cal F}'_s={\cal F}_s,
\ee
and the singular part $f_{b,s}(t)$ of the bulk-free energy density is transformed as
\begin{eqnarray}
\label{transfbsx}
f'_{b,s,\pm}(t)&=& \lim_{V'\to \infty}{\cal F}'_s/V'=(\det {\bf A})^{1/2}f_{b,s,\pm}(t).\;\;\;
\end{eqnarray}
This implies that the singular bulk part ${\cal F}_{b,s,\pm}$ of the free energy
\be
\label{singbulkstrich}
{\cal F}'_{b,s,\pm}= V'f'_{b,s}=Vf_{b,s,\pm}={\cal F}_{b,s,\pm}
\ee
is invariant under the shear transformation.

All of the above relations are exact and a consequence of the special shear transformation (\ref{shearalt})-(\ref{shearu}). They are valid for arbitrary $t$ above, at, and below $T_c$. So far the quantities ${\bf e}^{(\alpha)}$, ${\bar{\mbox {\boldmath$ \lambda$}}}$,
${\bar{\mbox {\boldmath$ \lambda$}}}$, ${\bf A}$, and ${\bf \bar A}$
governing the transformations  are nonuniversal quantities which are
defined as a function of the couplings $K_{i,j}$ according to (\ref{2i}). This was called parametrization (i) in \cite{dohm2018}.
They are independent of $n$ and $u_0$ and remain valid in the large-$n$ limit and for the Gaussian model.

For the purpose of a later extension of the anisotropic theory to systems
other than the $\varphi^4$ model we confine ourselves in the following
to the asymptotic critical scaling region where this parametrization in terms of
$K_{i,j}$ can be reformulated in favor of a parametrization in terms of
correlation lengths. This was called parametrization (ii) in Sec. V. of \cite{dohm2018}.
This critical scaling region is defined by large $|{\bf x'}|$, large $\xi'_\pm(t)$, but
finite $|{\bf x'}|/\xi'_\pm(t)\geq  0$, where
\be
\xi'_\pm(t)=\xi'_{0\pm} |t|^{-\nu}.
 \ee
is the asymptotic isotropic correlation length defined through the exponential decay of the correlation
function $G_\pm'({\bf x'},t)$ of the Hamiltonian (\ref{isotrH}). In this scaling region $G_\pm'({\bf x'},t)$ depends only on $|{\bf x'}|$ rather than  ${\bf x'}$ thus we shall make the replacement
\be
G_\pm'({\bf x'},t) \longrightarrow G_\pm'(|{\bf x'}|,t)
\ee
in the remainder of this paper, and correspondingly
\be
\hat G_\pm'({\bf k'},t) \longrightarrow \hat G_\pm'(|{\bf k'}|,t).
\ee
The amplitudes $\xi'_{0+}$
and  $\xi'_{0-}$ are universally related by
\be
\label{ratioampprime}
\xi'_{0+}/\xi'_{0-}=X_\xi
\ee
where $X_\xi$ is the same universal constant as in (\ref{ratioxi}). The principal correlation lengths above and below $T_c$ have been determined as \cite{cd2004}
\begin{eqnarray}
\label{xialphaxz}
\xi^{(\alpha)}_{\pm}(t)&=&\xi^{(\alpha)}_{0\pm}|t|^{-\nu}={\mbox {$\lambda$}_{\alpha}}^{1/2}\xi'_\pm(t).
\end{eqnarray}
As a consequence their amplitude ratios are independent of the direction $\alpha$ and are universal quantities given by
\begin{eqnarray}
\label{ratioamp}
\xi^{(\alpha)}_{0+}/\xi^{(\alpha)}_{0-}=\xi'_{0+}/\xi'_{0-}=X_\xi \;\; \text {for each}\;\; \alpha
\end{eqnarray}
where $X_\xi$ is independent of $\alpha$ and the same universal constant as in (\ref{ratioxi}). This holds for
arbitrary short-range interactions and lattice structures of weakly anisotropic
$\varphi^4$ models. Together with the fact that the critical exponents above, at,
and below $T_c$ of weakly anisotropic systems are the same as those of isotropic
systems in the same universality class we consider the result (\ref{ratioamp})
for the amplitude ratio $\xi^{(\alpha)}_{0+}/\xi^{(\alpha)}_{0-}$
to be an additional characteristic feature of  weakly anisotropic systems.

According to (\ref{xialphaxz}) the eigenvalues can be expressed in terms of ratios of
correlation lengths as
\begin{eqnarray}
\label{abc}
{\mbox {$\lambda$}_{\alpha}}^{1/2}&=&\xi^{(\alpha)}_{0+}/\xi'_{0+}
=\xi^{(\alpha)}_{0-}/\xi'_{0-}
\end{eqnarray}
which are the same  above and below $T_c$. This implies
\begin{eqnarray}
\label{detA}
(\det {\mbox {\boldmath$\lambda$}})^{1/2}&=& \prod^d_{\alpha=1}\big(\xi^{(\alpha)}_{0\pm}/\xi'_{0\pm}\big)=\big(\bar \xi_{0\pm}/\xi'_{0\pm}\big)^d
\end{eqnarray}
with the mean correlation length
\begin{eqnarray}
\label{ximeanx}
\bar \xi_{\pm}(t)&=&\big[\prod^d_{\alpha = 1} \xi_{\pm}^{(\alpha)}(t)\big]^{1/d}=\bar \xi_{0\pm}|t|^{-\nu},\\
 \bar \xi_{0\pm}&=&\big[\prod^d_{\alpha = 1} \xi_{0\pm}^{(\alpha)}\big]^{1/d},\\
\label{ratioampmean}
 \bar\xi_{0+}/ \bar\xi_{0-}&=&X_\xi .
\end{eqnarray}
Equations (\ref{detA}) and (\ref{detlambda}) yield the desired representation of $\det {\bf A}$ in terms of correlation lengths as
\begin{eqnarray}
\label{detAexpress1}
(\det {\bf A})^{1/2}&=& \prod^d_{\alpha=1}\big(\xi^{(\alpha)}_{0+}/\xi'_{0+}\big)=\big(\bar \xi_{0+}/\xi'_{0+}\big)^d\\\;\;\;\;
\label{detAexpress2}
&=&\prod^d_{\alpha=1}\big(\xi^{(\alpha)}_{0-}/\xi'_{0-}\big)=\big(\bar \xi_{0-}/\xi'_{0-}\big)^d.
\end{eqnarray}
This enables us to reformulate the special shear transformations (\ref{volumeprime})-(\ref{orderpar}) and (\ref{transfbsx}) in terms of correlation lengths as
\begin{eqnarray}
\label{volumeprimecor}
 V' &=&
\big(\bar \xi_{0\pm}/\xi'_{0\pm}\big)^{-d} \;V,\\
\label{isocorrfcor}
G'_\pm(|{\bf x'}|,t)&=& \big(\bar \xi_{0\pm}/\xi'_{0\pm}\big)^d\; G_\pm({\bf x},t),\;\;\;\;\;
\\
\label{orderparcor}
{\cal M'}(t)&=&\big(\bar \xi_{0\pm}/\xi'_{0\pm}\big)^{d/2}\; {\cal M}(t),\\
\label{transfbsxcor}
f'_{b,s}(t)&=&\big(\bar \xi_{0\pm}/\xi'_{0\pm}\big)^d\;f_{b,s}(t),
\end{eqnarray}
with the $T$-independent transformation factor $\big(\bar \xi_{0\pm}/\xi'_{0\pm}\big)^d$.
This reformulation is needed for the derivation of the anisotropic physical properties within the $\varphi^4$ theory in the subsequent sections.

The factor $\big(\bar \xi_{0\pm}/\xi'_{0\pm}\big)^d$ can be interpreted
as the ratio of the ellipsoidal correlation volume
\be
V_{\rm cor,\pm}=\prod^d_{\alpha = 1} \xi_{0\pm}^{(\alpha)}=\big(\bar \xi_{0\pm}\big)^d
\ee
of the anisotropic system and the spherical correlation volume
\be
 V'_{\rm cor,\pm}=\big(\xi'_{0\pm}\big)^d
\ee
of the transformed isotropic system, with the  ratio
\be
  V_{\rm cor,+}/ V_{\rm cor,-}=V'_{\rm cor,+}/ V'_{\rm cor,-}=(X_\xi)^d.
\ee
Eq. (\ref{volumeprimecor}) can be rewritten as
\begin{eqnarray}
\label{volinvar}
\frac{ V'}{V'_{\rm cor,+}} =\frac{V}{V_{\rm cor,+}},\;\; \;\frac{ V'}{V'_{\rm cor,-}} &=&\frac{V}{V_{\rm cor,-}}.
\end{eqnarray}
This means that the ratios of the geometric volumes and correlation volumes remain
invariant under the special shear transformation (\ref{shearalt}) above and below $T_c$. This implies that the quantity
\begin{eqnarray}
\label{scalinvariablephi}
t[V'/(\xi'_{0+})^{d}]^{1/(d\nu)}=t [V/(\bar \xi_{0+})^{d}]^{1/(d\nu)},
\end{eqnarray}
appearing as the scaling variable in the bulk and finite-size theory of the $\varphi^4$ theory, remains invariant under the special shear transformation.
Eqs. (\ref{detAexpress1}) and  (\ref{detAexpress2}) can be rewritten as
\begin{eqnarray}
\label{xibarxiprime}
\big(\xi'_{0\pm}\big)^d&=&(\det {\bf A})^{-1/2}\big(\bar\xi_{0\pm}\big)^d,
\end{eqnarray}
which describes the shear transformation of the  anisotropic correlation volume to
the isotropic correlation volume.

The invariance of the susceptibility (\ref{relationchiphi}) yields in the asymptotic critical region
\begin{eqnarray}
\chi_\pm(t)&=&\Gamma_\pm |t|^{-\gamma}=\chi_\pm'(t)=\Gamma'_\pm |t|^{-\gamma},\\
\label{Gamma}
\Gamma'_\pm &=&\Gamma_\pm .
\end{eqnarray}
This implies the invariance of the combination
\begin{eqnarray}
\label{invarianceG}
\big(\xi'_{0\pm}\big)^d G_\pm'(|{\bf x'}|,t)/\Gamma'&=&\big(\bar \xi_{0\pm}\big)^d G_\pm({\bf x},t)/\Gamma
\end{eqnarray}
under the special shear transformation. We shall show that the invariance of both the combination (\ref{invarianceG}) and the volume ratio (\ref{volinvar}) are universal features of weakly anisotropic systems.

The two independent nonuniversal parameters $\xi'_{0\pm}$ and $\Gamma'_\pm$ will not be specified further. They can be determined from the correlation function and the susceptibility as functions of the parameter $a_0$ in (\ref{rnullc}) and the coupling $u'_0$ of the isotropic Hamiltonian (\ref{2zxx}) where, according to (\ref{shearu}), $u'_0$ depends on the four-point coupling $u_0$ and on $K_{i,j}$ through the anisotropy matrix ${\bf A}$ of the original anisotropic $\varphi^4$ Hamiltonian. We note that $\xi'_{0\pm}$ plays the role as a reference length in the framework of renormalized perturbation theory \cite{dohm2008,dohm2018} based on the transformed $\varphi^4$ Hamiltonian (\ref{2zxx}).

It is important to note that the ratios of the eigenvalues can be expressed in terms of the ratio of the principal correlation lengths
\begin{eqnarray}
\label{abcratio}
{\mbox {$\lambda$}_{\alpha}}/{\mbox {$\lambda$}_{\beta}}=\Big(\xi^{(\alpha)}_{0\pm}/\xi^{(\beta)}_{0\pm}\Big)^2
\end{eqnarray}
as follows from (\ref{xialphaxz}).
This makes possible to reexpress the diagonal elements of the reduced rescaling matrix (\ref{lambdaquerspecial})  in terms of ratios of principal correlation lengths as
\begin{eqnarray}
\label{lambdaalphaxixz}
\bar \lambda_\alpha\big(\{\xi_{0 \pm}^{(\alpha)}\}\big)&=& \prod^d_{\beta=1,\; \beta \neq\alpha}\Big(\frac{\xi_{0+}^{(\alpha)}}{\xi_{0+}^{(\beta)}}\Big)^{2/d}
=\Big(\frac{\xi_{0+}^{(\alpha)}}{\bar \xi_{0+}}\Big)^2\\
\label{lambdaalphaxixzminus}
&=& \prod^d_{\beta=1,\; \beta \neq\alpha}\Big(\frac{\xi_{0-}^{(\alpha)}}{\xi_{0-}^{(\beta)}}\Big)^{2/d}
=\Big(\frac{\xi_{0-}^{(\alpha)}}{\bar \xi_{0-}}\Big)^2\;.\;\;\;\;
\end{eqnarray}
Here the isotropic correlation length $\xi'_{0\pm}$ has been canceled.
The rescaling and reduced rescaling matrices are related by
\begin{eqnarray}
\label{lambdaredlambdaspecial}
{\mbox {\boldmath$\lambda$}}^{1/2} \xi'_{0\pm}&=&{\mbox {\boldmath$\bar\lambda$}}^{1/2}\bar \xi_{0\pm}.
\end{eqnarray}
From the alternative definition (\ref{AquerUlambdaU}) and from (\ref{lambdaalphaxixz}) and (\ref{lambdaalphaxixzminus}) we then obtain the reduced anisotropy matrix in terms of ratios of principal correlation lengths as
\be
\label{barAxyy}
{\bf \bar A} ={\bf \bar A}\big(\{\xi_{0\pm}^{(\alpha)},{\bf e}^{(\alpha)}\}\big)={\bf U(\{{\bf e}^{(\alpha)}\})}^{-1}{\bf \bar{\mbox {\boldmath$\lambda$}}} \big(\{\xi_{0 \pm}^{(\alpha)}\}\big) {\bf U(\{{\bf e}^{(\alpha)}\})}
\ee
which is the parametrization (ii) of ${\bf \bar A}$ as given in Eqs. (5.12) and (5.30) of \cite{dohm2018}. Both ${\bf \bar{\mbox {\boldmath$\lambda$}}}$ and  ${\bf \bar A}$ are nonuniversal quantities since they depend on nonuniversal parameters $\xi_{0 \pm}^{(\alpha)}$ and  ${\bf e}^{(\alpha)}$.
We shall show in Sec.  IV.C, however, that in this parametrization (ii) the dependence of  the matrices ${\bf \bar{\mbox {\boldmath$\lambda$}}} \big(\{\xi_{0 \pm}^{(\alpha)}\}\big)$ and  ${\bf \bar A}\big(\{\xi_{0\pm}^{(\alpha)},{\bf e}^{(\alpha)}\}\big)$ on these parameters attains a universal structure, unlike the parametrization (i) in terms of the couplings.

All of the relations presented above for the $\varphi^4$ theory at finite $n$ remain applicable also in the large $n$-limit. Furthermore they remain applicable to the Gaussian lattice model for $T\geq T_c$,
\begin{eqnarray}
\label{2aGauss} H^G  &=&   v \Bigg[\sum_{i=1}^N \frac{r_0}{2}
\varphi_i^2
+ \sum_{i, j=1}^N \frac{K_{i,j}} {2} (\varphi_i -
\varphi_j)^2 \Bigg], \;\;\;\;\;\;\;\;
\end{eqnarray}
$r_0(T) =  a_0 t\geq 0$, which is obtained from (\ref{2a}) by setting $u_0=0$ and $r_{0c}=0$.
%
We shall consider the continuum version of this model in the form
\begin{eqnarray}
\label{continGauss}
&&H^G_{\text {field}} = \int_V d^d x \Big[\frac{r_0}{2}
\varphi^2 + \sum_{\alpha,
\beta=1}^d \frac{A_{ \alpha \beta}}{2} \frac{\partial \varphi}
{\partial x_\alpha} \frac{\partial \varphi} {\partial x_\beta}
  \Big]\;\;\;\;\;\;
\end{eqnarray}
with the same
anisotropy matrix ${\bf A}$ (\ref{2i}) as for the $\varphi^4$ model. Correspondingly the isotropic Gaussian Hamiltonian obtained after the special shear transformation (\ref{shearalt}) and (\ref{shearphi}) reads
\begin{eqnarray}
\label{2zxxGauss}
H'^{G}_{\text {field}}&=& \int_{V'} d^d x'
\Big[\frac{r_0} {2} \varphi'({\bf x}')^2 +  \frac{1} {2} (\nabla'
\varphi')^2
\Big].\;\;\;\;\;\;\;\;
\end{eqnarray}

The advantage of the special shear transformation (\ref{shearalt})-(\ref{shearu}) is that the transformed Hamiltonian (\ref{2zxx}) has the form of the standard isotropic field-theoretic Landau-Ginzburg-Wilson Hamiltonian. This implies that the same renormalization factors ($Z$ factors) can be employed for the transformed system as for the established isotropic $\varphi^4$ field theory.
This has permitted us to perform renormalized perturbation theory \cite{dohm2008,dohm2018} for the transformed isotropic system in $2<d<4$ dimensions.

The disadvantage of this shear transformation is, however, that it is not directly applicable to other weakly anisotropic systems such as the $n$-vector model where it is unknown how to construct an appropriate continuum Hamiltonian with a large-distance anisotropy matrix as a function of the couplings $E_{i,j}$ that plays the same role as ${\bf A}$ in the $\varphi^4$ theory. In the following we introduce a generalized shear transformation that does not need an anisotropy matrix and that is applicable to all weakly anisotropic systems.

\subsection{Generalized shear transformation of the $\varphi^4$ theory }
We aim at introducing a shear transformation of the $\varphi^4$ theory in the scaling region that differs from the special shear transformation in that the explicit use of the anisotropy matrix ${\bf A}$ is avoided.
The principal correlation lengths of the $\varphi^4$ model above, at,  and below $T_c$ are
directed along the $T$-independent principal unit vectors ${\bf e}^{(\alpha)}$ .
Accordingly we introduce the orthogonal set of $d$ {\it principal correlation vectors}
\begin{eqnarray}
\label{corrvectorphi}
{\mbox {\boldmath$\xi$}}_{\pm}^{(\alpha)}(t)={\mbox {\boldmath$\xi$}}_{0\pm}^{(\alpha)}|t|^{-\nu}=\xi^{(\alpha)}_{0\pm}|t|^{-\nu}{\bf e}^{(\alpha)},\;\;\alpha = 1,...,d\;\;\;
\end{eqnarray}
where $\xi^{(\alpha)}_{0\pm}$ are the amplitudes of the principal correlation lengths.
We use a Cartesian coordinate system with orthogonal unit vectors ${\mbox {\boldmath$\epsilon$}}^{(\alpha)}$, $\alpha= 1,...,d$ along the $d$ Cartesian axes.
The shear transformation is achieved by a rotation of the vectors ${\mbox {\boldmath$\xi$}}_{0\pm}^{(\alpha)}$ by means of the orthogonal matrix ${\bf U}\big(\{{\bf e}^{(\alpha)}\}\big)$
with matrix elements $U_{\alpha \beta} = e_{\beta}^{(\alpha)}$ (which are the same as in the special shear transformation)
such that these vectors point along the direction of the Cartesian axes,
\begin{eqnarray}
{\bf  U}{\bf e}^{(\alpha)}={\mbox {\boldmath$\epsilon$}}^{(\alpha)},\;\;{\bf  U}{\mbox {\boldmath$\xi$}}_{0\pm}^{(\alpha)}=\xi^{(\alpha)}_{0\pm}{\mbox {\boldmath$\epsilon$}}^{(\alpha)},
\end{eqnarray}
and by a subsequent rescaling of their lengths by means of a diagonal matrix ${\mbox {\boldmath$\widetilde\lambda$}}^{-1/2}$. This matrix can be chosen to be independent of the temperature, i.e., it can be chosen to be the same above, at, and below $T_c$ as will be specified below.
This yields the $d$ transformed correlation vectors
\be
\label{transcorrX}
{\mbox {\boldmath$\widetilde\xi$}}_{0\pm}^{(\alpha)}
={\mbox {\boldmath$\widetilde\lambda$}}^{-1/2}{\bf U}{\mbox {\boldmath$\xi$}}_{0\pm}^{(\alpha)}
=\widetilde\lambda_{\alpha}^{-1/2}\xi^{(\alpha)}_{0\pm}{\mbox {\boldmath$\epsilon$}}^{(\alpha)}
\ee
where $\widetilde\lambda_{\alpha}$
are the diagonal elements of ${\mbox {\boldmath$\widetilde\lambda$}}$. To obtain a system that is isotropic in the scaling region it is necessary and sufficient that $\widetilde\lambda_{\alpha}^{1/2} $ is proportional to $\xi^{(\alpha)}_{0\pm}$,
\be
\label{choiceyyxx}
\widetilde\lambda_{\alpha}^{1/2} = \widetilde c_{\pm}\; \xi^{(\alpha)}_{0\pm},
\ee
where $\widetilde c_{+}>0$ and  $\widetilde c_{-}>0$
are independent of the direction $\alpha$.
This rescaling guarantees that the lengths $|{\mbox {\boldmath$\widetilde\xi$}}_{0\pm}^{(\alpha)}|$ of the transformed correlation vectors (\ref{transcorrX})
\be
|{\mbox {\boldmath$\widetilde\xi$}}_{0\pm}^{(\alpha)}|
= |{\mbox {\boldmath$\widetilde\lambda$}}^{-1/2}{\bf U}{\mbox {\boldmath$\xi$}}_{0\pm}^{(\alpha)}|=\widetilde\lambda_\alpha^{-1/2}\xi^{(\alpha)}_{0\pm}=\widetilde c_{\pm}^{-1}
 \ee
become independent of the direction $\alpha$.
This implies that
isotropic correlations are obtained in the large-distance scaling regime where the isotropic lengths $\widetilde c_{\pm}^{-1}$ are as yet unspecified.
The transformation (\ref{shearalt})-(\ref{shearu}) corresponds to the special choice  $\widetilde c_{\pm}= (\xi'_{0 \pm})^{-1}$
where $\xi'_{ 0\pm}$ is the correlation-length amplitude of the isotropic Hamiltonian $H_{\text {field}}'$, (\ref{2zxx}) and thus depends on the parameters $a_0$ and $ (\det {\bf A})^{1/2} u_0$ of the original anisotropic $\varphi^4$ model through (\ref{shearu}). For the explicit expression of $\xi'_{0\pm}$ see Eq. (2.82) of \cite{dohm-arXiv}.

Here we make a more general and conceptually different choice of $\widetilde c_\pm$
{\it that is  independent of any parameter of the original anisotropic system}.
We choose
\begin{eqnarray}
\label{cplusminus}
\widetilde c_{+}^{-1}&=&\xi^{\text { iso}}_{0+},\;\;\;\label{cminus}
\widetilde c_{-}^{-1}=\xi^{\text { iso}}_{0-},
\end{eqnarray}
thus
\begin{eqnarray}
\label{xiplusminus}
{\mbox {\boldmath$\widetilde\xi$}}_{0\pm}^{(\alpha)}&=&\xi^{\text { iso}}_{0\pm}{\mbox {\boldmath$\epsilon$}}^{(\alpha)},\;\;\;
|{\mbox {\boldmath$\widetilde\xi$}}_{0\pm}^{(\alpha)}|=\xi^{\text { iso}}_{0\pm},\;\;\alpha = 1,...,d\;\;\;
\end{eqnarray}
where $\xi^{\text { iso}}_{0+}$ and  $\xi^{\text { iso}}_{0-}$ are free parameters, together with the requirement that
\be
\label{isoratio}
\xi^{\text { iso}}_{0+}/\xi^{\text { iso}}_{0-}
=X_\xi={\rm universal}.
\ee
This requirement is necessary in order to comply with the fact implied by two-scale-factor universality that the ratio of the correlation lengths above and below $T_c$ of any isotropic system near $T_c$ must satisfy the universal relation (\ref{ratioxi}).
According to (\ref{ratioamp}) this implies $\xi^{(\alpha)}_{0+}/ \xi^{(\alpha)}_{0-}=\xi^{\text { iso}}_{0+}/ \xi^{\text { iso}}_{0-}$ or
\be
\xi^{(\alpha)}_{0-}/ \xi^{\text { iso}}_{0-}=\xi^{(\alpha)}_{0+}/ \xi^{\text { iso}}_{0+},
\ee
thus we obtain from (\ref{choiceyyxx})-(\ref{isoratio}) $d$ different
constants
\begin{eqnarray}
\label{choicelambdax}
\widetilde\lambda_{\alpha}^{1/2}&=&\xi^{(\alpha)}_{0-}/ \xi^{\text { iso}}_{0-}=\xi^{(\alpha)}_{0+}/ \xi^{\text { iso}}_{0+}
\end{eqnarray}
which are the same above and below $T_c$. By continuity, this $T$-independent identification of $\widetilde\lambda_{\alpha}$  is applicable also at $T=T_c$. This property is analogous to the relation (\ref{abc}) where the $T$-independent eigenvalues $\lambda_{\alpha}(K_{i,j})$ of the anisotropy matrix  ${\bf A}(K_{i,j})$ can be expressed in terms of  $\xi^{(\alpha)}_{0\pm}/\xi'_{0\pm}$,
thus both diagonal matrices ${\mbox {\boldmath$\widetilde\lambda$}}$ and ${\mbox {\boldmath$\lambda$}}$ are  $T$-independent and are related by
\begin{eqnarray}
\label{choicelambda}
{\mbox {\boldmath$\widetilde\lambda$}}&=&\Bigg(\frac{\xi'_{0-}}{\xi^{\text { iso}}_{0-}}\Bigg)^2{\mbox {\boldmath$\lambda$}}=\Bigg(\frac{\xi'_{0+}}{\xi^{\text { iso}}_{0+}}\Bigg)^2{\mbox {\boldmath$\lambda$}}.
\end{eqnarray}
We note that the mean correlation length can be used to express $\det {\mbox {\boldmath$\widetilde\lambda$}}$ as
\begin{eqnarray}
\label{ratiomeaniso}
\det {\mbox {\boldmath$\widetilde\lambda$}}= (\bar \xi_{0\pm}/ \xi^{\rm iso}_{0\pm})^{2d}.
\end{eqnarray}
We apply this generalized shear transformation to the lattice vectors ${\bf x}$ of the anisotropic system
\begin{eqnarray}
\label{trans}
{\bf \widetilde x}= {\mbox {\boldmath$\widetilde\lambda$}}^{-1/2} {\bf U}{\bf x},
\end{eqnarray}
with ${\mbox {\boldmath$\widetilde\lambda$}}$ given by (\ref{choicelambdax}). This generates lattice vectors ${\bf \widetilde x}$ of
an isotropic system with the correlation-length amplitudes $\xi^{\text { iso}}_{0\pm}$ in the scaling region. This transformation is applicable not only in the presence of finite correlations lengths $\xi^{\text { iso}}_{\pm}(t)$ for $T\neq T_c$ but also right at $T=T_c$.
We may also consider the inverse shear transformation
\be
\label{inversexX}
 {\bf x}={\bf U}^{-1}{\mbox {\boldmath$\widetilde\lambda$}}^{1/2} {\bf \widetilde x}
\ee
of the isotropic system  to the anisotropic system with given principal axes and given principal correlation lengths. Eq. (\ref{inversexX}) describes  a rescaling of the coordinates of the isotropic system in the direction of the Cartesian axes followed by a rotation of these axes into the direction of the principal axes of the anisotropic system. Our generalized shear transformation differs significantly from the special transformation (\ref{shearalt})-(\ref{shearu}) in that our transformation is a pure coordinate transformation without transforming the field $\varphi$ and the coupling $u_0$.  As a consequence, this transformation leaves the amplitude of the order-parameter correlation function $G^{\rm sp}$ invariant (as discussed in Sec. II. B) but not of the susceptibility.

Within this generalized transformation, we are primarily interested in determining the reduced anisotropy matrix ${\bf \bar A}^{\rm gen}= {\bf U}^{-1}{\bar{\mbox {\boldmath$ \lambda$}}^{\rm gen}}{\bf U}$ via the alternative definition given in (\ref{AquerUlambdaU}) in terms of a
reduced rescaling matrix
${\mbox {\boldmath$\bar\lambda$}^{\rm gen}}$.
In this transformation the reduced rescaling matrix is defined by
\begin{eqnarray}
\label{lambdaquerxyy}
{\mbox {\boldmath$\bar\lambda$}^{\rm gen}}&=&\frac{{\mbox {\boldmath$\widetilde\lambda$}}}{(\det {\mbox {\boldmath$\widetilde\lambda$}})^{1/d}}\;\;,\\
\det {\mbox {\boldmath$\widetilde\lambda$}}&=&\prod^d_{\alpha=1}\widetilde\lambda_{\alpha},
\end{eqnarray}
where the diagonal elements $\widetilde\lambda_{\alpha}$ of ${\mbox {\boldmath$\widetilde\lambda$}}$ are given by (\ref{choicelambdax}). The rescaling matrix ${\mbox {\boldmath$\widetilde\lambda$}}$ and the reduced rescaling matrix ${\mbox {\boldmath$\bar\lambda$}^{\rm gen}}$ are related by
\begin{eqnarray}
\label{lambdaredlambda}
{\mbox {\boldmath$\widetilde\lambda$}}^{1/2} \xi^{\rm iso}_{0\pm}&=&\big({\mbox {\boldmath$\bar\lambda$}^{\rm gen}}\big)^{1/2}\bar \xi_{0\pm}.
\end{eqnarray}
We see that the dependence on  $\xi^{\rm iso}_{0\pm}$ is canceled in the diagonal matrix (\ref{lambdaquerxyy}). This implies that the matrix (\ref{lambdaquerxyy}) is the same as that defined in (\ref{lambdaquerspecial}),
\be
\label{lambdagen}
{\mbox {\boldmath$\bar\lambda$}^{\rm gen}}={\mbox {\boldmath$\bar\lambda$}},
\ee
with the same
 diagonal elements as given in (\ref{lambdaalphaxixz}) and (\ref{lambdaalphaxixzminus}).
As a consequence, the reduced anisotropy matrix in the generalized shear transformation
\be
\label{Aquergeneral}
{\bf \bar A}^{\rm gen}={\bf \bar A}\big(\{\xi_{0\pm}^{(\alpha)},{\bf e}^{(\alpha)}\}\big)
\ee
is identical with that given in (\ref{barAxyy}) within the special shear transformation (\ref{shearalt})-(\ref{shearu}). This result demonstrates that
the generalized shear transformation is capable deriving the structure
of the reduced anisotropy matrix ${\bf \bar A}$  of the anisotropic $\varphi^4$
model without explicit knowledge of the anisotropy matrix ${\bf A}$.
Furthermore the generalized shear transformation is simpler than the special shear transformation in that it is a pure coordinate transformation. For this reason this
transformation is applicable also to other weakly anisotropic systems.

The generalized shear transformation introduced above for the $\varphi^4$ theory at
finite $n$ remains applicable also in the large $n$-limit of the $\varphi^4$ theory
and to the Gaussian model for $T\geq T_c$. We note, however, that there is no advantage within the $\varphi^4$ theory to work with the generalized shear transformation (\ref{trans}) since the corresponding transformed Hamiltonian $\widetilde H_{\text {field}}$ would not have the form of a standard isotropic $\varphi^4$ theory. For this reason it is more convenient within the $\varphi^4$ theory to employ the special shear transformation (\ref{shearalt})-(\ref{shearu}) which allows one to use ordinary renormalized isotropic perturbation theory within the transformed system on the basis of the standard isotropic Landau-Ginzburg-Wilson Hamiltonian (\ref{2zxx}).
\subsection{Generalized shear transformation of the ${\bf n}$-vector model}
As a representative of weakly anisotropic systems other than the $\varphi^4$ model we take the fixed-length $n$-vector Hamiltonian $H^{\rm sp}$, (\ref{Hspin}), for general $d$ and $n$. We consider only the scaling region. Our only assumptions for this model are
(i) that there exist $d$ principal unit vectors ${\bf e}^{{\rm sp}(\alpha)}$  in the directions of $d$  principal axes together with $d$ principal correlation lengths above and below $T_c$
\be
\label{princcorr}
\xi^{{\rm sp}(\alpha)}_{\pm}(t)=\xi^{{\rm sp}(\alpha)}_{0\pm}|t|^{-\nu}
\ee
along these directions where $\nu$ is the critical exponent of the isotropic $n$-vector model, and (ii) that the amplitude ratios  satisfy
\be
\label{ratioalpha}
\xi^{{\rm sp}(\alpha)}_{0+}/\xi^{{\rm sp}(\alpha)}_{0-}= X_\xi \;\;\text {for each}\;\; \alpha
\ee
where $X_\xi$ is the same universal constant as in (\ref{ratioxi}) for isotropic systems of the $(d,n)$ universality class. These assumptions are supported by our exact results (\ref{ratioamp}) for the $\varphi^4$ model
for general $(d,n)$
and for the anisotropic two-dimensional Ising model \cite{dohm2019}, as discussed in Sec. VIII. B.
We shall show that for deriving the universal structures of the correlation function and of bulk amplitude relations it is not necessary to know the dependence of
$\xi^{{\rm sp}(\alpha)}_{0\pm}$
and of the principal unit vectors
${\bf e}^{{\rm sp}(\alpha)}$ on the microscopic couplings $E_{i,j}$. For most of our results and conclusions the assumption (\ref{ratioalpha}) is not needed.

Our introduction of the generalized shear transformation of the $n$-vector model is parallel to (\ref{corrvectorphi}) - (\ref{inversexX}).
Accordingly we introduce the orthogonal set of $d$  principal correlation vectors
\be
\label{corrvectorsp}
{\mbox {\boldmath$\xi$}}_{0\pm}^{{\rm sp}(\alpha)}=\xi^{{\rm sp}(\alpha)}_{0\pm}{\bf e}^{{\rm sp}(\alpha)}, \;\;\; \alpha=1,...,d
\ee
and perform a rotation by means of
\be
\label{Usp}
{\bf \widehat U}={\bf U}\big(\{{\bf e}^{{\rm sp}(\alpha)}\}\big)
\ee
with matrix elements  $\widehat U_{\alpha \beta} = e_{\beta}^{{\rm sp}(\alpha)}$
such that these vectors point along the direction of the Cartesian axes,
\begin{eqnarray}
{\bf \widehat U}{\bf e}^{{\rm sp}(\alpha)}={\mbox {\boldmath$\epsilon$}}^{(\alpha)},\;\;{\bf \widehat U}{\mbox {\boldmath$\xi$}}_{0\pm}^{{\rm sp}(\alpha)}&=&\xi^{{\rm sp}(\alpha)}_{0\pm}{\mbox {\boldmath$\epsilon$}}^{(\alpha)}.
\end{eqnarray}
As for the $\varphi^4$ model, the orthogonal matrix ${\bf \widehat U}$ contains $d(d-1)/2$ independent matrix elements which are needed to specify the directions ${\bf e}^{{\rm sp}(\alpha)}$ of the principal axes. These $d$ mutually orthogonal axes can be described by $d(d-1)/2$ independent angles $\Omega_i$, $i=1,2,..., d(d-1)/2$, i.e., one angle $\Omega$ in two dimensions, three angles in three dimensions, six angles in four dimensions, etc.,
\begin{eqnarray}
\label{Ud}
{\bf U}\big(\{{\bf e}^{{\rm sp}(\alpha)}\}\big)= {\bf U}\big(\Omega_1,\Omega_2, \Omega_3,...\big).
\end{eqnarray}
A subsequent rescaling of the lengths $\xi^{{\rm sp}(\alpha)}_{0\pm}$ by means of a diagonal matrix
${\mbox {\boldmath$\widehat\lambda$}}^{-1/2}$
 yields the $d$ transformed  vectors
\begin{eqnarray}
\label{transcorrsp}
{\mbox {\boldmath$\widehat\xi$}}_{0\pm}^{(\alpha)}
&=&{\mbox {\boldmath$\widehat\lambda$}}^{-1/2}{\bf \widehat U}{\mbox {\boldmath$\xi$}}_{0\pm}^{{\rm sp}(\alpha)}
=\widehat \lambda_{\alpha}^{-1/2}\xi^{{\rm sp}(\alpha)}_{0\pm}{\mbox {\boldmath$\epsilon$}}^{(\alpha)}.
\end{eqnarray}
Here we make the choice, similar to (\ref{choiceyyxx}),
\be
\label{choiceyy}
\widehat\lambda _{\alpha}^{1/2}= \widetilde c_{\pm}\; \xi^{{\rm sp}(\alpha)}_{0\pm},
\ee
together with the same choice for $\widetilde c_{\pm}$ as specified in (\ref{cplusminus}) and (\ref{isoratio}), with free parameters $\xi^{\text { iso}}_{0\pm}$. Because of (\ref{isoratio}) and (\ref{ratioalpha}) this implies
$\xi^{{\rm sp}(\alpha)}_{0+}/\xi^{{\rm sp}(\alpha)}_{0-}=\xi^{\text { iso}}_{0+}/ \xi^{\text { iso}}_{0-}$ or
\be
\xi^{{\rm sp}(\alpha)}_{0-}/ \xi^{\text { iso}}_{0-}=\xi^{{\rm sp}(\alpha)}_{0+}/ \xi^{\text { iso}}_{0+},
\ee
thus the diagonal elements of ${\mbox {\boldmath$\widehat\lambda$}}$ are defined to be $T$-independent parameters given by
\begin{eqnarray}
\label{choicelambdaxyy}
\widehat \lambda_{\alpha} &=&\Big(\xi^{{\rm sp}(\alpha)}_{0-}/ \xi^{\text { iso}}_{0-}\Big)^2=\Big(\xi^{{\rm sp}(\alpha)}_{0+}/ \xi^{\text { iso}}_{0+}\Big)^2,
\end{eqnarray}
similar to (\ref{abc}) and (\ref{choicelambdax}). This transforms the $d$ different principal correlation vectors (\ref{corrvectorsp}) to the $d$ vectors ${\mbox {\boldmath$\widehat\xi$}}_{0\pm}^{(\alpha)}
=\xi^{\text { iso}}_{0\pm}{\mbox {\boldmath$\epsilon$}}^{(\alpha)}$ with the angular-independent lengths $\xi^{\text { iso}}_{0+}$ and $\xi^{\text { iso}}_{0-}$ above and below $T_c$, respectively, representing an isotropic system.

Correspondingly we define the $T$-independent
shear transformation of the lattice points ${\bf x} \to { \bf \widehat x}$ at fixed couplings $E_{i,j}$ of the anisotropic $n$-vector model by
\begin{eqnarray}
\label{transyy}
{\bf \widehat x}= {\mbox {\boldmath$\widehat\lambda$}}^{-1/2} {\bf \widehat U}{\bf x},
\end{eqnarray}
with ${\mbox {\boldmath$\widehat\lambda$}}$ given by (\ref{choicelambdaxyy}) and ${\bf \widehat U}$ defined by (\ref{Usp}). This generates an $n$-vector model on a transformed lattice with lattice points ${\bf \widehat x}$ and  with isotropic correlations characterized by the correlation length $\xi^{\text { iso}}_{0\pm}$ for which two-scale-factor universality can be invoked.
In constructing this transformation no assumption has been made other than the existence of the principal correlation lengths (\ref{princcorr}) and, for the application to $T < T_c$, the universality of the amplitude relation (\ref{ratioalpha}).

Now we have arrived at the position to derive the structure of the reduced anisotropy matrix  ${\bf \bar A}^{\rm sp}$ that will enter the anisotropic correlation function $G^{\rm sp}({\bf x}, t)$ of the $n$-vector model to be derived in Sec. V. B. We define the reduced rescaling matrix
\begin{eqnarray}
\label{lambdaquerx}
{\mbox {\boldmath$\bar\lambda$}}^{\rm sp}\big(\{\xi_{0 \pm}^{{\rm sp}(\alpha)}\}\big)&=&{\mbox {\boldmath$\widehat\lambda$}}/\big(\det {\mbox {\boldmath$\widehat\lambda$}}\big)^{1/d}\;\;
\end{eqnarray}
which has the diagonal elements
\begin{eqnarray}
\label{lambdaalphaxixyy}
&&\bar \lambda^{\rm sp}_\alpha=\prod^d_{\beta=1,\; \beta \neq\alpha}\Big(\frac{\xi_{0\pm}^{{\rm sp}(\alpha)}}{\xi_{0\pm}^{{\rm sp}(\beta)}}\Big)^{2/d}
=\Big(\frac{\xi_{0\pm}^{{\rm sp}(\alpha)}}{\bar \xi^{\rm sp}_{0\pm}}\Big)^2\;\;\;
\end{eqnarray}
with the amplitude $\bar \xi^{\rm sp}_{0\pm}$ of the mean correlation length
\begin{eqnarray}
\label{ximeany}
&&\bar \xi^{\rm sp}_{\pm}(t)=\bar \xi^{\rm sp}_{0\pm}|t|^{-\nu},\;\; \bar \xi^{\rm sp}_{0\pm}=\big[\prod^d_{\alpha = 1} \xi_{0\pm}^{{\rm sp}(\alpha)}\big]^{1/d},\\
\label{barxiratio}
&&\bar \xi^{\rm sp}_{0+}/\bar \xi^{\rm sp}_{0-}= X_\xi.
\end{eqnarray}
In the ratios (\ref{lambdaalphaxixyy}) the dependence on the free parameters $\xi^{\rm iso}_{0\pm}$ is canceled.
Similar to (\ref{ratiomeaniso}) and (\ref{lambdaredlambda}), we have the relations
\begin{eqnarray}
\label{ratiomeanisox}
\det {\mbox {\boldmath$\widehat\lambda$}}= (\bar \xi^{\rm sp}_{0\pm}/ \xi^{\rm iso}_{0\pm})^{2d}
\end{eqnarray}
and
\begin{eqnarray}
\label{lambdaredlambdaspx}
{\mbox {\boldmath$\widehat\lambda$}}^{1/2} \xi^{\rm iso}_{0\pm}&=&\big({\mbox {\boldmath$\bar\lambda$}}^{\rm sp}\big)^{1/2}\bar \xi^{\rm sp}_{0\pm}.
\end{eqnarray}
Finally we obtain the reduced anisotropy matrix via the definition
\begin{eqnarray}
\label{Aquerxx}
{\bf \bar A}^{\rm sp}&=&\widehat{\bf U}^{-1}{\bf \bar{\mbox {\boldmath$\lambda$}}}^{\rm sp} \widehat{\bf U}\\
&=&{\bf \bar A}\big(\{\xi_{0\pm}^{{\rm sp}(\alpha)},{\bf e}^{{\rm sp}(\alpha)}\}\big)\\
\label{Aquerxxyy}
&=&{\bf U(\{{\bf e}^{{\rm sp}(\alpha)}\})}^{-1}{\bf \bar{\mbox {\boldmath$\lambda$}}}^{\rm sp} \big(\{\xi_{0 \pm}^{{\rm sp}(\alpha)}\}\big) {\bf U(\{{\bf e}^{{\rm sp}(\alpha)}\})}\;\;\;\;\;
\end{eqnarray}
where ${\bf \bar A}^{\rm sp}$ has the same structure as the matrix ${\bf \bar A}$ in (\ref{barAxyy}) for the $\varphi^4$ model but with the arguments
$\xi_{0 \pm}^{(\alpha)}$ and ${\bf e}^{(\alpha)}$
 replaced by $\xi_{0 \pm}^{{\rm sp}(\alpha)}$ and ${\bf e}^{{\rm sp}(\alpha)}$.
 This can be extended to any weakly anisotropic system beyond the $\varphi^4$ theory without explicit knowledge of an anisotropy matrix ${\bf A}$ as a function of the couplings.

This means that we have identified a temperature-independent universal structure of the reduced anisotropy matrix ${\bf \bar A}(\{\xi_{0\pm}^{(\alpha)},{\bf e}^{(\alpha)}\}\big)$ whose dependence either on the nonuniversal ratios $\xi_{0\pm}^{(\alpha)}/\xi_{0\pm}^{(\beta)}$ and nonuniversal principal unit vectors ${\bf e}^{(\alpha)}$ or on $\xi_{0\pm}^{{\rm sp}(\alpha)}/\xi_{0\pm}^{{\rm sp}(\beta)}$ and ${\bf e}^{{\rm sp}(\alpha)}$ has a universal functional form that is the same for all weakly anisotropic systems including the $\varphi^4$ model and the $n$-vector model. We have shown that for the construction of this reduced anisotropy matrix the existence and knowledge of an anisotropy matrix  ${\bf A}$ is not required, thus our construction of the structure of  ${\bf \bar A}$ is applicable to any weakly anisotropic system. This is an important ingredient for the general validity of multiparameter universality of weakly anisotropic systems.
\renewcommand{\thesection}{\Roman{section}}
\renewcommand{\theequation}{5.\arabic{equation}}
\setcounter{equation}{0}
\section{Multiparameter universality of the anisotropic
correlation function}
Besides the bulk free energy density the most fundamental physical quantity describing the critical fluctuations is the bulk order-parameter correlation function.
However, in the traditional theory of critical phenomena  \cite{fish-1,bre-1,pelissetto,priv,pri,hohenberg1976,cardybuch,zinn2007} including the nonperturbative functional renormalization group \cite{metzner2021,hassel2007,sinner2008} the issue of the general universality properties of weakly anisotropic correlation functions was not addressed. Only special anisotropic
models were studied whose bulk correlation functions \cite{bruce,Vaidya1976,WuCoy,CoyWu,Perk1,Perk2,Perk3,Perk4} were presented in a nonuniversal form.
Recently a general scaling form of the anisotropic bulk correlation function of the $O(n)$-symmetric $d$-dimensional $\varphi^4$ model has been presented \cite{dohm2018,dohm2019} in terms of a reduced anisotropy matrix
${\bf \bar A}$
which was found to violate two-scale-factor universality in both two and three dimensions.
Instead, within the $\varphi^4$ theory, these results exhibit the feature of multiparameter
universality \cite{dohm2018}
and it was hypothesized that this structure is valid quite generally for weakly anisotropic systems beyond $\varphi^4$ models.
In this section we prove the validity of this hypothesis for the bulk correlation function and point to the intrinsic diversity of this scaling structure.
\subsection{Anisotropic correlation function
of the $\varphi^4$ theory}
In \cite{dohm2008,dohm2018} the representation (\ref{3c}) of the isotropic correlation function was used  in order to derive the anisotropic correlation function $G_\pm({\bf x}, t)$ of the $\varphi^4$ theory in the form presented in Eq. (1.6) of \cite{dohm2018}.
In the following we shall work with the alternative representation (\ref{3calt}) and its Fourier transform (\ref{FourierG}) and present a more transparent derivation. Application of (\ref{3calt}) to the isotropic correlation function $G'_\pm(|{\bf x'}|, t)$ of the transformed isotropic Hamiltonian (\ref{2zxx}) yields
\begin{equation}
\label{3caltstrich} G'_\pm(|{\bf x'}|, t) = \frac{\Gamma'_+(\xi'_{0+})^{-2+\eta}}{ | {\bf x'} |^{d-2+\eta}} \;\Psi_\pm \Big(\frac{|{\bf x'}|}{ \xi'_{\pm}(t)} \Big) \;
\end{equation}
which can be rewritten as
\begin{equation}
\label{3caltstrichrewritten} G'_\pm(|{\bf x'}|, t) = \frac{\Gamma'_+}{(\xi'_{0+})^d} \Big(\frac{\xi'_{0+}}{ | {\bf x'} |} \Big)^{d-2+\eta} \;\Psi_\pm \Big(\frac{|{\bf x'}|}{ \xi'_{\pm}(t)} \Big) \; .
\end{equation}
Now we derive the anisotropic correlation function from  the isotropic correlation function (\ref{3caltstrichrewritten}) by inverting (\ref{isocorrfcor}),
\begin{eqnarray}
\label{Ginverted}
 &&G_\pm({\bf x},t)= \big(\xi'_{0\pm}/\bar \xi_{0\pm}\big)^d\;G_\pm'(|{\bf x'}|,t),
\end{eqnarray}
and expressing $|{\bf x'}|$ in terms of ${\bf x}$. Using the inverse of (\ref{AquerUlambdaU})
  \be
  \label{inverseAquer}
  {\mbox {\boldmath$\bar\lambda$}}^{-1}={\bf  U}{\bf \bar A}^{-1}{\bf  U}^{-1}
  \ee
we first rewrite $|{\bf x'}|$ as
\begin{eqnarray}
\label{trasformations}
&&|{\bf x'}|
    =\big[ {\bf x'}\cdot {\mbox {\boldmath$\bar\lambda$}}^{1/2}{\mbox {\boldmath$\bar\lambda$}}^{-1}{\mbox {\boldmath$\bar\lambda$}}^{1/2} {\bf x'}\big]^{1/2}\\
    &&=\big[ {\bf x'}\cdot {\mbox {\boldmath$\bar\lambda$}}^{1/2}{\bf  U}{\bf \bar A}^{-1}{\bf U}^{-1}{\mbox {\boldmath$\bar\lambda$}}^{1/2} {\bf x'}\big]^{1/2}\\
 &&=\big[({\mbox {\boldmath$\bar\lambda$}}^{1/2}{\bf U})^T {\bf x'}\cdot {\bf \bar A}^{-1}{\bf  U}^{-1}{\mbox {\boldmath$\bar\lambda$}}^{1/2} {\bf x'}\big]^{1/2}\;\;\;\;
 \\
 &&=\big[{\bf  U}^{-1}{\mbox {\boldmath$\bar\lambda$}}^{1/2} {\bf x'}\cdot {\bf \bar A}^{-1}{\bf  U}^{-1}{\mbox {\boldmath$\bar\lambda$}}^{1/2} {\bf x'}\big]^{1/2}\;\;\;\;
 \end{eqnarray}
 where we have used
\be
\label{transponiert}
({\mbox {\boldmath$\bar\lambda$}}^{1/2}{\bf U})^T= {\bf U}^T({\mbox {\boldmath$\bar\lambda$}}^{1/2})^T={\bf U}^{-1}{\mbox {\boldmath$\bar\lambda$}}^{1/2}.
\ee
 Then we express  ${\mbox {\boldmath$\bar\lambda$}}^{1/2}$ in terms of  ${\mbox {\boldmath$\lambda$}}^{1/2}$ according to (\ref{lambdaredlambdaspecial}),
\begin{eqnarray}
\label{lambdaredlambdaspecialnew}
{\mbox {\boldmath$\bar\lambda$}}^{1/2}=\frac{\xi'_{0\pm}}{\bar \xi_{0\pm}}\;{\mbox {\boldmath$\lambda$}}^{1/2}
\end{eqnarray}
which yields the identity
\begin{eqnarray}
|{\bf  x'}|&=&\frac{\xi'_{0\pm}}{\bar \xi_{0\pm}}\;[{\bf  U}^{-1}{\mbox {\boldmath$\lambda$}}^{1/2} {\bf  x'}\cdot {\bf \bar A}^{-1}{\bf  U}^{-1}{\mbox {\boldmath$\lambda$}}^{1/2} {\bf  x'}]^{1/2}.\;\;\;\;\;\;\;\;\;\;
 \end{eqnarray}
From the original shear transformation (\ref{shearalt}) we obtain the inverse shear transformation
\be
\label{inverseprime}
 {\bf U}^{-1}{\mbox {\boldmath$\lambda$}}^{1/2} {\bf  x'}= {\bf x}.
\ee
This leads to the relation
 \begin{eqnarray}
  \label{scalingarg}
 \frac{{\bf  x'}^2}{\big(\xi'_{0\pm}\big)^2}&=&
\frac{{\bf x}\cdot {\bf \bar A}^{-1}{\bf x}}{\big(\bar \xi_{0\pm}\big)^2}.
 \end{eqnarray}
Here the representation (\ref{barAxyy}) of the reduced anisotropy matrix ${\bf \bar A}$ is to be inserted as a function of the principal unit vectors and principal correlation lengths. Eq. (\ref{scalingarg}) describes how the isotropic structure is transferred to the anisotropic structure by means of the inverse shear transformation.
Using (\ref{Ginverted}) together with the invariance (\ref{Gamma}) and substituting  (\ref{scalingarg}) into  (\ref{3caltstrichrewritten}) we arrive at the anisotropic  bulk order-parameter correlation function above, at, and below $T_c$ of the $\varphi^4$ model for $2\leq d < 4$
\begin{eqnarray}
 &&G_\pm({\bf x},t)
\label{Gphi4aniso}
=\frac{\Gamma_+(\bar \xi_{0+})^{-2+\eta}}{ ({\bf x}\cdot {\bf \bar A}^{-1}{\bf x})^{(d-2+\eta)/2}} \Psi_\pm \Big(\frac{[{\bf x}\cdot {\bf \bar A}^{-1}{\bf x}]^{1/2}}{\bar \xi_\pm(t)}\Big)\;\;\;\;\;\;\;\;\;
 \end{eqnarray}
with the universal scaling function $ \Psi_\pm $, (\ref{3d}), of the isotropic system. At $T_c$ we obtain
\begin{eqnarray}
\label{GneuxxaltTcphi}
&&G_\pm({\bf x}, 0)
=\widetilde Q_3\frac{\Gamma_+\big(\bar \xi_{0+}\big)^{-2+\eta}}{ ({\bf x}\cdot {\bf \bar A}^{-1}{\bf x})^{(d-2+\eta)/2}} \;\;\;\;\;\;\;\;\;
 \end{eqnarray}
with the universal constant $\widetilde Q_3$ defined in (\ref{tildeQ}). These results are equivalent to the results of the  $\varphi^4$ theory in Eqs. (1.6) and (5.32) of \cite{dohm2018}.

An analogous anisotropic representation can be derived in Fourier space based on the shear transformation (\ref{sheark}).
Application of (\ref{FourierG}) to the Fourier transformed isotropic correlation function $\hat G'_\pm (|{\bf k'}|, t)$ of the transformed isotropic Hamiltonian (\ref{2zxx}) yields
\begin{eqnarray}
\label{3dak}\hat G'_\pm (|{\bf k'}|, t)
 &=& \frac{\Gamma'_+}{ \big(| {\bf k'} |\;\xi'_{0+}\big)^{2-\eta}} \;\hat \Psi_\pm \Big( | {\bf k'} |\xi'_{\pm}(t) \Big),\\
\label{atTcxiso} \hat G' (|{\bf k}|, 0) &=& Q_3\frac{\Gamma'_+}{\big(|{\bf k}|\;\xi'_{0+}\big)^{2-\eta}}\;.\;
\end{eqnarray}
Because of the sum rule (\ref{susceptx}) we have
\begin{eqnarray}
\label{susceptk}
\lim_ {| {\bf k'} |\to 0}\hat G'_\pm ( | {\bf k'} |, t)= \chi'_\pm(t)=\Gamma'_\pm |t|^{-\gamma}.
\end{eqnarray}
Using (\ref{AquerUlambdaU}) in the form
  \be
  \label{inverseAquer}
  {\mbox {\boldmath$\bar\lambda$}}={\bf  U}{\bf \bar A}{\bf  U}^{-1}
  \ee
we rewrite $|{\bf k'}|$ in a way analogous to $|{\bf x'}|$ and obtain the relation
\begin{eqnarray}
\label{quadratick}
|{\bf k}'|^2(\xi_{0\pm}')^{2}&=& ({\bf k}\cdot \;{\bf \bar A}{\bf k})(\bar \xi_{0\pm})^{2}.
\end{eqnarray}
Using the invariance (\ref{transshearFourier}) together with the invariance (\ref{Gamma}) and substituting  (\ref{quadratick}) into  (\ref{3dak}) we obtain the anisotropic correlation function of the $\varphi^4$ model in ${\bf k}$ space
\begin{eqnarray}
\label{Corrk}\hat G_\pm ({\bf k}, t)
 &=& \Gamma_+\;\frac{\hat \Psi_\pm \Big(  [{\bf k}\cdot \;{\bf \bar A}{\bf k}]^{1/2} \bar \xi_{\pm}(t) \Big)}{ \big( [{\bf k}\cdot \;{\bf \bar A}{\bf k}]^{1/2} \;\bar \xi_{0+}\big)^{2-\eta}} \;
\end{eqnarray}
with
\begin{eqnarray}
\label{atTcx} \hat G_\pm  ({\bf k}, 0) = Q_3\frac{ \Gamma_+ }{\big([{\bf k}\cdot \;{\bf \bar A}{\bf k}]^{1/2}\bar \xi_{0+}\big)^{2-\eta}}\;\;
\end{eqnarray}
at $T_c$, with the universal constant $Q_3$, (\ref{Qdrei}). A discussion of these results for the $\varphi^4$ theory is given in Sec. V.B. in the context of analogous results for the $n$-vector model.
\subsection{Proof of multiparameter universality  of the anisotropic correlation function}
Our proof is formulated for the anistropic $n$-vector model as an example for a system other than the anisotropic $\varphi^4$ model. It is based on the following properties:
(a) The universal structure of the scaling form (\ref{3c}) or (\ref{3calt})  of the isotropic correlation function,
(b) the general applicability of the generalized shear transformations (\ref{trans}) or (\ref{transyy}) to any weakly anisotropic system including the $n$-vector model,
(c) the universality of the structure of the reduced anisotropy matrix ${\bf \bar A}^{\rm sp} $, (\ref{Aquerxxyy}),
(d) the sum rule for the anisotropic correlation function near $T_c$
\begin{eqnarray}
\label{sumrulexsp}
\chi_\pm^{\rm sp}(t)=\int d^d{\bf x}\;\;G_\pm^{\rm sp} ({\bf x}, t)= \Gamma^{\rm sp}_\pm|t|^{-\gamma}
\end{eqnarray}
where $\Gamma^{\rm sp}_\pm$ is the nonuniversal critical amplitude of the susceptibility above and below $T_c$, respectively, and
(e) the invariance of the bulk correlation function $G_\pm^{\rm sp}({\bf x}, t)$ under our shear transformation.

The latter issue is due to the fact that, unlike the special shear transformation of the $\varphi^4$ theory \cite{cd2004,dohm2008}
where the transformation (\ref{shearphi})
 of the variable $\varphi$ is involved, our generalized shear transformation
(\ref{transyy}) transforms only the spatial coordinates ${\bf x} \to  {\bf \widehat x}$ of the lattice points without changing the topology of the interactions.
As discussed in Sec. II. B, this leaves the amplitude of the order-parameter correlation function $G_\pm^{\rm sp}({\bf x}, t)$ invariant. Thus our generalized shear transformation implies that the special transformations (\ref{isocorrf}), (\ref{isocorrfcor}), and  (\ref{Ginverted}) are replaced by the invariance
\begin{eqnarray}
\label{Gplussp}
G_\pm^{\rm sp}({\bf x}, t)
= G_\pm^{\rm sp,iso}(|{\bf \widehat x}|,t)
= G_\pm^{\rm sp, iso}(|{\mbox {\boldmath$\widehat\lambda$}}^{-1/2} {\bf \widehat U}{\bf x}|, t)
\end{eqnarray}
for the $n$-vector model where the transformed isotropic correlation function $ G_\pm^{\rm sp, iso}$ has the isotropic structure given in (\ref{3calt}). The susceptibility, however, is not invariant under the generalized shear transformation.
Eq. (\ref{sumrulexsp})  can be combined  with the sum rule
of the isotropic system  and with (\ref{transyy}), (\ref{ratiomeanisox}), and (\ref{Gplussp}) to obtain the susceptibility of the isotropic system as
\begin{eqnarray}
\label{relationchi}
\chi_\pm^{\rm sp, iso}(t)&=&\int d^d{\bf \widehat x}\;\; G_\pm^{\rm sp, iso} (|{\bf \widehat x}|, t)= \Gamma^{\rm sp, iso}_\pm\; |t|^{-\gamma}\;\;\;\;\;\;\\
&=& (\det {\mbox {\boldmath$\widehat\lambda$}})^{-1/2}\int d^d{\bf x}\;\;G_\pm^{\rm sp} ({\bf x}, t)\\
&=&(\xi^{\rm iso}_{0\pm}/\bar \xi^{\rm sp}_{0\pm})^{d}\chi_\pm^{\rm sp}(t)
\end{eqnarray}
where $\bar \xi^{\rm sp}_{0\pm}$ is the amplitude of the mean correlation length.
This implies the relation between the amplitudes of the susceptibilities of the isotropic and anisotropic systems
\be
\label{susdetermined}
\Gamma^{\rm sp, iso}_\pm=\Big(\xi^{\rm iso}_{0\pm}/\bar \xi^{\rm sp}_{0\pm}\Big)^{d}\;\Gamma_\pm^{\rm sp}
\ee
which can be interpreted as the invariance of the ratio of the correlation volume and the susceptibility amplitude
\be
\label{invarratio}
\big(\xi^{\rm iso}_{0\pm}\big)^d/ \Gamma^{\rm sp,iso}_\pm=\;\big(\bar \xi^{\rm sp}_{0\pm}\big)^d/\Gamma_\pm^{\rm sp}
\ee
under the generalized shear transformation. We note
that also the combination
\begin{eqnarray}
\label{invarcombi}
&&\big(\xi^{\rm iso}_{0\pm}\big)^d G_\pm^{\rm sp, iso}(|{\bf \widehat x}|,t)/ \Gamma^{\rm sp, iso}_\pm
=\;\big(\bar \xi^{\rm sp}_{0\pm}\big)^d G_\pm^{\rm sp}({\bf x}, t)/\Gamma_\pm^{\rm sp}\;\;\;\;\;\;\;\;\;\;
\end{eqnarray}
remains invariant which shows the universality of the same feature (\ref{invarianceG}) of the special shear transformation of the $\varphi^4$ theory.
From (\ref{3calt}) we have the representation of the isotropic bulk correlation function
\begin{equation}
\label{3caltx} G_\pm^{\rm sp, iso} (|{\bf \widehat x}|, t) = \frac{\Gamma_+^{\rm sp, iso}\big( \xi^{\rm iso}_{0+}\big)^{-2+\eta}}{ | {\bf \widehat x} |^{(d-2+ \eta)/2}} \;\Psi_\pm \Big(\frac{|{\bf \widehat x}|}{ \xi^{\rm iso}_{\pm}(t)} \Big) .\;
\end{equation}
or
\begin{equation}
\label{3caltxyz}  G_\pm^{\rm sp, iso}({\bf \widehat x}, t) = \frac{\Gamma_+^{\rm sp, iso}}{\big( \xi^{\rm iso}_{0+}\big)^{d}}\Bigg(\frac{
\xi_{0+}^{\rm iso}  }{ | {\bf \widehat x} |}\Bigg)^{d-2+ \eta} \;\Psi_\pm \Big(\frac{|{\bf \widehat x}|}{ \xi^{\rm iso}_{\pm}(t)} \Big) .\;
\end{equation}
The derivation of the anisotropic bulk correlation function $G_\pm^{\rm sp}({\bf x}, t)$, (\ref{Gplussp}) from (\ref{3caltxyz}) is analogous to that in (\ref{3caltstrichrewritten})-(\ref{Gphi4aniso}) for the $\varphi^4$ model. Thus we rewrite
\begin{eqnarray}
\label{trasformations}
&&|{\bf \widehat x}|
    =\big[ {\bf \widehat  x}\cdot \big({\mbox {\boldmath$\bar\lambda$}}^{\rm sp}\big)^{1/2}\big({\mbox {\boldmath$\bar\lambda$}}^{\rm sp}\big)^{-1}\big({\mbox {\boldmath$\bar\lambda$}}^{\rm sp}\big)^{1/2} {\bf \widehat  x}\big]^{1/2}\\
    &&=\big[ {\bf \widehat x}\cdot \big({\mbox {\boldmath$\bar\lambda$}}^{\rm sp}\big)^{1/2}{\bf \widehat U}\big({\bf \bar A}^{\rm sp}\big)^{-1}{\bf \widehat U}^{-1}\big({\mbox {\boldmath$\bar\lambda$}}^{\rm sp}\big)^{1/2} {\bf \widehat x}\big]^{1/2}\\
 &&=\big[{\bf \widehat U}^{-1}\big({\mbox {\boldmath$\bar\lambda$}}^{\rm sp}\big)^{1/2} {\bf \widehat x}\cdot \big({\bf \bar A}^{\rm sp}\big)^{-1}{\bf \widehat U}^{-1}\big({\mbox {\boldmath$\bar\lambda$}}^{\rm sp}\big)^{1/2} {\bf \widehat x}\big]^{1/2}\;\;\;\;\\
 \label{xwidehatxx}
 &&=\frac{\xi^{\rm iso}_{0\pm}}{\bar \xi^{\rm sp}_{0\pm}}\;[{\bf \widehat U}^{-1}{\mbox {\boldmath$\widehat\lambda$}}^{1/2} {\bf \widehat x}\cdot \big({\bf \bar A}^{\rm sp}\big)^{-1}{\bf \widehat U}^{-1}{\mbox {\boldmath$\widehat\lambda$}}^{1/2} {\bf \widehat x}]^{1/2}
 \end{eqnarray}
 where we have used
 \be
 \label{inverseAquersp}
 \big({\mbox {\boldmath$\bar\lambda$}}^{\rm sp}\big)^{-1}={\bf \widehat U}\big({\bf \bar A}^{\rm sp}\big)^{-1}{\bf \widehat U}^{-1}
 \ee
  which is the inverse of (\ref{Aquerxx}), and
\begin{eqnarray}
\label{lambdaredlambdaspecialnew}
\big({\mbox {\boldmath$\bar\lambda$}}^{\rm sp}\big)^{1/2}=\frac{\xi^{\rm iso}_{0\pm}}{\bar \xi^{\rm sp}_{0\pm}}\;{\mbox {\boldmath$\widehat\lambda$}}^{1/2}
\end{eqnarray}
according to (\ref{lambdaredlambdaspx}).
 Finally we employ the inverse of the transformation (\ref{transyy}),
\be
\label{inversewidehat}
 {\bf \widehat U}^{-1}{\mbox {\boldmath$\widehat\lambda$}}^{1/2} {\bf \widehat x}= {\bf x}
\ee
and obtain from (\ref{xwidehatxx}) the relation
 \begin{eqnarray}
 \label{xwidehatbetrag}
 \frac{|{\bf \widehat x}|}{\xi^{\text {iso}}_{0\pm}}
 &=&\frac{[{\bf x}\cdot \big({\bf \bar A}^{\rm sp}\big)^{-1}{\bf x}]^{1/2}}{\bar \xi^{\rm sp}_{0\pm}}
\;,
 \end{eqnarray}
compare (\ref{scalingarg}). A similar transformation holds for
\be
\label{transfactor}
\frac{
\xi^{\rm iso}_{\rm T}(t)}{ | {\bf \widehat x} |}
=\frac{ \bar \xi^{\rm sp}_{\rm T}(t)}{ \big[{\bf x}\cdot \big({\bf \bar A}^{\rm sp}\big)^{-1}{\bf x}\big]^{1/2}}
\ee
in (\ref{trans correl large}).
Together with (\ref{susdetermined})
this leads to the scaling form of the anisotropic  bulk order-parameter correlation function (\ref{corrfctsp}) above, at, and below $T_c$ of the $n$-vector model (\ref{Hspin}) for $2\leq d < 4$
\begin{eqnarray}
&&G_\pm^{\rm sp}({\bf x}, t)=\nonumber
\\
\label{Gneuxxalt}
&&\frac{\Gamma_+^{\rm sp}\big(\bar \xi^{\rm sp}_{0+}\big)^{-2+\eta}}{ \big[{\bf x}\cdot \big({\bf \bar A}^{\rm sp}\big)^{-1}{\bf x}\big]^{(d-2+\eta)/2}} \Psi_\pm \Bigg(\frac{\big[{\bf x}\cdot \big({\bf \bar A}^{\rm sp}\big)^{-1}{\bf x}\big]^{1/2}}{\bar \xi^{\rm sp}_\pm(t)}\Bigg)\;\;\;\;\;\;\;\;\;
 \end{eqnarray}
with the universal scaling function $ \Psi_\pm $, (\ref{3d}), of the
isotropic system. Here ${\bf \bar A}^{\rm sp}$ is to be inserted in the form of (\ref{Aquerxxyy}) as a function of the principal unit vectors and principal correlation lengths.
Right at $T_c$ we have the purely algebraic behavior
\begin{eqnarray}
\label{GneuxxaltTc}
&&G_\pm^{\rm sp}({\bf x}, 0)
=\widetilde Q_3\frac{\Gamma_+^{\rm sp}\big(\bar \xi^{\rm sp}_{0+}\big)^{-2+\eta}}{ \big[{\bf x}\cdot \big({\bf \bar A}^{\rm sp}\big)^{-1}{\bf x}\big]^{(d-2+\eta)/2}} .\;\;\;\;\;\;\;\;\;
 \end{eqnarray}
Eqs.  (\ref{Gneuxxalt}) and (\ref{GneuxxaltTc}) are the anisotropy-dependent generalizations of Eqs. (\ref{3calt}) and (\ref{3caltTc}). Similarly we obtain the anisotropic counterpart of the transverse correlation function (\ref{trans correl large}) below $T_c$ for $n>1,d>2$
 \begin{eqnarray}
\label{Ganisotrans}
G^{\rm sp}_{\rm T} ({\bf x}, t) &=& {\cal C}_{\rm T}  [{\cal M}^{\rm sp}(t)]^2
\Bigg(\frac{ \bar \xi^{\rm sp}_{\rm T}(t)}{ \big[{\bf x}\cdot \big({\bf \bar A}^{\rm sp}\big)^{-1}{\bf x}\big]^{1/2}}\Bigg)^{d-2}\;\;\;\;
 \end{eqnarray}
where ${\cal M}^{\rm sp}(t)={\cal M}^{\rm iso}(t)$ is invariant under the generalized
shear transformation
and where ${\bf \bar A}^{\rm sp}$ is represented as a function of
$\xi^{{\rm sp}(\alpha)}_{0\rm T}/\xi^{{\rm sp}(\beta)}_{0\rm T}$ and ${\bf e}^{{\rm sp} (\alpha)}$.
In a similar way we obtain from (\ref{3c}), with $D^{\rm iso}_1$ replaced by
\be
\label{D1}
\widehat D_1=  \Big(\xi^{\rm iso}_{0+}/\bar \xi^{\rm sp}_{0+}\Big)^{d-2+\eta}\;D^{\rm sp}_1,
\ee
the representation above, at, and below $T_c$
\begin{eqnarray}
\label{Gneuxx}
&&G_\pm^{\rm sp}({\bf x}, t) = \nonumber\\
&&\frac{D_1^{\rm sp}}{ [{\bf x}\cdot \big({\bf \bar A}^{\rm sp}\big)^{-1}{\bf x}]^{(d-2+\eta)/2}}\;
 \Phi_\pm \Bigg(\frac{\big[{\bf x}\cdot \big({\bf \bar A}^{\rm sp}\big)^{-1}{\bf x}\big]^{1/2}}{\bar \xi^{\rm sp}_\pm(t)}\Bigg)\;\;\;\;\;\;\;\;\;
\end{eqnarray}
where the universal scaling function $ \Phi_\pm $ is the same as in (\ref{3c}) for the isotropic system.

The analysis in Fourier space is parallel to that in Sec. V. A. Defining the generalized shear transformation in ${\bf k}$ space [compare (\ref{sheark})]
\begin{eqnarray}
\label{ktransyy}
{\bf \widehat k}= {\mbox {\boldmath$\widehat\lambda$}}^{1/2} {\bf \widehat U}{\bf k},
\end{eqnarray}
and the Fourier transforms
\begin{eqnarray}
\label{3daspiso}  \hat G^{\rm sp, iso}_\pm (|{\bf \widehat k}|, t)&=&\int d^d {\bf \widehat x}\;e^{- i {\bf \widehat k} \cdot {\bf \widehat x} }G^{\rm sp, iso}_\pm (|{\bf \widehat x}|, t),\\
\label{3dasp} \hat G^{\rm sp}_\pm ({\bf k}, t)&=&\int d^d {\bf x}\;e^{- i {\bf k} \cdot {\bf x} }G^{\rm sp}_\pm ({\bf x}, t),
\end{eqnarray}
with ${\bf \widehat k} \cdot {\bf \widehat x}={\bf k} \cdot {\bf x}$ and
\begin{eqnarray}
\hat G^{\rm sp}_\pm ({\bf k}, t)&=& (\det {\mbox {\boldmath$\widehat\lambda$}})^{1/2}\hat G^{\rm sp, iso}_\pm (|{\bf \widehat k}|, t),\\
\label{isoanisorelation}
|{\bf \widehat k}|^2(\xi_{0\pm}^{\rm iso})^{2}&=& ({\bf k}\cdot \;{\bf \bar A^{\rm sp}}{\bf k})(\bar \xi_{0\pm}^{\rm sp})^{2},
\end{eqnarray}
we obtain the isotropic and anisotropic correlation function in ${\bf k}$ space for the $n$-vector model
\begin{eqnarray}
\label{isoCorrksp}\hat G_\pm^{\rm sp, iso} (|{\bf \widehat k}|, t)
 &=& \Gamma^{\rm sp, iso}_+\;\frac{ \hat \Psi_\pm \Big(  |{\bf \widehat k}|\xi^{\rm sp, iso}_{\pm}(t) \Big)}{ \big( |{\bf \widehat k}|\xi^{\rm sp, iso}_{0,+}\big)^{2-\eta}} \;,
\end{eqnarray}
\begin{eqnarray}
\label{Corrkgeneral}\hat G_\pm^{\rm sp} ({\bf k}, t)
 &=& \Gamma^{\rm sp}_+\;\frac{\hat \Psi_\pm \Big(  [{\bf k}\cdot \;{\bf \bar A}^{\rm sp}{\bf k}]^{1/2} \bar \xi^{\rm sp}_{\pm}(t) \Big)}{ \big( \big[{\bf k}\cdot \;{\bf \bar A}^{\rm sp}{\bf k}\big]^{1/2} \;\bar \xi^{\rm sp}_{0+}\big)^{2-\eta}}, \;
\end{eqnarray}
where $\hat \Psi_\pm$ is the same function as in (\ref{FourierG}) and (\ref{3dak}),
with
\begin{eqnarray}
\label{atTcxgeneral} \hat G^{\rm sp} _\pm  ({\bf k}, 0) = Q_3 \frac{ \Gamma^{\rm sp}_+ }{\big(\big[{\bf k}\cdot \;{\bf \bar A}^{\rm sp}{\bf k}\big]^{1/2}\bar \xi^{\rm sp}_{0+}\big)^{2-\eta}}\;\;
\end{eqnarray}
at $T_c$, with the same universal constant $Q_3$ as in (\ref{atTcx}).

 The effect of the inverse shear transformations on the arguments of the correlation functions is condensed into the exact relations (\ref{scalingarg}), (\ref{quadratick}), (\ref{xwidehatbetrag}), and (\ref{isoanisorelation}) which describe how the isotropic structure is transferred to the anisotropic structure.
No specific properties of the $n$-vector model have been used, thus the derivation given above
can be extended to any weakly anisotropic system beyond the $\varphi^4$ theory without explicit knowledge of an anisotropy matrix ${\bf A}$ as a function of the couplings.
Our results (\ref{Gneuxxalt}), (\ref{Ganisotrans}), (\ref{Gneuxx}), and  (\ref{Corrkgeneral}) for the anisotropic $n$-vector model have the same form as derived for the anisotropic $\varphi^4$ model given in (\ref{Gphi4aniso}) and (\ref{Corrk}) and in Eqs. (1.6) and (5.42) of \cite{dohm2018}, with the same critical exponents $\nu$ and $\eta$, the same universal structure of the reduced anisotropy matrix ${\bf \bar A}$ or ${\bf \bar A}^{\rm sp}$, the same universal scaling functions $\Psi_\pm$ (or $\Phi_\pm$) and $\hat \Psi_\pm$, and with the same universal constants $Q_3$ and $\widetilde Q_3$ for a given $(d,n)$ universality class. This proves the validity of multiparameter universality for the anisotropic bulk correlation function for general $d$ and $n$ with up to $d(d+1)/2+1$ independent nonuniversal parameters. This is the central result of this section. Here we have not made any assumptions other than the validity of two-scale-factor universality for isotropic systems and the existence of principal correlation lengths and principal axes for weakly anisotropic systems together with (\ref{ratioalpha}). Our derivation cannot, of course, make any prediction about the dependence of the nonuniversal parameters on the couplings $E_{i,j}$ of the $n$-vector model, unlike the dependence on the couplings $K_{i,j}$ within the $\varphi^4$ model \cite{dohm2019,dohm2018}.

Among the $d(d+1)/2+1$ nonuniversal parameters of weakly anisotropic systems there are only two independent parameters that can be determined by thermodynamic measurements, namely the amplitude $\Gamma^{\rm sp}_+$ of the susceptibility and the amplitude $\bar \xi_{0+}$ of the mean correlation length which enters the amplitude of the specific heat above $T_c$, as will be shown in Sec. VI. B. While two-scale-factor universality implies that isotropic correlation functions can be expressed in terms of purely thermodynamic amplitudes this property is destroyed by weak anisotropy. The remaining $d(d+1)/2-1$ independent nonuniversal parameters $\xi_{0\pm}^{{\rm sp}(\alpha)}/\xi_{0\pm}^{{\rm sp}(\beta)}$ or $\xi_{0{\rm T}}^{{\rm sp}(\alpha)}/\xi_{0{\rm T}}^{{\rm sp}(\beta)}$ and ${\bf e}^{{\rm sp}(\alpha)}$  are contained in the reduced anisotropy matrix ${\bf \bar A}^{\rm sp}$. Although this matrix has a universal structure in terms of ratios of principal correlation lengths and in terms of principal angles the latter are nonuniversal quantities describing
the nonuniversal angular dependence of the critical correlations.  Thus there exists a high degree of intrinsic diversity in the asymptotic critical region of weakly anisotropic systems with up to five intrinsic parameters in three dimensions.
On the experimental side, the determination of these parameters requires spatially resolved scattering measurements.
On the theoretical side, it has not been widely recognized in the literature \cite{cardy1987,cardy1983,cardybuch,Indekeu,barber1984,zia,diehl-chamati,diehl2010}
that in most cases the directions of the principal axes depend in a generically unknown way on the anisotropic couplings \cite{dohm2019,night1983}. Thus, in practice, it is by no means simple to identify the appropriate principal axes of anisotropic two- and three-dimensional $n$-vector models or of real systems before the invoked anisotropic scale transformations can be performed.
For example, even for the two-dimensional anisotropic Ising model the principal axes and correlation lengths are known only in a few special cases. We shall substantiate our findings in the next sections by exact results in the spherical and Gaussian universality classes and by an approximate result derived from the FRG for $d=3,n=1$ \cite{hassel2007,sinner2008}
as well as in Sec. VII by exact results for the $d=2$ Ising universality class.
\subsection{Exact anisotropic correlation function in the large-${\bf n}$ limit}
Exact analytic results can be derived for systems belonging to the spherical universality class
which includes the $O(n)$-symmetric $\varphi^4$ model in the large-$n$ limit, the $n$-vector model in the large-$n$ limit,
the spherical model, and the mean spherical model \cite{berlin-kac,joyce,brankov,Baxter1982,stanley1968}. So far the exact scaling form of the anisotropic bulk correlation function of these models has not been given in the literature. As a representative of this universality class
we take the large-$n$ limit of the continuum version of the $\varphi^4$ model (\ref{contin}). We consider the anisotropic bulk correlation function per component $G_\infty$ for $T\geq T_c$ in the large-$n$ limit at fixed $u_0n$ \cite{cd2000-2,cd1998}
\begin{eqnarray}
\label{largen}
&&G_\infty({\bf x},t)=\lim_{n\to\infty}\frac{1}{n}<\varphi({\bf x})\varphi'({\bf 0})>\\
\label{integr}
&&= \int _{\bf k} \;\hat G_\infty({\bf k},t)e^{i{\bf k}\cdot{\bf x}}.
\end{eqnarray}
After the shear transformation (\ref{shearalt})-(\ref{shearu})
the isotropic $\varphi^4$ Hamiltonian (\ref{2zxx}) and
the correlation functions
\begin{eqnarray}
\label{isocorrfninftyz}
G'_\infty ({\bf x'}, t)&=& (\det {\bf A})^{1/2} G_\infty({\bf x},t),\\
\label{isocorrfninftykspace}
\hat G'_\infty(|{\bf k}'|, t)&=& \hat G'_\infty(|{\mbox {\boldmath$\lambda$}}^{1/2} {\bf U}{\bf k}|, t)=\hat G_\infty({\bf k},t),
\end{eqnarray}
are obtained. They are given by \cite{cd2000-2,cd1998}
\begin{eqnarray}
\hat G'_\infty(|{\bf k}'|, t)&=&\{[\chi'_{\infty}(t)]^{-1}+({\bf k'})^2\}^{-1}\;\;\;,\\
\label{intxstrichz}
G'_\infty({\bf x}',t)&=& \int _{\bf k'} \;\frac{e^{i{\bf k'}\cdot{\bf x}'}}{[\chi'_{\infty}(t)]^{-1}+({\bf k'})^2}\;\;,
\end{eqnarray}
where the inverse of the bulk susceptibility per component is determined implicitly by
\be
\label{xiunendlich}
[\chi'_{\infty}(t)]^{-1}=r_0+4u'_0 n\int _{\bf k'}\{[\chi'_{\infty}(t)]^{-1}+({\bf k'})^2\}^{-1}.
\ee
The same equation determines  the square of the bulk correlation length above $T_c$ because \cite{cd1998}
\begin{eqnarray}
\chi'_\infty(t)&=&[\xi'_{\infty}(t)]^2.
\end{eqnarray}
Here $\int _{\bf k'}$ stands for $(2\pi)^{-d}\int d^d k'$ with a transformed cutoff.
This cutoff dependence becomes negligible in the asymptotic region near $T_c$ where $G'_\infty ({\bf x'}, t) \to G'_\infty (|{\bf x'}|, t)$ becomes isotropic.
For $t>0$ and $2<d<4$ the asymptotic behavior is \cite{cd1998}
\begin{eqnarray}
\chi'_\infty(t) &=&\Gamma'_{\infty}\;t^{-\gamma_\infty},\\
\label{largenxi}
\xi'_{\infty}(t)&=&\xi'_{\infty,0}t^{-\nu_\infty},\\
 \label{xinfin}
 \xi'_{\infty,0}&=&[4 u'_0 n A_d/(\varepsilon a_0)]^{1/(d-2)},\\
 \label{Ad}
 A_d \;&=&\; \frac{\Gamma(3-d/2)}{2^{d-2} \pi^{d/2}
(d-2)}
\end{eqnarray}
with $\varepsilon=4-d$, with the critical exponents
\begin{eqnarray}
\label{expoentsninfinity}
\gamma'_\infty=2\nu_\infty, \;\nu_\infty=1/(d-2),\;\eta_\infty=0.\;\;
\end{eqnarray}
In particular the amplitude of the susceptibility
\be
\label{gammastrich}
\Gamma'_\infty=\big(\xi'_{\infty,0}\big)^2
\ee
is independent of the cutoff of the $\varphi^4$ Hamiltonian. We note that this simple relation is not valid for all members of the spherical universality class, e.g., it is not valid for the spherical model \cite{dan-2}.

The universal isotropic scaling form for $t\geq0$ in ${\bf k'}$ space is simply
\begin{eqnarray}
\label{ppxx}
\hat G'_\infty(|{\bf k}'|,t)&=&
({\bf k'})^{-2}\hat \Psi_\infty(|{\bf k'}|\xi'_\infty(t)),\\
\label{psidachunendlich}
\hat \Psi_\infty(y)&=& (1+y^{-2})^{-1},\\
Q_{\infty,3}&=&\hat \Psi_\infty(\infty)=1.
\end{eqnarray}
To derive $G'_\infty(|{\bf x}'|,t)$ in real space we use the integral representation $w^{-1}=\int_0^\infty ds e^{-w s}$ of the
quantity $w=(\xi'_{\infty})^{-2}+({\bf k'})^2$. Then we obtain from (\ref{intxstrichz})
\be
\label{GaussIntegral}
G'_\infty(|{\bf x}'|,t)= \int_0^\infty ds e^{-s/[\xi'_\infty(t)]^2}\int _{\bf k'}\; e^{-s({\bf k'})^2+i{\bf k'}\cdot{\bf x}'}.
\ee
Performing the Gaussian ${\bf k'}$-integration
at infinite cutoff \cite{cd1998,dohm2008} we obtain the scaling form
\begin{eqnarray}
\label{GaussH}
G'_\infty(|{\bf x}'|,t)
&=&\int_0^\infty \frac{d s}{(4\pi s)^{d/2}}e^{-s/(\xi'_\infty)^2-|{\bf x}'|^2/(4s)}\;\;\;\;\;\;\;\;\;
\\
\label{Psixz}
&=&\frac{1}{|{\bf x}'|^{d-2}}\Psi_{\infty}(|{\bf x}'|/\xi'_{\infty})\;\;\;\;\;\;\;\;\;
\end{eqnarray}
where the exact universal scaling function
is given by
\begin{eqnarray}
\label{GaussPsiz}
\Psi_{\infty}(y)&=&\frac{y^{d-2}}{(4\pi)^{d/2}}
\int_0^\infty d s s^{-d/2}e^{-s-y^2/(4s)},\;\;\;\\
\label{tildeQetanull}
 \widetilde Q_{\infty,3}&=&
 \Psi_{\infty}(0)= \frac{\Gamma[(d-2)/2]}{4\pi^{d/2}}\;.\;\;\;\;
\end{eqnarray}
This function
depends only on the scalar scaling variable $y=|{\bf x}'|/\xi'_{\infty}$
 because of isotropy.
$G'_\infty$ has indeed the form of (\ref{3calt}) with $\eta=0$ since, according to (\ref{gammastrich}), we may replace 1 in (\ref{Psixz}) by $1=\Gamma'_\infty (\xi'_{\infty,0})^{-2}$. Only owing to the simple relation (\ref{gammastrich}) the isotropic correlation function in the large-$n$ limit has a single-parameter scaling form which is not universally valid for finite $n$ and $\eta\neq 0$.

On the basis of the exact isotropic results (\ref{ppxx})-(\ref{GaussPsiz}) it is straight forward to determine the anisotropic correlation functions in real space and ${\bf k}$ space from (\ref{isocorrfninftyz}) and (\ref{isocorrfninftykspace}) by substituting the general results of Secs. IV and V.
Using (\ref{detlambda}), (\ref{detA}), (\ref{Gamma}) and the general relations   (\ref{scalingarg}) and (\ref{quadratick})
we obtain the
exact anisotropic correlation function of the $\varphi^4$ model in the large-$n$ limit  for $d>2$ and $t\geq 0$
\begin{eqnarray}
\label{3nbarAhNullx} &&G_\infty ({\bf x}, t)  \nonumber\\
 \label{large-n-G}
 &&\;=\frac{\Gamma_{\infty} \;(\bar \xi_{\infty,0})^{-2}}{ [{\bf x}\cdot ({\bf \bar A}^{-1}{\bf x})]^{( d -2 )/2}}\;
 \Psi_{\infty} \Big(\frac{[{\bf x}\cdot ({\bf \bar A}^{-1}{\bf x})]^{1/2}}{\bar \xi_{\infty}(t)}\Big),\;\;\;\;\;\;\;\\
\label{large-n-Corrk}
&&\hat G_\infty ({\bf k}, t)
 = \frac{\Gamma_\infty\;\;(\bar \xi_{\infty,0})^{-2}}{ {\bf k}\cdot \;{\bf \bar A}{\bf k}}\hat \Psi_\infty \Big(  [{\bf k}\cdot \;{\bf \bar A}{\bf k}]^{1/2} \bar \xi_{\infty}(t) \Big), \;\;\;\;\;\;\;\;\;\\
 &&\Gamma_\infty\;\;(\bar \xi_{\infty,0})^{-2}=\Big(\prod^d_{\alpha = 1} {\mbox {$\lambda$}_{\alpha}}\Big)^{-1/d}= (\det {\bf A})^{-1/d},
 \end{eqnarray}
where $\Psi_{\infty}$ and $\hat \Psi_{\infty}$ are the same scaling functions as in the isotropic case. Here $\Gamma_{\infty}=\Gamma_{\infty}'=(\bar \xi_{\infty,0})^2$ and $\bar \xi_{\infty,0}$ are the amplitudes of the
susceptibility per component and mean correlation length of the anisotropic system above $T_c$,
\begin{eqnarray}
\chi_\infty(t) &=&\Gamma_{\infty}\;t^{-\gamma_\infty},\\
\label{meaninfinity}
\bar \xi_{\infty}(t)&=&\bar \xi_{\infty,0}\;t^{-\nu_\infty},\\
\label{largenximeanx}
\bar \xi_{\infty,0}&=&\big[\prod^d_{\alpha = 1} \xi_{\infty,0}^{(\alpha)}\big]^{1/d}
\end{eqnarray}
where the amplitudes  $\xi_{\infty,0}^{(\alpha)}$ of the principal correlation lengths
\begin{eqnarray}
\label{xialphaxinf}
\xi^{(\alpha)}_{\infty 0}
={\mbox {$\lambda$}_{\alpha}}^{1/2}\xi'_{\infty,0}
\end{eqnarray}
are determined by the eigenvalues ${\mbox {$\lambda$}_{\alpha}}$ of the anisotropy matrix ${\bf A}$ and by the amplitude $\xi'_{\infty,0}$ of the isotropic correlation length above $T_c$ of the Hamiltonian (\ref{2zxx}).

Below $T_c$ the algebraic large-distance behavior of the transverse correlation function of the isotropic $\varphi^4$ Hamiltonian (\ref{2zxx}) in the large-$n$ limit reads
 for $d>2$,
\begin{eqnarray}
\label{trans correl large infty}
G'_{\infty,\rm T} (|{\bf x}'|, t) &=& {\cal C}_{\rm T}  \big[{\cal M}'_\infty(t)\big]^2 \big[\xi'_{\infty, \rm T }(t)/|{\bf x}'|\big]^{d-2},\;\;\;\;\\
\big[{\cal M}'_\infty(t)\big]^2
&=&\frac{r'_{\infty,0c}-r_0}{4u'_0n}=(B'_\infty)^2 |t|^{2\beta_\infty},\\
{\beta_\infty}&=&1/2,\;\;\;\;\;\\
\label{Binfinstrich}
(B'_\infty)^2&=& a_0/(4u'_0n),\\
r'_{\infty,0c}&=&-4u'_0n \int_{\bf k'}{\bf k'}^{-2},
\end{eqnarray}
according to (\ref{trans correl large}) and to Eq. (28) of \cite{cd1998}. The transverse correlation length $\xi'_{\infty, \rm T }(t)=\xi'_{\infty, \rm T ,0}|t|^{-\nu_\infty}$ below $T_c$ is universally related to
the correlation length $\xi'_{\infty}(t)$ above $T_c$
\be
\xi'_{\infty, \rm T ,0}/\xi'_{\infty,0}  = X_{\infty \rm T}=\text {universal},
\ee
with a known universal constant $X_{\infty \rm T}$ \cite{priv}. By means of the inverse special shear transformation the anisotropic transverse correlation function $G_{\infty,\rm T} ({\bf x}, t)$ of the $\varphi^4$ model in the large-$n$ limit is obtained as
\begin{eqnarray}
\label{inverstrans correl large infty}
G_{\infty,\rm T} ({\bf x}, t)&=& (\det {\bf A})^{-1/2}
G'_{\infty,\rm T} (|{\bf x}'|, t) \\
&=& {\cal C}_{\rm T}  \big[{\cal M}_\infty(t)\big]^2 \Big(\frac{\bar \xi_{\infty,\rm T}}{ [{\bf x}\cdot ({\bf \bar A}^{-1}{\bf x})]^{1/2} }\Big)^{d-2},\;\;\;\;\;\;\;\;\;\;\\
\big[{\cal M}_\infty(t)\big]^2
&=&\frac{r_{\infty,0c}-r_0}{4u_0n}=(B_\infty)^2 |t|, \\
\label{Binfin}
(B_\infty)^2&=& a_0/(4u_0n),\\
r_{\infty,0c}&=&-4u_0n \int_{\bf k}{\bf k}^{-2}=r'_{\infty,0c},\\
\label{meancorinf}
\bar \xi_{\infty,\rm T}(t)&=&\Big[\prod^d_{\alpha = 1} \xi_{\infty,\rm T}^{(\alpha)}(t)\Big]^{1/d},\\
\xi_{\infty,\rm T}^{(\alpha)}(t)&=&\xi_{\infty,{\rm T},0}^{(\alpha)}|t|^{-\nu_\infty},\;\;\\
 \xi_{\infty,{\rm T},0}^{(\alpha)}&=&\lambda_\alpha^{1/2} \xi_{\infty,{\rm T},0}'.\;\;\;\;\;\;
\end{eqnarray}
where ${\bf \bar A}$ must be expressed in terms of $\xi_{\infty,{\rm T},0}^{(\alpha)}$, compare Eqs. (5.41)-(5.44) of \cite{dohm2018} for finite $n>1$.

Our proof of multiparameter universality in Sec. IV. C. implies that the
correlation functions of the anisotropic $n$-vector model (\ref{Hspin})
in the large-$n$ limit have the same structure as those derived for the anisotropic $\varphi^4$ model, with the same universal scaling functions $\Psi_{\infty}$ and $\hat \Psi_{\infty}$ and
with the reduced anisotropy matrix ${\bf \bar A^{\rm sp}}$ and two
different nonuniversal parameters $\Gamma^{\rm sp}_{\infty}$ and
$ \bar \xi^{\rm sp}_{\infty,0}$ that depend on the couplings $E_{i,j}$. The same statement applies to other models of the spherical universality class.
\subsection{Exact anisotropic Gaussian correlation function }
 The following is based on the continuum version of the anisotropic Gaussian model
(\ref{continGauss}) and  the isotropic Gaussian Hamiltonian (\ref{2zxxGauss}) obtained
after the special shear transformation (\ref{shearalt})-(\ref{shearu}).
The structure of the anisotropic and isotropic Gaussian correlation functions $G^{\rm G}({\bf x},t)$, $\hat G^{\rm G}({\bf k},t)$ and $G'^{\rm G}(|{\bf x}'|,t)$, $\hat G'^{\rm G}(|{\bf k}'|,t)$, respectively, are closely related to those in the large-$n$ limit with essentially the same scaling functions but with Gaussian critical exponents
\begin{eqnarray}
 \nu^{\rm G}=1/2,\;  \gamma^{\rm G}=1, \;\eta^{\rm G}=0,
\end{eqnarray}
and different nonuniversal parameters $\Gamma^{\rm G}_+$ and $\bar \xi^{\rm G}_{0+}$. These correlation functions are defined for $d\geq 2$ by
\begin{eqnarray}
\label{Glargen}
G^{\rm G}({\bf x},t)
&=& n <\varphi({\bf x})\varphi({\bf 0})>
\label{integr}
= \int _{\bf k} \;\hat G^{\rm G}({\bf k},t)e^{i{\bf k}\cdot{\bf x}},\;\;\;\;\;\;\;\;\;\\
\label{intxstrichzz}
G'^{\rm G}(|{\bf x}'|,t)&=& \int_{\bf k'} \hat G'^{\rm G}(|{\bf k}'|,t)
\;e^{i{\bf k'}\cdot{\bf x}'},
\\
\hat G'^{\rm G}(|{\bf k}'|,t)&=&n\;[r_0+({\bf k'})^2]^{-1}\;\;\;,
\end{eqnarray}
where
\be
r_0=a_0 t = [\xi_+'^{\rm G}(t)]^{-2}, \;\;\; t\geq 0,
\ee
with the isotropic Gaussian correlation length \cite{dohm2008}
\be
\label{corrGauss}
\xi_+'^{\rm G}(t)=r_0^{-1/2}=\xi_{0+}'^{\rm G}t^{-1/2}, \;\;\xi_{0+}'^{\rm G}= a_0^{-1/2}.
\ee
The evaluation of the integral (\ref{intxstrichzz}) in the asymptotic region is analogous to that of (\ref{intxstrichz}) and leads to
\begin{eqnarray}
\label{Gaussscalingx}
G'^{\rm G}(|{\bf x}'|,t)&=&\frac{1}{|{\bf x}'|^{d-2}}\Psi^{\rm G}(|{\bf x}'|/\xi_+'^G(t)),\\
\label{Gaussscalingzusatz}
\Psi^{\rm G}(y)&\equiv& n\Psi_{\infty}(y),\\
\label{tildeQetanullGauss}
 \widetilde Q^{\rm G}_3&=&
 \Psi^{\rm G}(0)= n\frac{\Gamma[(d-2)/2]}{4\pi^{d/2}},\;\;\;\;
\end{eqnarray}
where the Gaussian scaling function (\ref{Gaussscalingzusatz}) divided by $n$ is identical with the scaling function (\ref{GaussPsiz}) in the large-$n$ limit. There exists agreement with the structure of (\ref{3calt}) because of the identity $1=\Gamma'^{\rm G}_+ (\xi'^{\rm G}_{0+})^{-2}$ where $\Gamma'^{\rm G}_+$ is the amplitude of the isotropic Gaussian  susceptibility
\begin{eqnarray}
\chi'^{\rm G}(t)&=&\Gamma'^{\rm G}_+t^{-1}=r_0^{-1}=[\xi_+'^{\rm G}(t)]^2,\\
\label{GammaXi}
\Gamma'^{\rm G}_+&=&(\xi_{0+}'^{\rm G})^2.
\end{eqnarray}
The universal isotropic scaling form for $t\geq0$ in ${\bf k'}$ space is  analogous to (\ref{ppxx}),
\begin{eqnarray}
\label{Gaussscaling}
\hat G'^{\rm G}(|{\bf k}'|,t)&=&
({\bf k'})^{-2}\hat \Psi^{\rm G}(|{\bf k'}|\xi_+'^{\rm G}(t)),\\
\hat \Psi^{\rm G}(y)&=& (1+y^{-2})^{-1},\\
Q^{\rm G}_3&=&\hat \Psi^{\rm G}(\infty)=1.
\end{eqnarray}
Correspondingly we obtain the
exact anisotropic correlation function of the Gaussian model for $d\geq2$ and $t\geq 0$
\begin{eqnarray}
\label{Gaussaniso}
G^{\rm G}({\bf x}, t) &=&
\frac{\Gamma_+^{\rm G}\;(\bar \xi^{\rm G}_{0+})^{-2}}{ [{\bf x}\cdot {\bf \bar A}^{-1}{\bf x}]^{(d-2)/2}}
\Psi^{\rm G} \Big(\frac{[{\bf x}\cdot {\bf \bar A}^{-1}{\bf x}]^{1/2}}{\bar \xi^G_+(t)}\Big),\;\;\;\;\;\;\;\;\;\;\\
\label{GaussCorrk}
\hat G^{\rm G} ({\bf k}, t)
 &=&\frac{ \Gamma_+^{\rm G}\;(\bar \xi^{\rm G}_{0+})^{-2}}{[{\bf k}\cdot \;{\bf \bar A}{\bf k}]} \hat \Psi^{\rm G} \Big(  [{\bf k}\cdot \;{\bf \bar A}{\bf k}]^{1/2} \bar \xi_+^{\rm G}(t) \Big), \;\\
 \Gamma_+^{\rm G}\;(\bar \xi^{\rm G}_{0+})^{-2}&=&\Big(\prod^d_{\alpha = 1} {\mbox {$\lambda$}_{\alpha}}\Big)^{-1/d}= (\det {\bf A})^{-1/d},
\end{eqnarray}
 where ${\bf \bar A}$ must be expressed in terms of $\xi_{0+}^{{\rm G}(\alpha)}$.
Here $\Gamma^{\rm G}_+=\Gamma'^{\rm G}_+$ and $\bar \xi^{\rm G}_{0+}$ are the amplitudes of the
susceptibility per component and mean correlation length of the anisotropic system above $T_c$,
\begin{eqnarray}
\chi^{\rm G}_+(t) &=&\Gamma^{\rm G}_+\;t^{-1},\\
\bar \xi^{\rm G}_+(t)&=&\bar \xi^{\rm G}_{0+}t^{-1/2},\\
%
%
\label{Gaussximeanx}
\bar \xi^{\rm G}_{0+}&=&\big[\prod^d_{\alpha = 1} \xi_{0+}^{{\rm G}(\alpha)}\big]^{1/d}
\end{eqnarray}
where the amplitudes  $\xi_{0+}^{{\rm G}(\alpha)}$ of the principal correlation lengths
$\xi^{{\rm G}(\alpha)}_{0+}
={\mbox {$\lambda$}_{\alpha}}^{1/2}\xi'^{\rm G}_{0+}$
are determined by the eigenvalues ${\mbox {$\lambda$}_{\alpha}}$ of the anisotropy matrix ${\bf A}$ and the amplitude $\xi'^{\rm G}_{0+}= a_0^{-1/2}$, (\ref{corrGauss}), of the isotropic correlation length above $T_c$ of the Gaussian Hamiltonian (\ref{2zxxGauss}). As in the large-$n$ limit, the single-parameter isotropic scaling forms (\ref{Gaussscalingx}) and (\ref{Gaussscaling}) are due to the special relation (\ref{GammaXi}).

Owing to two-scale-factor universality all isotropic systems of the Gaussian universality class have an asymptotic correlation function that has the same structure  as that of (\ref{Gaussscalingx}) and (\ref{Gaussscaling}) with two nonuniversal
amplitudes $\Gamma^{\rm G, iso}_+$ and $\xi^{\rm G, iso}_{0+}$ which in general are independent of each other. This statement applies to the Gaussian version of the $n$-vector model (\ref{Hspin})
(with spin variables $- \infty \leq S_i^{(\mu)} \leq \infty$). Thus,
according to the proof of multiparameter universality in Sec. IV.C.,
the same structure of the anisotropic Gaussian correlation function as
in (\ref{Gaussaniso}) and (\ref{Corrk})
is predicted for the Gaussian version of the isotropic $n$-vector model where the reduced anisotropy matrix ${\bf \bar A}^{\rm sp}$ comes into play
 as well as two nonuniversal parameters
 $\Gamma^{\rm sp}_+$ and $\bar \xi^{\rm sp}_{0+}$  which need to be determined as functions of $E_{i,j}$.
 \subsection{Application to three dimensions:\\Determination of $\hat \Psi_\pm$ from the FRG}
Application of the preceding analysis to three dimensions is of interest in view of the various anisotropic three-dimensional systems in condensed matter physics.
Thereby one encounters, in general, three problems:

(i) the determination of the three-dimensional universal isotropic scaling functions $\Psi_\pm, \hat \Psi_\pm$ or, equivalently, $\hat D_\pm$, of the $O(n)$-symmetric universality classes,

(ii) the determination of up to seven independent nonuniversal parameters $\Gamma_+,\xi^{(1)}_{0+},\xi^{(2)}_{0+},\xi^{(3)}_{0+}, \Omega_1,\Omega_2,\Omega_3$
where the principal angles $\Omega_\alpha$ describe the orientation of the three principal directions, and

(iii) the construction of the reduced anisotropy matrix
\begin{eqnarray}
\label{AquerOmega}
{\bf \bar A} ={\bf \bar A}\big(\{\xi_{0\pm}^{(\alpha)},{\bf e}^{(\alpha)}\}\big)= {\bf \bar A}\Big(\frac{\xi^{(1)}_{0\pm}}{\bar \xi_{0\pm}},\frac{\xi^{(2)}_{0\pm}}{\bar \xi_{0\pm}}, \frac{\xi^{(3)}_{0\pm}}{\bar \xi_{0\pm}},  \Omega_1,\Omega_2,\Omega_3\Big)\nonumber\\
\end{eqnarray}
where ${\bf e}^{(\alpha)}$ are the three principal unit vectors and where $\bar \xi_{0\pm}$ is the amplitude of the mean correlation length. Several examples of ${\bf \bar A}$ for three-dimensional anisotropic $\varphi^4$ models have been presented
\cite{cd2004,dohm2008,dohm2018,DWKS2021}, in particular with general principal angles $\Omega$ and $\Omega+ \pi/2$ describing planar anisotropies in a three-dimensional environment \cite{DW2021}. This matrix appears not only in the bulk correlation function but also in the finite-size properties of the excess free energy and the Casimir force of weakly anisotropic systems \cite{cd2004,dohm2008,dohm2018,DW2021}. A determination of the nonuniversal parameters (ii) in terms of the couplings is a nontrivial problem for three-dimensional anisotropic $n$-vector models and real systems that we shall not further discuss in this paper.

Here we confine ourselves to a discussion of the problem (i) in the context of the isotropic order-parameter correlation function $\hat G_\pm^{\rm FRG}(k,t)$ derived in the framework of the FRG \cite{hassel2007,sinner2008}.
The calculation was performed within the isotropic $\varphi^4$ theory with a finite cutoff $\Lambda_0$ and a four-point coupling $u_0$ in $d<4$ dimensions for $n=1$. 
On the basis of truncated FRG flow equations the following approximate result
was obtained above $(+)$ and below $(-)$ $T_c$
\begin{eqnarray}
\label{CorrkFRG} \hat G_\pm^{\rm FRG} (k, t) &=& k_c^{-2}g^\pm(k\xi^{\rm FRG}_\pm,k/k_c),
\\
\label{sigmaFRG}
g^\pm(x_\pm,y)&=& [y^2+\sigma^\pm(x_\pm,y)]^{-1}
\end{eqnarray}
with $x_\pm=k\xi^{\rm FRG}_\pm$, $y=k/k_c$, $k=|{\bf k}|$ where ${\bf k}$ is the wave vector, $\xi^{\rm FRG}_\pm$
is the correlation length, and $k_c$
is a finite nonuniversal wavenumber that depends on the cutoff and the four-point coupling. For the definition of the functions $g^\pm(x_\pm,y)$
and $\sigma^\pm(x_\pm,y)$ 
we refer to \cite{hassel2007,sinner2008}.
This result is supposed to be applicable to the region
$ k \ll k_c$
and to the crossover 
to a nonasymptotic critical region $ k_c \ll k\ll \Lambda_0$ as well as in the limit $k\to 0$ at fixed $\xi^{\rm FRG}_\pm <\infty$.  The implications of the exact sum rule (\ref{susceptx}) were not discussed. It was claimed \cite{hassel2007}, without justification, that the scaling function $\sigma^\pm(x_\pm,y)$ is universal. Below we shall refute this claim by showing that $g^\pm(x_\pm,y)$ is not universal.

We first discuss (\ref{CorrkFRG}) in the standard asymptotic scaling region $k_c\xi^{\rm FRG}_\pm\gg1$ and $k/k_c\ll1$ at fixed  $ k\xi^{\rm FRG}_\pm$, including the case $ \xi^{\rm FRG}_\pm= \infty$ and the limit $k\to 0$ at $\xi^{\rm FRG}_\pm <\infty$. We shall show that $\hat G_\pm^{\rm FRG} (k, t)$ can be written in the two-parameter scaling forms (\ref{FourierG}),  (\ref{atTcxisoiso}), and (\ref{overallplusminus}) and that the universal scaling function $\hat \Psi_\pm$ is not determined by $g^\pm$ or by $\sigma^\pm$ but by a certain ratio of  $g^\pm$ and a ratio of its overall amplitudes.

Comparison of (\ref{CorrkFRG}) with (\ref{FourierG}) in the limit $k_c \to \infty$ at fixed $k\xi^{\rm FRG}_\pm$ yields the universal ratio
\begin{eqnarray}
\label{CorrFRGratio}
\lim_{k_c\to \infty} \frac{\hat G_\pm^{\rm FRG} (k, t)}{\hat G_+^{\rm FRG} (k, 0)}&=&\lim_{k_c\to \infty} \frac{g^\pm(k\xi^{\rm FRG}_\pm, k/k_c)}{g^+ (\infty, k/k_c)}\;\;\;\;\;\;\\
\label{CorrFRGratioPsi}
&=&\frac{\hat \Psi_\pm(k\xi^{\rm FRG}_\pm)}{\hat \Psi_+(\infty)}\;.
\end{eqnarray}
Next we determine the universal constant $Q_3$. We first consider the case $T=T_c$, $\xi^{\rm FRG}_\pm=\infty$,
\begin{eqnarray}
\label{CorrkFRGTc} \hat G_\pm^{\rm FRG} (k, 0) &=& k_c^{-2}g^\pm(\infty,k/k_c).
\end{eqnarray}
Since $\hat G_\pm^{\rm FRG} (k, 0)$ must behave as $\simeq k^{-2+\eta}$ for small $k$ we obtain for $k/k_c\ll 1$
\begin{eqnarray}
\label{gTc} g^\pm(\infty,k/k_c) &=&A^{\rm FRG} \; (k_c/k)^{2-\eta},\\
\label{GTc}\hat G_\pm^{\rm FRG} (k, 0) &=& A^{\rm FRG}\frac{k_c^{-\eta}}{k^{2-\eta}}.
\end{eqnarray}
The $d$-dependent constant $A^{\rm FRG}> 0$ is denoted by $A_D^{-1}$ in Eq. (58) of \cite{sinner2008}, with $A_3 \approx 1.075$ in three dimensions, i.e.,
\begin{eqnarray}
\label{AFRGnumerisch} A^{\rm FRG} \approx  0.930 \;, \;\; d=3.
\end{eqnarray}
Furthermore, compatibility of (\ref{CorrkFRG}) with the free-energy density of the $\varphi^4$ model requires that (\ref{CorrkFRG}) satisfies the exact sum rule (\ref{susceptx}) which yields the susceptibility $\chi^{\rm FRG}_\pm$
\begin{eqnarray}
\label{sumruleFRG} \hat G_\pm^{\rm FRG} (0, t) &=& \chi^{\rm FRG}_\pm\\
&=&k_c^{-2}\lim_{k\to 0}g^\pm(k\xi^{\rm FRG}_\pm,k/k_c)\;\;\;\;\;\;\;\;\;\;\;\;\\
\label{exactfplus}
&=& k_c^{-2}f_\chi^\pm(k_c\xi^{\rm FRG}_\pm).
\end{eqnarray}
As discussed in Sec. II. A, Eqs. (\ref{sumruleFRG})-(\ref{exactfplus}) are exactly valid in the entire range where (\ref{CorrkFRG}) is valid.
For $k_c\xi^{\rm FRG}_\pm\gg 1$, $\chi^{\rm FRG}_\pm$ must behave as
\begin{eqnarray}
\label{chiasym}
\chi^{\rm FRG}_\pm &=&\Gamma^{\rm FRG}_\pm t^{-\gamma},
\end{eqnarray}
with $\gamma=\nu (2-\eta)$ and $\xi^{\rm FRG}_\pm=\xi^{\rm FRG}_{0\pm} t^{-\nu}$. This implies
\begin{eqnarray}
\label{fchi}
f_\chi^\pm(k_c\xi^{\rm FRG}_\pm)&=&C^{\rm FRG}_\pm(k_c\xi^{\rm FRG}_\pm)^{2-\eta} \;,
\\
\label{GammaFRG}
\Gamma^{\rm FRG}_\pm&=& C^{\rm FRG}_\pm k_c^{-\eta}  (\xi^{\rm FRG}_{0\pm})^{2-\eta},
\end{eqnarray}
with the $d$-dependent constant $C^{\rm FRG}_\pm$. The factor $k_c^{-\eta}$ in  (\ref{GammaFRG}) and (\ref{GTc}) can be eliminated. This yields at $T_c$
\begin{eqnarray}
\label{Q3formula}
\hat G_+^{\rm FRG} (k, 0)&=&\frac{A^{\rm FRG}}{C^{\rm FRG}}\frac{\Gamma^{\rm FRG}_+}{(k\xi^{\rm FRG}_{0+})^{2-\eta}}.
\end{eqnarray}
From a comparison with the exact asymptotic isotropic scaling form at $T_c$, (\ref{Qdrei}) and
(\ref{atTcxisoiso}), we obtain the identification of the universal quantity $Q_3$ in terms of the ratio
\begin{eqnarray}
\label{Qdreiratio}
Q_3= \frac{A^{\rm FRG}}{C_+^{\rm FRG} }=\hat \Psi_+ (\infty)=\hat \Psi_- (\infty)= \;\;\text {universal}.\;\;\;\;\;
\end{eqnarray}
Together with the determination of the ratio (\ref{CorrFRGratioPsi}) this completes our determination of the isotropic two-parameter scaling form (\ref{FourierG}),
\begin{eqnarray}
\label{FRGFourierG}
\hat G_\pm^{\rm FRG} (k, t)=\frac{\Gamma^{\rm FRG}_+}{ \big( k\;\xi^{\rm FRG}_{0+}\big)^{2-\eta}} \;\hat \Psi_\pm \Big( k\xi^{\rm FRG}_\pm \Big)\;
\end{eqnarray}
with the universal scaling function $\hat \Psi_\pm$ for $d<4$ in the regime $k/k_c\ll1, k_c\xi^{\rm FRG}_\pm\gg1$, $k\xi^{\rm FRG}_{0+}>0$,
\begin{eqnarray}
\label{scalfunc}
\hat \Psi_\pm(k\xi^{\rm FRG}_\pm)= \frac{A^{\rm FRG}}{C_+^{\rm FRG} }\lim_{k_c\to \infty} \frac{g^\pm(k\xi^{\rm FRG}_\pm, k/k_c)}{g^+ (\infty, k/k_c)}.
\end{eqnarray}
We note that in (\ref{FRGFourierG}) $\hat G_\pm^{\rm FRG} (k, t)$ is uniquely divided into nonuniversal and universal parts.
The corresponding  two-parameter Fisher-Aharony scaling form  for $t\neq 0,k\geq0,$
\begin{eqnarray}
\label{FRGoverallplusminus}
\hat G_\pm^{\rm FRG}( k,t)&=& \Gamma^{\rm FRG}_\pm|t|^{-\gamma}\;\hat D_\pm\Big( k\xi^{\rm FRG}_\pm(t) \Big)\;\;
 \end{eqnarray}
follows from (\ref{overallplusminus})-(\ref{Dplusnull}). The structure of (\ref{FRGFourierG}) and (\ref{FRGoverallplusminus}) has been confirmed by exact analytic results for isotropic Ising and  $\varphi^4$ models on two-dimensional lattices \cite{dohm2019,DKW2023}. We see that, unlike the original formulation (\ref{CorrkFRG}) in terms of the nonuniversal model parameter $k_c$, this parameter has been completely eliminated in (\ref{FRGFourierG}) and (\ref{FRGoverallplusminus}) in favor of the observable thermodynamic amplitude $\Gamma^{\rm FRG}_\pm$ of the susceptibility, as expected on general grounds. But there still exist two independent nonuniversal thermodynamic parameters $\Gamma^{\rm FRG}_+$ and $\xi^{\rm FRG}_{0+}$ in the asymptotic scaling forms (\ref{FRGFourierG}) and (\ref{FRGoverallplusminus}) including  $T=T_c$. This refutes the existence of a "one-parameter scaling picture" and "one-parameter scaling hypothesis" \cite{hassel2007} for systems with $\eta\neq 0$ for which the principle of two-scale-factor universality with two independent thermodynamic parameters is well established.

After the identification of $\hat \Psi\pm$ in terms of the ratios $A^{\rm FRG}/ C_+^{\rm FRG}$ and $g^\pm(x_\pm,y)/g^+(\infty,y)$ from \cite{hassel2007,sinner2008} we substitute (\ref{scalfunc}) into (\ref{Corrk}) which yields the bulk correlation function $\hat G_\pm$ of the anisotropic $\varphi^4$ model in the  asymptotic critical region in three dimensions for $n=1$. Due to multiparameter universality this also yields the prediction of the  correlation function of all other weakly anisotropic systems in the $(d=3,n=1)$ universality class provided that the nonuniversal parameters (ii) specified above are known.

The value of $C_+^{\rm FRG}$ was not calculated in \cite{hassel2007,sinner2008}.  It would be worthwhile to determine this value in order to obtain an estimate for  $Q_3$ which then can be compared with the known estimates
$Q_3 \approx 0.922$ ($\varepsilon$ expansion) or $0.90 \pm 0.01$ (numerical studies)
for $d=3,n=1$ \cite{priv}. (Note that these values are defined in combination with the "true" (exponential) correlation length $\xi_{0+}$ which differs from the second-moment correlation length \cite{pelissetto,tarko,brezin1974}.)

So far our analysis confirms the universality only of the
{\it ratios} of the amplitudes $A^{\rm FRG}$ and $C_+^{\rm FRG}$ and of the functions $g^\pm(x_\pm,y)$ and $g^+(\infty,y)$ appearing in (\ref{CorrFRGratio}), (\ref{Qdreiratio}), and (\ref{scalfunc}) in the asymptotic scaling regime. Thus there is no compelling reason for the assertion \cite{hassel2007,sinner2008} that the functions $\sigma^\pm(x_\pm,y)$ and $g^\pm(x_\pm,y)$ themselves are universal. No argument in support of this assertion was given in \cite{hassel2007,sinner2008}. This claim would imply the existence of new universal amplitudes $A^{\rm FRG}$ or $C_+^{\rm FRG}$ that do not exist in the established theory of bulk critical phenomena in the asymptotic region of isotropic systems \cite{priv,pelissetto,zinn2007,fish-1,bre-1}.

In the following we show that the function  $g^\pm(x_\pm,y)$ is not universal. From (\ref{GTc}), (\ref{Q3formula}), and (\ref{Qdreiratio}) we obtain the relation between amplitudes
\begin{eqnarray}
A^{\rm FRG}k_c^{-\eta} = Q_3 \;\frac{\Gamma^{\rm FRG}_+}{(\xi^{\rm FRG}_{0+})^{2-\eta}}\;\;.
\end{eqnarray}
The right-hand side is uniquely divided into a universal  part $Q_3$ and a nonuniversal amplitude ratio of thermodynamic quantities. The nonuniversal parameter
 $k_c$ on the left-hand side characterizes the nonasymptotic range  $k>k_c$ of wave numbers of the correlation function of the $\varphi^4$ theory and is determined by non-thermodynamic quantities, i. e., the cutoff $\Lambda_0$ and the four-point coupling $u_0$ of the $\varphi^4$ Hamiltonian \cite{hassel2007,sinner2008},
\begin{eqnarray}
\label{kcmicrox}
k_c&=&  \Lambda_0 \exp (-{\it l}_c)
\end{eqnarray}
where ${\it l}_c( u_0 \Lambda_0^{d-4} )$ is a dimensionless flow parameter. Thus $k_c^{-\eta}$ is different from the ratio $\Gamma^{\rm FRG}_+/(\xi^{\rm FRG}_{0+})^{2-\eta}$ and is not universally related to this ratio. We conclude that the dimensionless amplitude $A^{\rm FRG}$ must be different from $Q_3$ and must contain a nonuniversal part, i.e., $A^{\rm FRG}$ is a nonuniversal quantity. Since $A^{\rm FRG}$ governs the small-$k$ behavior of $g^\pm(\infty,k/k_c)$ according to (\ref{gTc}) we conclude that the function $g^\pm(\infty,k/k_c)$ is nonuniversal in the asymptotic region at $T_c$ for small $k/k_c$. By continuity and analyticity requirements this nonuniversal dependence for small $k/k_c$ cannot turn smoothly into a universal dependence for larger $k/k_c$ in the nonasymptotic regime $k>k_c$.

Furthermore, because of the relation
\begin{eqnarray}
C_+^{\rm FRG}&=&A^{\rm FRG}/Q_3\\
 &=&k_c^\eta \Gamma^{\rm FRG}_+/(\xi^{\rm FRG}_{0+})^{2-\eta}
\end{eqnarray}
we conclude that $C_+^{\rm FRG}$ must be a nonuniversal quantity. This conclusion is supported by exact analytic results for the ratio $\Gamma^{\rm iso}_+/(\xi^{\rm iso}_{0+})^{2-\eta}$ of isotropic Ising and $\varphi^4$ models on two-dimensional square and parallelogram lattices \cite{dohm2019,DKW2023}.

We also consider the case $k=0$ in the asymptotic range $k_c \xi^{\rm FRG}_+ \gg 1$ where we have from (\ref{sumruleFRG})-(\ref{GammaFRG})
\begin{eqnarray}
\label{gknull}
k_c^2\hat G^{\rm FRG}_+(0,t)&=& \lim_{k\to 0}g^+(k\xi^{\rm FRG}_+,k/k_c)\\
&=&C^{\rm FRG}_+(k_c\xi^{\rm FRG}_+)^{2-\eta}.
\end{eqnarray}
The nonuniversality of $C_+^{\rm FRG}$ implies that the function $g^+(k\xi^{\rm FRG}_+,k/k_c)$ is a nonuniversal quantity for $k\to 0$ and large $k_c \xi^{\rm FRG}_+$. A similar argumentation holds for $g^-(k\xi^{\rm FRG}_-,k/k_c)$ for $k\to 0$. Together  with the nonuniversality of $g^\pm(\infty,k/k_c)$ this means that the function $g^\pm(x_\pm,y)$ is not universal along the two vertical and horizontal axes  $(\xi^{\rm FRG}_\pm)^{-1}=0$ and $k=0$ of the $k-(\xi^{\rm FRG}_\pm)^{-1}$ plane.  By continuity and analyticity requirements this nonuniversal dependence cannot turn smoothly into a universal dependence in the plane away from these axes. This implies the nonuniversality of $g^\pm(x_\pm,y)$ in the region between these axes including both the asymptotic and nonasymptotic regions, thus the curves for $\Delta \sigma^\pm$ shown in Fig. 5 of \cite{hassel2007} and in Fig. 3 of \cite{sinner2008} are nonuniversal.

We briefly comment on the consequences of the structure of (\ref{CorrkFRG}) for $t\neq 0$.
We rewrite (\ref{sumruleFRG}) and (\ref{exactfplus}) in the form
\begin{eqnarray}
\label{rewritechiFRGx} \chi^{\rm FRG}_\pm
&=&(\xi^{\rm FRG}_\pm)^2 {\cal F}_\pm(k_c\xi^{\rm FRG}_\pm)
\end{eqnarray}
with
\begin{eqnarray}
\label{calFFRG}
{\cal F}_\pm(k_c\xi^{\rm FRG}_\pm)&=&(k_c\xi^{\rm FRG}_\pm)^{-2}\;f_\chi^\pm(k_c\xi^{\rm FRG}_\pm)
\end{eqnarray}
where $f_\chi^\pm$ is the nonuniversal function defined in (\ref{exactfplus}). This constitutes an implicit equation that determines the parameter
$k_c$ as a nonuniversal function of $\chi^{\rm FRG}_\pm$ and $\xi^{\rm FRG}_\pm$ in the entire range where (\ref{CorrkFRG}) is valid. Thus $k_c$ can be exactly eliminated in favor of $\chi^{\rm FRG}_\pm$ and $\xi^{\rm FRG}_\pm$, and (\ref{CorrkFRG}) can be expressed completely in terms of these two nonuniversal temperature-dependent thermodynamic quantities, without any explicit cutoff dependence even outside the standard asymptotic critical region. We consider this nonuniversal structure to be due to the approximations made in \cite{hassel2007,sinner2008} within the $\varphi^4$ model which are not expected to remain universally valid for all systems in the Ising universality class.
Nevertheless, after the identification of the universal part of (\ref{CorrkFRG}) through (\ref{FRGFourierG}) and (\ref{scalfunc}), this result is a prediction that is of substantial interest for a comparison with other systems of the $d=3$ Ising universality class in the asymptotic critical region.

\renewcommand{\thesection}{\Roman{section}}
\renewcommand{\theequation}{6.\arabic{equation}}
\setcounter{equation}{0}
\section{Multiparameter universality of critical bulk amplitude relations}
The universal critical point bulk amplitude relations play an important role in the traditional theory of critical
phenomena \cite{priv} where, however, no attention has been paid to amplitude
relations in the subclass of weakly anisotropic systems within a universality class.
The failure of two-scale-factor universality in this subclass was pointed out in \cite{cd2004}. Subsequently several bulk amplitude relations have been shown to be valid within the weakly anisotropic bulk $\varphi^4$ theory in $2<d<4$ dimensions \cite{dohm2006,dohm2008,dohm2018,dohm2019} with the same universal constants as for the isotropic case, such as $Q_c, Q_1, R^+_\xi, Q_2,\widetilde Q_3,P_2,P_3,W_1,X_-(0)$ but where, in general, up to $d(d+1)/2+1$ independent nonuniversal parameters are involved rather than only two nonuniversal parameters. These properties were called multiparameter universality \cite{dohm2008} for bulk amplitude relations. It was hypothesized \cite{dohm2018} that these properties are valid not only for $\varphi^4$ models but also for all weakly anisotropic bulk systems other than $\varphi^4$ models. So far no proof exists for this hypothesis. Our generalized shear transformation introduced in Sec. IV enables us to present such a proof. As a representative of a weakly anisotropic system other than the $\varphi^4$ model  we take the $n$-vector model.
\subsection{Proof of multiparameter universality}
Application of the generalized shear transformation (\ref{transyy}) to the volume $V$ of the anisotropic system yields the transformed volume of the isotropic system
\begin{eqnarray}
\label{Vtranssp}
\widehat V\equiv V^{\rm iso} &=&(\det {\mbox {\boldmath$\widehat\lambda$}})^{-1/2}\;V=\big(\xi^{\text {iso}}_{0\pm}/\bar \xi^{\rm sp}_{0\pm}\big)^d\;V
\end{eqnarray}
where we have used (\ref{ratiomeanisox}). This implies the invariance of the volume ratios
\begin{eqnarray}
\label{volinvarsp}
\frac{ V^{\rm iso}}{V^{\rm iso}_{\rm cor,+}} &=&\frac{V}{V_{\rm cor,+}}, \;\frac{ V^{\rm iso}}{V^{\rm iso}_{\rm cor,-}} =\frac{V}{V_{\rm cor,-}}
\end{eqnarray}
where
$V^{\rm iso}_{\rm cor,\pm}=\big(\xi^{\text {iso}}_{0\pm}\big)^d$
and
$V_{\rm cor\pm}=\prod^d_{\alpha = 1} \xi_{0\pm}^{{\rm sp}(\alpha)}= \big(\bar \xi^{\rm sp}_{0\pm}\big)^d$
are the isotropic (spherical) and anisotropic (ellipsoidal) correlation volumes above and below $T_c$, respectively. This is analogous to the invariance  (\ref{volinvar}) in the $\varphi^4$ theory.

We consider the singular part ${ \cal F}^{\rm sp}_s$ of total free energy
(\ref{freesingnonsing}) of the anisotropic $n$-vector model in the volume $V$.
The generalized shear transformation (\ref{transyy}) generates
 the singular part ${\cal F}^{\rm iso}_{s}$
of the transformed isotropic system. We have argued in Sec. II.B. that
since this transformation involves only a smooth change of the positions ${\bf x} \to \widehat {\bf x}$ of the lattice points 
it does not change
its singular part, thus  we have the invariance
\begin{eqnarray}
\label{spinvar}
\label{Fiso}
{\cal F}^{\rm iso}_{s}={ \cal F}^{\rm sp}_{s}.
\end{eqnarray}
This is parallel to (\ref{totalsingshearphi}) of the special shear transformation.

From (\ref{spinvar}) and (\ref{Vtranssp}) we obtain
the relation between the singular parts of the bulk free-energy densities of the isotropic and anisotropic $n$-vector model
\begin{eqnarray}
\label{bulk free sp}f^{\rm sp}_{b,s,\pm}&=&\lim_{ V \to \infty}\frac{{ \cal F}^{\rm sp}_{s}}{V}
=
\Big(
\frac{\xi^{\text {iso}}_{0\pm}}{\bar \xi^{\rm sp}_{0\pm}}
\Big)^d\;
\lim_{ V^{\rm iso} \to \infty}\frac{{\cal F}^{\rm iso}_{s}}{V^{\rm iso}}\nonumber\\
&=&\Big(
\frac{\xi^{\text {iso}}_{0\pm}}{\bar \xi^{\rm sp}_{0\pm}}
\Big)^d\;\;f^{\text {iso}}_{b,s,\pm}.
\end{eqnarray}
Together with the isotropic relation (\ref{3a}) this yields the singular part of the bulk free-energy density of the anisotropic system
\begin{eqnarray}
\label{3afsbaniso}
f^{\rm sp}_{b,s,\pm} (t) =\left\{
\begin{array}{r@{\quad \quad}l}
                         \; A_\pm |t|^{d\nu}\quad          & \mbox{for} \;2<d<4\;, \\
                         \; \frac{1}{2}A_\pm |t|^{2}\ln |t| & \mbox{for} \;d=2, \;
                \end{array} \right.
\end{eqnarray}
where the amplitudes $A_\pm$ of the anisotropic system are given by
\begin{eqnarray}
A_\pm &=&\Big(
\frac{\xi^{\text {iso}}_{0\pm}}{\bar \xi^{\rm sp}_{0\pm}}
\Big)^d\;  A^{\text {iso}}_\pm, \;\;\; d\geq2,\\
\label{ratiofsp}
\frac{f^{\rm sp}_{b,s,+} (t)}{f^{\rm sp}_{b,s,-} (t)}&=&\frac{A_+}{A_-}=\frac{A^{\rm iso}_+}{A^{\rm iso}_-}= \;\;\rm {universal}, \;\;d\geq 2.\;\;.
\end{eqnarray}
Substituting the isotropic relation (\ref{3fy}) we arrive at the universal relation
for the anisotropic amplitude  $A_+$ above $T_c$ of the $n$-vector model
\begin{eqnarray}
\label{3fx}
 \left(\bar \xi^{\rm sp}_{0+}\right)^d A_+  =
 \left(\xi_{0+}^{\text {iso}}\right)^d A_+^{\text {iso}}  = Q_1, \;         d\geq 2,
\end{eqnarray}
where in the anisotropic case the same universal constant $Q_1$ appears as in the isotropic case.  Thus we obtain the singular part of the free-energy density  of the anisotropic $n$-vector model for $t>0$
\begin{eqnarray}
\label{3azneuzzz}
 f^{\rm sp}_{b,s,+} (t) =\left\{
\begin{array}{r@{\quad \quad}l}
                         \; Q_1\left(\bar \xi^{\rm sp}_{0+}\right)^{-d}t^{d \nu},\quad          &  \;2<d<4\;, \\
                         \; \frac{1}{2}Q_1\left(\bar \xi^{\rm sp}_{0+}\right)^{-2} \;t^{2}\ln t,&  \;d=2, \;
                \end{array} \right.
\end{eqnarray}
and obtain $f^{\rm sp}_{b,s,-} (t)$ for $t<0$ from (\ref{ratiofsp}) as
\begin{eqnarray}
\label{minus3azneuzzz}
 f^{\rm sp}_{b,s,-} (t) =\left\{
\begin{array}{r@{\quad \quad}l}
                         \; \frac{A_-}{A_+}Q_1\left(\bar \xi^{\rm sp}_{0+}\right)^{-d}|t|^{d \nu},\quad          &  \;2<d<4\;, \\
                         \; \frac{1}{2}Q_1\left(\bar \xi^{\rm sp}_{0+}\right)^{-2} \;t^{2}\ln |t|,&  \;d=2, \;
                \end{array} \right.
\end{eqnarray}
where we have used (\ref{aplusminus}) for $d=2$. We see that for $d=2$ dimensions $f^{\rm sp}_{b,s,\pm} $ depends only on $|t|$ rather than $t$.

Our final step is to reformulate our result for the free-energy density in a scaling form for the singular bulk part of the free energy of the anisotropic $n$-vector model
\begin{eqnarray}
\label{anisofreenvector} {\cal F}^{\rm sp}_{b,s,\pm}(t,V) &=&V f^{\rm sp}_{b,s,\pm} (t).
\end{eqnarray}
This is achieved by  means of the observation that the scaling variable $\widetilde x$, (\ref{scalinvariablezz}), introduced for the isotropic system can be expressed in two different ways as
\begin{eqnarray}
\label{scalinvariable}
\widetilde x&=&t[V^{\rm iso}/(\xi_{0+}^{\rm iso})^{d}]^{1/(d\nu)}\\
\label{scalinvariablesp}
&=&t [V/(\bar \xi_{0+}^{\rm sp})^{d}]^{1/(d\nu)}
\end{eqnarray}
where in (\ref{scalinvariablesp}) we have used the invariance of the volume ratios above $T_c$ in (\ref{volinvarsp}) under the shear transformation (\ref{transyy}). A corresponding observation was already made for the finite-size scaling variable $\hat x$ defined in Eq. (6.12) of \cite{dohm2018} in the context of renormalized perturbation theory for the anisotropic $\varphi^4$ model in a finite geometry in $2<d<4$ dimensions. From (\ref{ratiofsp})-(\ref{scalinvariablesp}) and
(\ref{FISOzz})-(\ref{Fisominuszz})
we derive the scaling form for the singular bulk part of the free energy of the anisotropic $n$-vector model in $2<d<4$ dimensions
\begin{eqnarray}
\label{FISO}
{\cal F}^{\rm sp}_{b,s,+}(t,V)& =&Q_1\; \widetilde x^{d\nu}={ \cal F}^{\rm iso}_{b,s,+}(t,V^{\rm iso}) , \,\, t>0,\;\;\;\;\;
\\
{\cal F}^{\rm sp}_{b,s,-}(t,V)& =&\frac{A^{\text {iso}}_-}{ A^{\rm iso}_+}Q_1\; |\widetilde x|^{d\nu}={ \cal F}^{\rm iso}_{b,s,-}(t,V^{\rm iso}) , \,\, t<0,\;\;\;\;\;\;\;\;\;
\end{eqnarray}
and the corresponding expression for the ($d=2,n=1$) universality class
\begin{eqnarray}
\label{Fisominus}
{\cal F}^{\rm sp}_{b,s,\pm}(t,V)& =&\frac{1}{2}Q_1\; |\widetilde x|^2 \ln |t|={ \cal F}^{\rm iso}_{b,s,\pm}(t,V^{\rm iso}), \,\,\,\,\,\,\,\,
\end{eqnarray}
with a nonscaling logarithmic factor $\ln |t|$.  The nonuniversality due to weak anisotropy enters only the scaling variable $\widetilde x$. For $d=2$, ${\cal F}^{\rm sp}_{b,s,\pm}$ is a function of $|t|$.

Eqs. (\ref{FISO})-(\ref{Fisominus}) explicitly reflect the invariance of the singular bulk parts under the shear transformation of the $n$-vector model and constitute the anisotropic extension of the singular bulk parts of the free energy of isotropic systems presented in (\ref{FISOzz})-(\ref{Fisominuszz}). Multiparameter universality manifests itself by the fact that the corresponding result for the anisotropic  $\varphi^4$ model  is obtained from (\ref{scalinvariable})-(\ref{Fisominus}) simply by the substitutions $\bar \xi_{0+}^{\rm sp}\to \bar \xi_{0+},V^{\rm iso}\to V', \xi_{0+}^{\rm iso}\to \xi_{0+}'$ in $\widetilde x$, (\ref{scalinvariablesp}) which corresponds to the scaling variable (\ref{scalinvariablephi}) of the $\varphi^4$ theory, with the same universal constants $Q_1$ and $A^{\rm iso}_-/A^{\rm iso}_+$, where $\bar \xi_{0+}$ is defined in (\ref{ximeanx}). This was shown already previously in Eq. (3.33) of \cite{dohm2008} and Eq. (5.16) of \cite{dohm2018}. Since $\bar \xi^{\rm sp}_{0+}$ and $\bar \xi_{0+}$  depend on $d$ independent principal correlation lengths $\xi^{\rm sp (\alpha)}_{0+}$ and $ \xi_{0+}^{(\alpha)}$, respectively, our results for $f^{\rm sp}_{b,s,\pm}$ and ${\cal F}^{\rm sp}_{b,s,\pm}$ violate two-scale-factor universality.

In a similar way one can prove  the validity of multiparameter universality for the result of the singular bulk part of the free energy of the anisotropic  Gaussian model  for $t>0$
\begin{eqnarray}
\label{anisoGaussFISO}
{ \cal F}^{\rm G}_{b,s,+}(t,V)  =\left\{
\begin{array}{r@{\quad \quad}l}
                         \; Q^{\rm G}_1\; (\widetilde x^{\rm G})^{d/2}\quad          & \mbox{for} \;d>2, \\
                         \; \frac{1}{2} Q^{\rm G}_1\; \widetilde x^{\rm G} \ln t & \mbox{for} \;d=2, \;
                \end{array} \right.
\end{eqnarray}
with the Gaussian scaling variable
\begin{eqnarray}
\label{anisoGaussscalinvariable}
\widetilde x^{\rm G}&=&t[V^{\rm iso}/(\xi_{0+}^{\rm G, iso})^{d}]^{2/d}=t [V/(\bar \xi_{0+}^{\rm G})^{d}]^{2/d},
\end{eqnarray}
with the Gaussian mean correlation length (\ref{Gaussximeanx}), and with the same universal constant $Q^{\rm G}_1$, (\ref{Q1Gaussd})  and (\ref{Q1Gauss2}), as in the isotropic case. Eqs. (\ref{anisoGaussFISO}) and (\ref{anisoGaussscalinvariable}) are the anisotropic extension of (\ref{GaussFISO}) and (\ref{Gaussscalinvariable}) for the isotropic Gaussian model.

There exist further universal bulk amplitude relations of isotropic systems that involve the correlation length which are affected by anisotropy. As an example we consider the relation (\ref{amplrelisoxxy}).
The amplitude of the order  parameter ${\cal M}^{\rm sp}=B^{\rm sp}|t|^\beta$ of the anisotropic system is left invariant under the generalized shear transformation (\ref{transyy}),
\be
B^{\rm sp}=B^{\text {iso}},
\ee
in contrast to (\ref{orderpar}) and (\ref{orderparcor}) for the special shear transformation.
Together with the amplitude relation $\widehat \Gamma_+\equiv\Gamma_+^{\rm iso}=\Big(\xi^{\rm iso}_{0+}/\bar \xi^{\rm sp}_{0+}\Big)^{d}\;\Gamma_+^{\rm sp}$, (\ref{susdetermined}), this yields
\begin{eqnarray}
\label{amplrelisox}
(B^{\text {iso}})^2 (\Gamma_+^{\text {iso}})^{-1}(\xi^{\text {iso}}_{0+})^d=\big(B^{\rm sp}\big)^2 \big(\Gamma^{\rm sp}_+\big)^{-1}\big(\bar\xi^{\rm sp}_{0+}\big)^d= Q_c,\nonumber\\
\end{eqnarray}
with the same universal constant $Q_c$ for both isotropic and anisotropic systems. On the basis of the special shear transformation (\ref{shearalt})-(\ref{shearu}) the same relation with the same constant $Q_c $ was recently proven within the anisotropic $\varphi^4$ model in $d=2$ dimensions \cite{dohm2019} and was verified to be valid also
in the exactly solvable anisotropic Ising model \cite{WuCoy,Vaidya1976} where $Q_c$ was identified as given in Eq. (76) of \cite{dohm2019}. From  (\ref{xinfin}), (\ref{gammastrich}), and (\ref{Binfinstrich}) we confirm that (\ref{amplrelisox}) is also valid in the large-$n$ limit, with the universal constant
\be
Q_c= A_d/(4-d),\;\;\;\; n=\infty, \;\;2<d<4,
\ee
with $ A_d$ given by (\ref{Ad}).

In deriving the exact critical bulk amplitude relations (\ref{volinvarsp}), (\ref{3azneuzzz}),   and (\ref{amplrelisox}) for the $n$-vector model we have not made any assumptions other than the validity of two-scale-factor universality for isotropic systems and the existence of principal axes and correlation lengths for weakly anisotropic systems. No specific properties of the $n$-vector model were needed in the derivation. Thus these results are proven to be valid for arbitrary weakly anisotropic systems.
In a similar way, the validity of all the other bulk amplitude relations considered in \cite{dohm2008,dohm2018} for general $n$ and $d$ within anisotropic $\varphi^4$ model including those at finite external field (such as Eq. (3.35) of \cite{dohm2008}) can be proven for all weakly anisotropic systems. This feature is called multiparameter universality since these relations contain the same universal constants that appear already in the isotropic case but in the anisotropic case up to $d(d+1)/2+1$ independent nonuniversal parameters (rather than only two independent parameters)
are involved whose definition depends on the nonuniversal orientation of the principal axes and on the nonuniversal amplitudes of the principal correlation lengths.
\subsection{Bulk specific heat of anisotropic systems}
The universal bulk amplitude relations for the free energy  derived above are relevant for the analysis of the singular part
\begin{eqnarray}
\label{singbulkheat}
C^{\rm sp}_{b,s\pm}(t)&=&-\partial^2f^{\rm sp}_{b,s\pm}(t)/\partial t^2,
\end{eqnarray}
of the bulk specific heat per unit volume (divided by $k_B$) of anisotropic systems above and below $T_c$. Above $T_c$ it can be expressed in terms of the  mean correlation length $\bar \xi^{\rm sp}_{0+}$ defined in (\ref{ximeany}). We obtain from (\ref{bulk free sp})-(\ref{3azneuzzz})
\begin{eqnarray}
\label{specheatbaraniso}
C^{\rm sp}_{b,s+}(t) & =&\frac{\big(R^+_\xi\big)^d}{\alpha\; (\bar \xi^{\rm sp}_{0+})^d} \;t^{-\alpha},\;\;\;d>2,\\
\label{specheatdzweibaraniso}
C^{\rm sp}_{b,s+}(t) & =&-\;\frac{Q_1}{ \big(\bar \xi^{\rm sp}_{0+}\big)^2} \;\ln t,\;\;d=2,\;\;\;\;
\end{eqnarray}
above $T_c$ and 
\begin{eqnarray}
\label{specheatminus}
C^{\rm sp}_{b,s-}(t)  & =& \frac{A_-}{A_+}\;C^{\rm sp}_{b,s+}(t),\;\;d\geq 2,
\end{eqnarray}
below $T_c$.
Eqs. (\ref{specheatbaraniso}) and (\ref{specheatdzweibaraniso}) are the anisotropic extensions of (\ref{specheat}) and (\ref{specheatdzwei}).

Eqs. (\ref{specheatbaraniso})-(\ref{specheatminus}) demonstrate that thermodynamic measurements of the amplitude of specific heat of anisotropic systems can determine the mean correlation length $\bar \xi^{\rm sp}_{0+}$ of real anisotropic systems. Together with thermodynamic measurements of the amplitude $\Gamma_+$ of the susceptibility of anisotropic systems one obtains the two amplitudes determining the overall amplitude of the bulk correlation function (\ref{Gneuxxalt}). Together with the universal ratio (\ref{barxiratio}) this also determines  $\bar \xi^{\rm sp}_{\pm}(t)$ in the argument of $\Psi_\pm$ in (\ref{Gneuxxalt}).

These predictions are valid for any weakly anisotropic bulk system, e.g., for superconductors and magnetic materials. Relations equivalent to (\ref{specheatbaraniso}) have been employed previously \cite{schneider2004} in the analysis of experimental data of superconductors whose critical behavior belongs to the $(d=3,n=2)$ $XY$ universality class. In the earlier work \cite{schneider2004} the anisotropic fluctuations were treated within a Gaussian approximation for the case of a diagonal anisotropy matrix (effective mass tensor). Here we have provided a general and exact foundation for the bulk critical specific heat in weakly anisotropic systems.
\renewcommand{\thesection}{\Roman{section}}
\renewcommand{\theequation}{7.\arabic{equation}}
\setcounter{equation}{0}
\section{Correlation function of the two-dimensional Ising universality class}
The proof of multiparameter universality presented in the preceding sections has a significant impact on the
anisotropic bulk correlation function of the $(d=2,n=1 )$ universality class \cite{dohm2019}
with the exact critical exponents
\be
\nu=1, \;\;\eta=1/4.
\ee
 This includes both anisotropic Ising and  anisotropic scalar $\varphi^4$ models.
\subsection{Isotropic scaling form and exact universal bulk scaling function $\Psi_\pm(y_\pm)$ }
There exists a large variety of interactions and lattice structures of $(d=2,n=1)$ systems that have an isotropic bulk correlation function in the scaling region near $T_c$. For example, for the two-dimensional $\varphi^4$ lattice model (\ref{2a}) with short-range pair interactions the condition of isotropy is
\begin{eqnarray}
{\bf A}^{\rm iso}=
\left(\begin{array}{ccc}
 a & c \\
 c & b \\
\end{array}\right)= c^{\rm iso}_0 \left(\begin{array}{ccc}
 1 & 0\\
 0 &1 \\
\end{array}\right)
\end{eqnarray}
with $c^{\rm iso}_0>0$ which, for given lattice structure, is a condition for the  couplings $K_{i,j}$ according to (\ref{2i}). The same condition holds for the two-dimensional Gaussian lattice model (\ref{2aGauss}).
A corresponding general condition of isotropy for the couplings $E_{i,j}$ of two-dimensional Ising models (with the Hamiltonian $H^{\rm Is}$ given in (\ref{HspinIsing}) below) is unknown. If the principal correlation lengths $\xi^{\rm Is (\alpha)}_{0+}$ of Ising models are known as a function of the couplings $E_{i,j}$ the condition of isotropy reads
\be
\xi^{\rm Is (1)}_{0+}=\xi^{\rm Is (2)}_{0+}.
\ee
There exists an unlimited number of lattice structures and couplings satisfying these conditions of isotropy. There exist even two-dimensional systems with short-range multi-spin interactions within the Ising universality class that have isotropic correlation functions.
Owing to the principle of two-scale-factor universality all of these isotropic systems have the same structure of the bulk order-parameter correlation function above and below $T_c$ \cite{dohm2019}
\begin{eqnarray}
\label{3calt2d} G^{\rm iso} ({\bf x}, t) &=& \frac{\Gamma^{\rm iso}_+(\xi_{0+}^{\rm iso})^{-7/4}}{ | {\bf x} |^{1/4}} \;\Psi_\pm \Big(\frac{|{\bf x}|}{ \xi^{\rm iso}_{\pm}(t)} \Big), \;\\
\label{isocorr2d}
\xi^{\rm iso}_\pm(t)&=&\xi^{\rm iso}_{0\pm}\;|t|^{-1},
\end{eqnarray}
with the universal scaling function $\Psi_\pm$. To determine this function $\Psi_\pm $ it suffices to consider the simplest nontrivial example of this universality class. This is not the isotropic two-dimensional $\varphi^4$ model but the exactly solvable isotropic Ising model with equal nearest-neighbor (NN) couplings $E>0$ on a square lattice with the Hamiltonian
\be
\label{Isingiso}
H^{\rm Is,iso}= - E \sum_{j,k} [ \sigma_{j,k}  \sigma_{j,k+1}+ \sigma_{j,k}  \sigma_{j+1,k}], \;\; \sigma_{j,k}=\pm 1,
\ee
and with the lattice spacing $\tilde a=1$. The bulk correlation function of this model is
isotropic in the scaling limit (but not in the range
$|{\bf x}| / \xi^{\rm iso}_{\pm} \gg 1$ \cite{cd2000-1,cd2000-2,dan-2,dohm2008}).
It was calculated exactly in \cite{WuCoy}
where it was presented in a nonuniversal scaling form  without identifying
the universal part of the scaling function.
Recently \cite{dohm2019} we have written this correlation function in the
universal form (\ref{3calt2d}) and
have identified the universal scaling function  $\Psi_\pm(y_\pm)$ as well as
the arguments $y_\pm$,
together with the two nonuniversal amplitudes, specialized to the square lattice
with equal NN couplings,
\begin{eqnarray}
\label{xiamplsuare}
\xi_{0+}^{\rm squ, iso}&=& 2\big[\ln (1 + 2^{1/2})\big]^{-1},\\
\label{Gammaamplsuare}
\Gamma^{\rm squ, iso}_+&=& 2^{19/8}\pi \big(\xi_{0+}^{\rm squ, iso}\big)^{7/4} p_+,
\end{eqnarray}
where the constant $p_+$ is expressed in terms of a Painlev\'e function of the third kind,
as defined in Eqs. (30)-(36) of \cite{dohm2019} and in the text after these equations.
The exact universal amplitude ratio of the exponential correlation lengths is
\cite{WuCoy,pelissetto}
\be
\label{ratioxid2}
\xi^{\rm squ,iso}_{0 +}/\xi^{\rm squ,iso}_{0 -}=X_\xi= 2.
\ee
The scaling function
$\Psi_\pm(y_\pm)$
applies to all systems in the subclass of isotropic systems in the $(d=2,n=1)$
universality class,
in particular,
to Ising models with
isotropic bulk correlation functions on two-dimensional lattices other than the square lattice.
So far the exact results for such Ising models on "isotropic lattices" \cite{Perk2} (e.g. honeycomb lattice, see Fig. 1 in \cite{pri})
have not been discussed in the light
of the predictions of two-scale-factor universality with only two (rather than
four \cite{Perk2}) nonuniversal parameters, e.g., two independent nonuniversal amplitudes
corresponding
to (\ref{xiamplsuare}) and (\ref{Gammaamplsuare}) which are different for different
lattice structures.
\subsection{ Exact anisotropic bulk scaling form}
We apply our general analysis of Secs. IV and V to systems in the $(d=2,n=1)$ universality class. Examples for these systems are Ising models on two-dimensional Bravais lattices with short-range pair interactions described by the Hamiltonian
\begin{eqnarray}
\label{HspinIsing}
H^{\rm Is} = - \sum_{i,j} E_{i,j} \sigma_i \cdot \sigma_j, \;\; \sigma_i=\pm 1.
\end{eqnarray}
In the following all equations are given for these examples as indicated by the superscript $"\rm Is"$ for the nonuniversal quantities but these
equations apply to all weakly anisotropic systems of the $(d=2,n=1)$ universality class.
We assume the existence of two principal axes and two principal correlation lengths
in a range of couplings $E_{i,j}$ where the model (\ref{HspinIsing}) displays
weakly anisotropic critical behavior. The orientation of the principal axes in
the direction of the angles $\Omega^{\rm Is}$ and $\Omega^{\rm Is} +\pi/2$ is
described by the principal unit vectors
\begin{equation}
 \label{2pp1}
 {\bf e}^{(1){\rm Is}}={\bf e}(\Omega^{\rm Is}) = \left(\begin{array}{c}
  \cos \Omega^{\rm Is} \\
  \sin \Omega^{\rm Is} \\
 \end{array}\right)  ,
\end{equation}
 \begin{equation}
 \label{2pp2}
 {\bf e}^{(2){\rm Is}}={\bf e}(\Omega^{\rm Is}+\pi/2) = \left(\begin{array}{c}
  -\sin \Omega^{\rm Is} \\
   \cos \Omega^{\rm Is}\\
  \end{array}\right) .\;
\end{equation}
Correspondingly the rotation matrix of the generalized shear transformation for a clockwise rotation is
\begin{eqnarray}
\label{Urotz}
&&{\bf{ U}}^{\rm Is}={\bf{ U}}(\Omega^{\rm Is})=\left(\begin{array}{ccc}
 \cos\; \Omega^{\rm Is}& \sin\; \Omega^{\rm Is} \\
  -\sin\; \Omega^{\rm Is} &  \cos\; \Omega^{\rm Is} \\
\end{array}\right).\;\;\;\;
\end{eqnarray}
A counterclockwise rotation provided by the matrix
\begin{eqnarray}
\label{Urotcc}
&&{\bf{ U}}^{\rm Is}_{cc}={\bf{ U}}_{cc}(\Omega^{\rm Is})=\left(\begin{array}{ccc}
 \sin\; \Omega^{\rm Is}& -\cos\; \Omega^{\rm Is} \\
  \cos\; \Omega^{\rm Is} &  \sin\; \Omega^{\rm Is} \\
\end{array}\right)\;\;\;\;
\end{eqnarray}
would lead to equivalent results. The two principal correlation lengths
\begin{eqnarray}
\label{correl2}
\xi_{\pm}^{(\alpha){\rm Is}}(t)&=&\xi^{(\alpha){\rm Is}}_{0\pm}|t|^{-1}, \alpha=1,2,
\end{eqnarray}
have the nonuniversal ratio
\be
\label{qIs}
q^{\rm Is}=\xi^{(1){\rm Is}}_{0+}/\xi^{(2){\rm Is}}_{0+}=\xi^{(1){\rm Is}}_{0-}/\xi^{(2){\rm Is}}_{0-}.
\ee
The mean correlation length is
\begin{eqnarray}
\label{ximeanx2z}
\bar \xi^{\rm Is}_{\pm}(t)&=&\bar \xi^{\rm Is}_{0\pm}|t|^{-1},\\
\label{ximean0x2z}
\bar \xi^{\rm Is}_{0\pm}&=&\big[ \xi_{0\pm}^{(1){\rm Is}}\xi_{0\pm}^{(2){\rm Is}}\big]^{1/2},
\end{eqnarray}
which can be used to express the principal correlation lengths as
\be
\label{einszwei}
\xi^{(1){\rm Is}}_{0\pm}=
\bar \xi^{\rm Is}_{0\pm}(q^{\rm Is})^{1/2},\;\;\; \xi^{(2){\rm Is}}_{0\pm}=
\bar \xi^{\rm Is}_{0\pm}(q^{\rm Is})^{-1/2}.
\ee
The reduced rescaling matrix of the generalized shear transformation reads
\begin{eqnarray}
\label{lambdascal}
{\mbox {\boldmath$\bar\lambda$}^{\rm Is}}
=\left(\begin{array}{ccc}
 \bar\lambda^{\rm Is}_1& 0 \\
 0 & \;\bar\lambda^{\rm Is}_2 \\
\end{array}\right)
=\left(\begin{array}{ccc}
 q^{\rm Is}& 0 \\
 0 & \;(q^{\rm Is})^{-1} \\
\end{array}\right)\;\;\;\;\;\;\;
\end{eqnarray}
with the diagonal elements
\be
\label{widetildelambda2}
\bar\lambda^{\rm Is}_1=[\xi^{(1){\rm Is}}_{0\pm}/\bar\xi^{\rm Is}_{0\pm}]^2= q^{\rm Is},\;\;\;\;
\bar\lambda^{\rm Is}_2=[\xi^{(2){\rm Is}}_{0\pm}/\bar\xi^{\rm Is}_{0\pm}]^2=(q^{\rm Is})^{-1}\;.
\ee
The exact scaling form of the anisotropic correlation function of the model (\ref{HspinIsing})
is obtained from the general result (\ref{Gneuxxalt}) as
\begin{eqnarray}
\label{GneuxxaltIsingz}
G^{\rm Is}({\bf x}, t) &=& \frac{\Gamma^{\rm Is}_+\;(\bar \xi^{\rm Is}_{0+})^{-7/4}}{\big [{\bf x}\cdot \big({\bf \bar A}^{\rm Is}\big)^{-1}{\bf x}\big]^{1/8}}\;
 \Psi_\pm \Big(\frac{[{\bf x}\cdot \big({\bf \bar A}^{\rm Is}\big)^{-1}{\bf x}]^{1/2}}{\bar \xi^{\rm Is}_\pm(t)}\Big)\;\;\;\;\;\;\;\;\;
 \end{eqnarray}
with the same universal scaling function $\Psi_\pm$ as in
(\ref{3calt2d}). According to (\ref{Aquerxx})-(\ref{Aquerxxyy})
the reduced anisotropy matrix is given by
\begin{eqnarray}
\label{AquerIsing}
{\bf \bar A}^{\rm Is}(q^{\rm Is},\Omega^{\rm Is})&=&({\bf U}^{\rm Is})^{-1}{\bf \bar{\mbox {\boldmath$\lambda$}}}^{\rm Is} {\bf U}^{\rm Is}\\
&=&{\bf \bar A}(q^{\rm Is},\Omega^{\rm Is})
\end{eqnarray}
where ${\bf \bar A}(q,\Omega)$ has a universal structure given by
\begin{eqnarray}
\label{Aquer2z}
{\bf \bar A}(q,\Omega)
&=&\left(\begin{array}{ccc}
 q \;c_\Omega^2+q^{-1}s_\Omega^2 & \;\;\;(q-q^{-1})\;c_\Omega \;s_\Omega\\
(q-q^{-1})\; c_\Omega\; s_\Omega& q \;s_\Omega^2 +q^{-1}\;c_\Omega^2
\end{array}\right)\nonumber\\
\end{eqnarray}
with the abbreviations $c_\Omega\equiv\cos\Omega,s_\Omega\equiv\sin\Omega$. The result (\ref{GneuxxaltIsingz})-(\ref{Aquer2z}) has the same form as derived recently \cite{dohm2019} by means of the special shear transformation (\ref{shearalt})-(\ref{shearu}) for the two-dimensional anisotropic scalar $\varphi^4$ model (\ref{2a}). The universality of $\Psi_\pm$ and of the structure of ${\bf \bar A}(q,\Omega)$ confirms explicitly the validity of multiparameter universality for  the bulk correlation functions of all weakly anisotropic systems in the $(d=2,n=1)$ universality class.

We add the following comments. Two-scale-factor universality is violated as
$G^{\rm Is}({\bf x}, t)$, (\ref{GneuxxaltIsingz}), depends on the four independent nonuniversal parameters $\Gamma^{\rm Is}_+ , \bar \xi^{\rm Is}_{0+},q^{\rm Is},\Omega^{\rm Is}$. The angle $\Omega^{\rm Is}$ of the principal axes of Ising models  depends in an unknown way on the microscopic interactions $E_{i,j}$ and needs to be determined for each special Ising model under consideration.  This is in contrast to the known dependence \cite{dohm2019} of the corresponding angle
$\Omega$ of the anisotropic $d=2$ $\varphi^4$ model (\ref{2a}) on the couplings $K_{i,j}$ through the matrix elements of the anisotropy matrix ${\bf  A}$ (\ref{2i}). A corresponding anisotropy matrix for the Ising model is unknown.
Likewise a general condition for weak anisotropy for the Ising model (\ref{HspinIsing})
analogous to (\ref{condition}) for the $\varphi^4$ model is unknown.
The significance of our general result (\ref{GneuxxaltIsingz})-(\ref{Aquer2z}) is that the universal validity of the structure of the correlation function (\ref{GneuxxaltIsingz}) and of the reduced anisotropy matrix ${\bf \bar A}^{\rm Is}(q^{\rm Is},\Omega^{\rm Is})$ no longer rests upon exact calculations within special models on special lattices (such as square, triangular, or honeycomb lattices) or upon the hypothesis of multiparameter universality but is a proven fact that applies to all two-dimensional weakly anisotropic systems
of the $(d=2,n=1)$ universality class. This implies that the calculation of the correlation function of any two-dimensional weakly anisotropic system of the Ising universality class no longer requires a new calculation of a scaling function and of an anisotropy matrix but can be restricted to the much simpler task of determining four nonuniversal parameters, namely the thermodynamic amplitude $\Gamma^{\rm Is}_+$ of the susceptibility above $T_c$,  the two principal correlation lengths, and the angle $\Omega^{\rm Is}$ of the principal axes.
This conclusion constitutes a fundamental simplification in the analytic theory of two-dimensional anisotropic systems as well as in the analysis of numerical or experimental data of anisotropic correlation functions.

We substantiate our findings by the correlation function of the anisotropic
triangular-lattice Ising model \cite{stephenson,WuCoy,Vaidya1976} for which the exact
universal and nonuniversal properties have been identified in most cases \cite{dohm2019} from the explicit exact results that were given in \cite{WuCoy,Vaidya1976} as a function of the couplings $E_i$.
The Hamiltonian of this model reads
\be
\label{Htr}
H^{\rm tr}= \sum_{j,k} [-E_1 \sigma_{j,k}  \sigma_{j,k+1}-E_2 \sigma_{j,k}  \sigma_{j+1,k}-E_3 \sigma_{j,k}  \sigma_{j+1,k+1}]
\ee
with horizontal, vertical, and diagonal couplings $E_1,E_2,E_3$ on a square lattice (see Fig. 2 of \cite{dohm2019}). The condition for a ferromagnetic critical point with weak anisotropy is \cite{Berker}
\be
\label{range}
E_1+E_2 > 0, E_1+E_3 > 0, E_2+E_3 > 0.
\ee
In this range of the couplings the applicability of our proof of multiparameter universality is
guaranteed because of the existence of the principal axes and principal correlation lengths,
as shown in \cite{dohm2019}. Specifically, (i) the exact form of
${\bf \bar A}^{\rm tr}(q^{\rm tr},\Omega^{\rm tr})$  was identified
in \cite{dohm2019} in terms of $\hat S_i=\sinh 2 \beta_c^{\rm tr}E_i$ for general $E_i$, $i=1,2,3$, (ii) the existence of the two principal
axes was verified by the exact
determination of the angle $\Omega^{\rm Is}$ from the two extrema of the angular-dependent
correlation length, (iii) the exact ratio of $q^{\rm tr}$ of the principal correlation lengths was determined in \cite{dohm2019} in terms of $\hat S_i$
for general $E_i$,
and (iv) the existence of the mean correlation length $\bar \xi^{\rm tr}_{\pm}(t)$  follows from its identification through Eqs. (52)-(54) of \cite{dohm2019} in terms of the exact "scaled variable" $t^{\rm Vai}$ of Vaidya \cite{Vaidya1976} (denoted
by $t$ in Eq. (10) of \cite{Vaidya1976}).
The determination of $\bar \xi^{\rm tr}_{\pm}(t)$ of the Ising model in the full range of (\ref{range}) can be obtained by expanding the scaled variable in Eq. (10) of \cite{Vaidya1976} around $T^{\rm tr}_c$ to leading order in $|t|=|T-T^{\rm tr}_c|/T^{\rm tr}_c$, as noted in \cite{dohm2019}.
More explicitly, the scaled variable of Vaidya is identified according to Eq. (53) of \cite{dohm2019} for small $0<t\ll 1$ as
\be
\label{tVai}
 t^{\rm Vai}= \frac{R^{\rm tr}(t)}{\bar \xi^{\rm tr}_+(t)} \longrightarrow \frac{R^{\rm tr}(0)}{\bar \xi^{\rm tr}_{0+}}t
 \ee
where the distance variable $R^{\rm tr}(t)$ in Eq. (11) of \cite{Vaidya1976}
has a finite limit $R^{\rm tr}(0)=\lim_{t\to 0}R^{\rm tr}(t)$.
We have verified that the mean correlation  length $\bar \xi^{\rm tr}_{0+}$ is determined uniquely as a function of $E_i$ from (\ref{tVai}) and Eq. (10) of \cite{Vaidya1976} for general $E_i$.
This proves the existence of the two principal correlation lengths $\xi^{(\alpha){\rm tr}}_{0+ }$ according to (\ref{einszwei}) via the combination of $(q^{\rm tr})^{1/2}$, $(q^{\rm tr})^{-1/2}$, and $\bar \xi^{\rm tr}_{0+}$
 for general $E_i$. The amplitudes below $T_c$ follow from $\bar \xi^{\rm tr}_{0+}/\bar \xi^{\rm tr}_{0-}=2$. The expressions of $\bar \xi^{\rm tr}_{0+}$ and of $\xi^{(\alpha){\rm tr}}_{0+ }$ in terms of $\hat S_i$ are derived in \cite{DKW2023}.
 In retrospect, the verification of multiparameter universality of the structure of the correlation function of the special Ising model (\ref{Htr}) achieved in  \cite{dohm2019} was to be expected in view of the general proof of the present paper.
\renewcommand{\thesection}{\Roman{section}}
\renewcommand{\theequation}{8.\arabic{equation}}
\setcounter{equation}{0}
\section{Angular-dependent correlation vector}
In \cite{dohm2019} the notion of an angular-dependent correlation length was introduced.
In the following we further develop this notion by introducing the {\it vector of the angular-dependent correlation length}.
Since such vectors exists in all weakly $d$-dimensional anisotropic systems including $\varphi^4$, Gaussian, and fixed-length spin models we suppress the superscripts $"{\rm Is}"$, $"{\rm sp}"$, and "G" in the following. Here we confine ourselves to two dimensions.
It is of interest to introduce a correlation length as a measure of the spatial range of the critical correlations in a certain spatial direction with an angle $\theta$ described by a unit vector
\begin{eqnarray}
 {\bf e}(\theta) = \left(\begin{array}{c}
  \cos \theta \\
  \sin \theta \\
 \end{array}\right)  . \;
\end{eqnarray}
This can be done by introducing polar coordinates ${\bf x} = r\;{\bf e}(\theta) $
for the argument ${\bf x}$ of the anisotropic bulk correlation function $G({\bf x},t)$. As expected on physical grounds such a correlation length should not depend on the absolute value of $\theta$ but on the angle $\theta-\Omega$ relative to $\Omega$ describing the direction of the principal correlation lengths. This is indeed the case. By rewriting the argument of the scaling function $\Psi_\pm$ in (\ref{GneuxxaltIsingz}) as
\begin{eqnarray}
\frac{[{\bf x}\cdot {\bf \bar A}(q,\Omega)^{-1}{\bf x}]^{1/2}}{\bar \xi_\pm(t)}= \frac{r}{\xi_\pm(t,\theta-\Omega,q)}
\end{eqnarray}
we obtain from (\ref{Aquer2z}) the angular-dependent correlation length
\begin{eqnarray}
\label{anguldefin}
&&\xi_{\pm}(t,\theta-\Omega,q)=\xi_{0\pm}(\theta-\Omega,q)|t|^{-1},\\
\label{angulthetaz}
&&\xi_{0\pm}(\theta-\Omega,q)=\frac{\bar \xi_{0\pm}}{f(\theta-\Omega,q)},\\
\label{aaa}
&&f(\theta-\Omega,q)=[q\sin^2(\theta-\Omega)+q^{-1}\cos^2(\theta-\Omega)]^{1/2}.\;\;\;\;\;\;\;\;\;\;\;\;
\end{eqnarray}
Similarly the prefactor in (\ref{GneuxxaltIsingz}) can be rewritten in this form.
This yields the alternative representation of the correlation function in polar coordinates
\begin{eqnarray}
\label{GneuxxaltIsingr}
&&G({\bf x}, t) = \nonumber\\
&&\frac{\Gamma_+ \big(\bar \xi_{0+}\big)^{-7/4}}{ [r f(\theta-\Omega,q)]^{1/4}}\;
 \Psi_\pm \Big( \frac{r f(\theta-\Omega,q)}{\bar \xi_\pm(t)}\Big)=\nonumber\\
 &&\frac{\Gamma_+}{\big(\bar \xi_{0+}\big)^{2}}\Bigg(\frac{\xi_{0\pm}(\theta-\Omega,q)}{ r }\Bigg)^{1/4}\;
 \Psi_\pm \Big( \frac{r}{\xi_\pm(t,\theta-\Omega,q)}\Big).\;\;\;\;\;\;\;\;\;\;
 \end{eqnarray}
The exact angular dependence described by the function $f(\theta-\Omega,q)$ was first found  within the $\varphi^4$ model for general $K_{i,j}$ and the triangular-lattice Ising model (\ref{Htr}) in terms of $\hat S_i$ for general $E_i$ \cite{dohm2019}. Because of the universality of the structure of the reduced anisotropy matrix ${\bf \bar A}$, the function $f(\theta-\Omega,q)$ describes a universal structure of the $(\theta-\Omega)$-dependence for all weakly anisotropic two-dimensional systems including $\varphi^4$, Gaussian, and Ising models.
The usefulness of this correlation length is its property of having two extrema  with respect to $\theta$ at the angles
\be
\theta^{(1)}=\Omega, \; \theta^{(2)}=\Omega+\pi/2
\ee
which determine the two principal directions, i. e.,
\begin{eqnarray}
\label{corrvectoreins}
\xi_{0\pm}(0,q)&=&\bar \xi_{0\pm}q^{1/2}=\xi_{0\pm}^{(1)} ,\\
\label{corrvectorzwei}
\xi_{0\pm}(\pi/2,q)
&=&\bar \xi_{0\pm}q^{-1/2}=\xi_{0\pm}^{(2)}.
\end{eqnarray}
Here we extend the definition of the amplitude $\xi_{0\pm}(\theta-\Omega,q)$ to a definition of
an angular-dependent correlation vector for two-dimensional weakly anisotropic systems
\begin{eqnarray}
\label{corrvectorangult}
{\mbox {\boldmath$\xi$}}_\pm(t,\theta,\Omega,q)&=&{\mbox {\boldmath$\xi$}}_{0\pm}(\theta,\Omega,q)|t|^{-\nu},\\
\label{corrvectorangul}
{\mbox {\boldmath$\xi$}}_{0\pm}(\theta,\Omega,q)&=&\xi_{0\pm}(\theta-\Omega,q)\;{\bf e}(\theta)
\end{eqnarray}
that is oriented in the direction of the angle $\theta$.
This is a generalization of the principal correlation vectors (\ref{corrvectorphi}) and (\ref{corrvectorsp}) which are obtained from (\ref{angulthetaz})-(\ref{corrvectorangul}) for $\theta=\theta^{(1)}$ and $\theta=\theta^{(2)}$.
It can be verified that the generalized shear transformation
of Sec. IV. C with ${\bf \widehat U}(\Omega)={\bf U}(\Omega)$ defined in (\ref{Urotz})
indeed transforms the angular-dependent correlation vector
(\ref{corrvectorangul}) to a vector with a rescaled isotropic length $\xi^{\rm iso}_{0\pm}$,
\begin{eqnarray}
\label{transangular}
&{\mbox {\boldmath$\widehat\lambda$}}^{-1/2}& {\bf \widehat U}(\Omega)\;{\mbox {\boldmath$\xi$}}_{0\pm}(\theta,\Omega,q)=\xi^{\rm iso}_{0\pm}\;{\bf e}(\theta-\Omega,q)
\end{eqnarray}
with a unit vector ${\bf e}(\theta-\Omega,q)$ whose orientation depends on the relative angle $\theta-\Omega$,
\begin{eqnarray}
 \label{2pp1}
 {\bf e}(\theta-\Omega,q)=\frac{1}{\big[1+q^2\tan^2(\theta-\Omega)\big]^{1/2}} \left(\begin{array}{c}
  1  \\
  q\tan(\theta-\Omega)
 \end{array}\right),\nonumber\\
  \end{eqnarray}
\be
|{\bf e}(\theta-\Omega,q)|=1.
\ee
This agrees with the transformation of the principal correlation vectors discussed in Sec. IV.
An analogous result is obtained within the $\varphi^4$ and Gaussian models if the special shear transformation (\ref{shearalt}) is applied to ${\mbox {\boldmath$\xi$}}_{0\pm}(\theta,\Omega,q)$ where  $\xi^{\rm iso}_{0\pm}$ is replaced by $\xi'_{0\pm}$ (or $\xi_{0+}'^G$, respectively), i.e.,
\begin{eqnarray}
\label{transangularphi}
&{\mbox {\boldmath$\lambda$}}^{-1/2}& {\bf U}(\Omega)\;{\mbox {\boldmath$\xi$}}_{0\pm}(\theta,\Omega,q)=\xi'_{0\pm}\;{\bf e}(\theta-\Omega,q)
\end{eqnarray}
with the same unit vector ${\bf e}(\theta-\Omega,q)$.
The same angular-dependent representation can be employed in the shear transformation applied to any lattice point ${\bf x}$  of the anisotropic system.
For the generalized shear transformation (\ref{transyy}) this yields the transformed vector
\begin{eqnarray}
\label{transangularx}
{\bf \widehat x}=
&{\mbox {\boldmath$\widehat\lambda$}}^{-1/2}& {\bf \widehat U}(\Omega)\;{\bf x}=\frac{\xi^{\rm iso}_{0\pm}}{\xi_{0\pm}(\theta-\Omega,q)}\;|{\bf x}|\;{\bf e}(\theta-\Omega,q)\;\;\;\;\;\;\;\;\;\;\;\;
\\
\label{transangularxzz}
&=&\frac{\xi^{\rm iso}_{0\pm}}{\bar \xi_{0\pm}}\;|{\bf x}|\;f(\theta-\Omega,q)\;{\bf e}(\theta,q,\Omega)
\end{eqnarray}
where the factor $\xi^{\rm iso}_{0\pm}/\xi_{0\pm}(\theta-\Omega,q)$ describes the amount of rescaling in the direction of $\theta$. For the special shear transformation (\ref{shearalt}) of the $\varphi^4$ and Gaussian models the corresponding representation of the vector ${\bf x'}$ reads
\begin{eqnarray}
\label{transangularxprime}
{\bf x'}=
&{\mbox {\boldmath$\lambda$}}^{-1/2}& {\bf U}(\Omega)\;{\bf x}=\frac{\xi'_{0\pm}}{\xi_{0\pm}(\theta-\Omega,q)}\;|{\bf x}|\;{\bf e}(\theta-\Omega,q)\;\;\;\;\;\;\;\;\;\;\;\;
\\
\label{transangularxzzprime}
&=&\frac{\xi'_{0\pm}}{\bar \xi_{0\pm}}\;|{\bf x}|\;f(\theta-\Omega,q)\;{\bf e}(\theta-\Omega,q)
\end{eqnarray}
where the rescaling factor is $\xi'_{0\pm}/\xi_{0\pm}(\theta-\Omega,q)$ (or  $\xi'^G_{0\pm}/\xi^G_{0\pm}(\theta-\Omega,q)$ for the Gaussian model).

Although the function $f(\theta-\Omega,q)$ has a universal form describing the angular dependence of the correlation length of all weakly anisotropic two-dimensional systems one should keep in mind that it contains a substantial source of an intrinsic diversity through the dependence on the nonuniversal angle $\Omega$ and the ratio $q$. Both $\Omega$ and $q$ depend on all microscopic details. Within the $\varphi^4$ theory $q$ and $\Omega$ are well defined via the eigenvalues and  the orientation of the eigenvectors of the anisotropy matrix ${\bf A}$ \cite{dohm2019} but they are unknown for, e. g., an unlimited number of weakly anisotropic Ising models on various lattices with short range interactions.

We are not aware of an analytic approach to determining the principal axes of such systems. This serious lack of knowledge is a nontrivial source of nonuniversality in the physics of weakly anisotropic systems. It causes a directional nonuniversality of two- and three-dimensional correlation functions \cite{dohm2019} that has not been anticipated in the traditional theory where it was believed that weak anisotropy enters only the amplitudes of power laws \cite{zia}. As long as the principal axes and correlation lengths are unknown for an anisotropic system no appropriate rotation matrix can be defined, and not only the required {\it amount} of rescaling is unknown but also the {\it directions} are unknown along which the rescaling should be performed. Thus, in general, the effects of weak anisotropy cannot be simply "transformed away" by a "trivial rescaling". This problem has not been adequately addressed in the earlier literature on weakly anisotropic bulk and confined systems \cite{cardy1987,cardybuch,zia,diehl-chamati,diehl2010}. In particular in confined systems these anisotropy effects may become unexpectedly complex \cite{DW2021}.

The notion of an angular-dependent representation of correlation vectors and
lattice points is applicable to both bulk and confined systems with arbitrary
boundary conditions. This is of relevance to the application of the shear
transformation to the boundaries of finite systems as will be further discussed in Sec. IX. A.
\renewcommand{\thesection}{\Roman{section}}
\renewcommand{\theequation}{9.\arabic{equation}}
\setcounter{equation}{0}

\section{Critical Casimir forces in anisotropic systems}
\subsection{Proof of multiparameter universality of the critical Casimir amplitude in two dimensions}
Recently exact results have been derived for the critical free energy
and the ensuing critical Casimir amplitude
of the anisotropic $\varphi^4$ model at $T_c$ on a finite  rectangle with periodic boundary conditions \cite{DW2021}. Surprisingly complex finite-size effects were found near the instability where weak anisotropy breaks down. These exact results were
based on  conformal field theory  \cite{franc1997} and the principle of two-scale-factor universality \cite{priv,pri} for finite isotropic systems combined with the special shear transformation (\ref{shearalt})-(\ref{shearu}) of the anisotropic $\varphi^4$ model on a rectangle to an isotropic $\varphi^4$ model on a parallelogram.
Corresponding predictions were presented for the finite anisotropic
triangular-lattice Ising model \cite{Vaidya1976,dohm2019} on the basis of the assumption that multiparameter universality \cite{dohm2018} is valid for this model but no proof was given. A quantitative test was performed \cite{DWKS2021} by high-precision Monte Carlo simulations for a special Ising model on a square with a diagonal anisotropic coupling and remarkable agreement was found with the predicted critical amplitude ${\cal F}_c^{\rm Is}$ of the free energy at $T_c$.

It was noted \cite{DW2021} that the assumption of multiparameter universality for the anisotropic triangular-lattice Ising model is equivalent to an "effective shear transformation" between the isotropic Ising model on a parallelogram and the anisotropic Ising model on a rectangle but this effective shear transformation was not specified.
Here we shall identify this effective shear transformation together with a proof for the validity of multiparameter universality for critical free energy ${\cal F}_c^{\rm Is}$ of the anisotropic Ising model. Our proof is based on (a) the generalized shear transformations (\ref{transyy}) and  (\ref{transangularx}) applied to the boundaries of the Ising model, (b) the invariance of the critical free energy under this shear transformation as discussed in Sec. II. B, (c) the inversion of this shear transformation. The strategy is to perform the generalized shear transformation from the anisotropic rectangle to an isotropic parallelogram, to take the exact critical free energy on the parallelogram from isotropic CFT \cite{franc1997}, and to determine the critical free energy on the anisotropic rectangle by means of inverting the shear transformation.  This can be done without recourse to the $\varphi^4$ model and no assumptions are needed other than the existence of weakly anisotropic behavior, i.e., the existence of principal axes with the angle $\Omega^{\rm Is}$ and the ratio $q^{\rm Is}$ of the principal correlation lengths. We recall that both  $\Omega^{\rm Is}$ and $q^{\rm Is}$ are known exactly for the triangular-lattice Ising model (\ref{Htr}) \cite{dohm2019}.

We assume a finite $L_\parallel \times L_\perp$ rectangle spanned by the vectors ${\bf L}_\parallel=L_\parallel\;(1,0)$  and ${\bf L}_\perp=L_\perp\;(0,1)$ in the horizontal and vertical directions with the aspect ratio
\be
\rho=L_\perp/L_\parallel.
\ee
We apply our generalized shear transformation (\ref{transyy}) directly to the anisotropic Ising model on the rectangle in order to obtain an isotropic Ising model on a parallelogram.
The transformation applied to the boundaries of the rectangle reads
\begin{eqnarray}
{\bf L}_{p \parallel}&=&({\mbox {\boldmath$\lambda$}}^{\rm Is})^{-1/2} {\bf U}(\Omega^{\rm Is}){\bf L_\parallel},
\\
{\bf L}_{p \perp}&=&({\mbox {\boldmath$\lambda$}}^{\rm Is})^{-1/2} {\bf U}(\Omega^{\rm Is}){\bf L}_\perp,
\end{eqnarray}
where ${\bf U}(\Omega^{\rm Is})$ is given by (\ref{Urotz}) and
\begin{eqnarray}
\label{lambdascalz}
{\mbox {\boldmath$\lambda$}^{\rm Is}}
=\left(\begin{array}{ccc}
 \lambda^{\rm Is}_1& 0 \\
 0 & \;\lambda^{\rm Is}_2 \\
\end{array}\right)
\end{eqnarray}
with the diagonal elements
\be
\label{widetildelambda2z}
\lambda^{\rm Is}_1=[\xi^{{\rm Is}(1)}_{0\pm}/\xi^{\rm iso}_{0\pm}]^2,\;\;\;\;
\lambda^{\rm Is}_2=[\xi^{{\rm Is}(2)}_{0\pm}/\xi^{\rm iso}_{0\pm}]^2\;,
\ee
(compare (\ref{choicelambdaxyy})). This
generates an isotropic Ising model with the bulk correlation-length amplitude $\xi^{\rm iso}_{0\pm}$ on a parallelogram spanned by the vectors
${\bf L}_{p \parallel}$ and ${\bf L}_{p \perp} $. From the general transformation formulae (\ref{transangularx}), (\ref{angulthetaz}), (\ref{aaa}), and  (\ref{2pp1}) we obtain for $\theta=0$ and $\theta = \pi/2$
\begin{eqnarray}
\label{Lparaz}
{\bf L}_{p \parallel}&=&L_\parallel\; \frac{\xi^{\rm iso}_{0\pm}/\xi^{\rm Is}_{0\pm}(-\Omega^{\rm Is},q^{\rm Is})}{\big[1+(q^{\rm Is})^2\tan^2\Omega^{\rm Is}\big]^{1/2}} \left(\begin{array}{c}
  1  \\
 - q^{\rm Is}\tan\Omega^{\rm Is}
 \end{array}\right),  \;\;\;\;\;\;\;\;
\\
\label{Lz}
{\bf L}_{p \perp }&=&L_\perp\;\frac{\xi^{\rm iso}_{0\pm}/\xi^{\rm Is}_{0\pm}(\pi/2-\Omega^{\rm Is},q^{\rm Is})}{\big[1+(q^{\rm Is})^2\cot^2\Omega^{\rm Is}\big]^{1/2}} \left(\begin{array}{c}
  1  \\
  q^{\rm Is}\cot\Omega^{\rm Is}
 \end{array}\right),   \;\;\;\;\;\;
\end{eqnarray}
with
\begin{eqnarray}
|{\bf L}_{p \parallel}|&=&L_\parallel\; \xi^{\rm iso}_{0\pm}/\xi^{\rm Is}_{0\pm}(-\Omega^{\rm Is},q^{\rm Is}),\\
|{\bf L}_{p\perp }|&=&L_\perp\;\xi^{\rm iso}_{0\pm}/\xi^{\rm Is}_{0\pm}(\pi/2-\Omega^{\rm Is},q^{\rm Is}),
\end{eqnarray}
where we have used $\tan(\pi/2-\Omega)=\cot \Omega$. We see that the ratio $\xi^{\rm iso}_{0\pm}/\xi^{\rm Is}_{0\pm}(\theta-\Omega^{\rm Is},q^{\rm Is}))$ determines the rescaling of the lengths in the directions $\theta=0$ and $\theta=\pi/2$.
The  parallelogram is characterized by the transformed aspect ratio
\be
\label{rhoparaaspect}
\rho_{\rm p}=|{\bf L}_{p\perp }|/|{\bf L}_{p \parallel}|
\ee
and by the angle $\alpha$ between the vectors ${\bf L}_{p\perp }$ and ${\bf L}_{p \parallel}$ (Fig.2 of \cite{DW2021}). This angle is related to ${\bf L}_{p\perp }$ and ${\bf L}_{p \parallel}$ by
\be
\label{alphapara}
\cos \alpha=\frac{{\bf L}_{p \perp}\cdot{\bf L}_{p \parallel}}{|{\bf L}_{p\perp }||{\bf L}_{p \parallel}|}.
\ee
 Substituting (\ref{Lparaz}) and
(\ref{Lz}) into (\ref{rhoparaaspect}) and
(\ref{alphapara}) and  using (\ref{angulthetaz}) and (\ref{aaa})
we obtain the transformed aspect ratio
\begin{eqnarray}
\label{ratioparazz}
\rho_{\rm{p}}(\rho,q^{\rm{Is}},\Omega^{\rm{Is}})&=&\frac{L_\perp/\xi^{\rm{Is}}_{0+}(\pi/2-\Omega^{\rm{Is}},q^{\rm{Is}})}{L_\parallel/\xi^{\rm{Is}}_{0+}(-\Omega^{\rm{Is}},q^{\rm{Is}})} \\
\label{lengthratioPRL}
&=& \rho\; \Bigg[\frac{\tan^2\Omega^{\rm Is}+(q^{\rm Is})^2}{1+(q^{\rm Is})^2\tan^2\Omega^{\rm Is}}\Bigg]^{1/2}\;\;\;
\\
\label{elementratio}
&=&\rho\;
\Bigg[\frac{{\bf \bar A}(q^{\rm Is},\Omega^{\rm Is})_{11}}{{\bf \bar A}(q^{\rm Is},\Omega^{\rm Is})_{22}}\Bigg]^{1/2}
\end{eqnarray}
and the angle $\alpha$ determined by
\begin{eqnarray}
\label{alphaising}
\cot \alpha(q^{\rm Is},\Omega^{\rm Is})
&=& \frac{ (q^{\rm Is})^{-1}-q^{\rm Is}\tan \Omega^{\rm Is} \;\tan(\pi/2 - \Omega^{\rm Is})}{\tan \Omega^{\rm Is}+ \tan(\pi/2-\Omega^{\rm Is})}\;\;\;\;\;\;\;\;\;\;
\\
\label{alphaalt}
&=&[(q^{\rm Is})^{-1}-q^{\rm Is}]\cos \Omega^{\rm Is} \sin \Omega^{\rm Is}
\\
\label{element21}
&=& - {\bf \bar A}(q^{\rm Is},\Omega^{\rm Is})_{12}
\end{eqnarray}
where the matrix elements ${\bf \bar A}(q^{\rm Is},\Omega^{\rm Is})_{\alpha \beta}$ are defined in (\ref{AquerIsing})-(\ref{Aquer2z}). We note that the free parameter  $\xi^{\rm iso}_{0\pm}$ in (\ref{Lparaz}) and
(\ref{Lz}) is canceled in all subsequent equations, as expected.

No specific properties of the weakly anisotropic Ising model were needed in the derivation of (\ref{ratioparazz})-(\ref{element21}), thus these relations have a universal structure. In Eqs. (7) and (8) of \cite{DW2021} analogous
formulae corresponding to (\ref{lengthratioPRL}),
(\ref{elementratio}) and (\ref{alphaalt}),
(\ref{element21}), but with $q^{\rm Is},\Omega^{\rm Is}$ replaced by $q,\Omega$, were first presented for the $\varphi^4$ model. These formulae were obtained on the basis of geometric considerations as an extended version of the  derivation in the context of Fig. 2 in \cite{dohm2006} for the $\varphi^4$ model (see also Eqs. (16) and (17) of \cite{dohm2006} for the special shear transformation from an anisotropic square to an isotropic rhombus for $\Omega=\pi/4$). These formulae for the $\varphi^4$ model were then adopted in \cite{DW2021} for the Ising model by the substitution $q \to q^{\rm Is}, \Omega \to \Omega^{\rm Is}$ on the basis of the assumption that multiparameter universality is valid for the Ising model. Here we have provided an exact analytic derivation for this substitution directly within the Ising model, without recourse to the $\varphi^4$ model, through the general transformation formula (\ref{transangularx}) derived from our generalized shear transformation (\ref{transyy}). Thus this transformation between anisotropic and isotropic Ising models specifies what was called "effective shear transformation" in Fig. 1 of \cite{DW2021}.

The critical free energy of the transformed isotropic Ising model on the parallelogram is denoted by ${\cal F}^{\rm Is,iso}_c$. As noted in Sec. III. D, two-scale-factor universality implies that ${\cal F}^{\rm Is,iso}_c$ is universal, i.e., it has a universal dependence on the geometric parameters $\alpha$ and $\rho_{\rm p}$ of the parallelogram. Since the critical free energy ${\cal F}^{\rm Is}_c$ on the anisotropic  rectangle remains invariant under this pure coordinate transformation (see Sec. II. B) we obtain
\begin{eqnarray}
\label{FcIs}
{\cal F}^{\rm Is}_c={\cal F}^{\rm Is,iso}_c(\alpha,\rho_{\rm p}).
\end{eqnarray}
As pointed out in \cite{DW2021} the exact result for ${\cal F}^{\rm Is,iso}_c(\alpha,\rho_{\rm p})$ for periodic boundary conditions can be taken directly from the critical free energy ${\cal F}_c^{\rm{CFT}}(\tau) $ of conformal field theory for the isotropic Ising model on a torus \cite{franc1997}
\begin{eqnarray}
\label{FcZCFT}
{\cal F}^{\rm Is,iso}_c(\alpha,\rho_{\rm p})&=&{\cal F}_c^{\rm{CFT}}(\tau)= -\ln Z^{\rm{CFT}}(\tau),
\\
\tau(\alpha,\rho_{\rm{p}})& =& \mbox{Re}\; \tau + i\; \mbox{Im} \;\tau=\rho_{\rm{p}}\exp(i\; \alpha)\;\;\;\;\;
\end{eqnarray}
which is characterized by the complex torus modular parameter $\tau(\alpha,\rho_{\rm{p}})$. The $\tau$-dependence of ${\cal F}_c^{\rm{CFT}}(\tau) $ is universal. The partition function $Z^{\rm{CFT}}(\tau)$ is expressed in terms of Jacobi theta functions $\theta_i(0|\tau)\equiv \theta_i(\tau)$ as \cite{franc1997}
\begin{equation}
\label{ZIsing}
Z^{\rm{CFT}}(\tau)=\big({|\theta_2(\tau)|+|\theta_3(\tau)|+|\theta_4(\tau)|}\big)/\big({2|\eta(\tau)|}\big),
\end{equation}
with
$\eta(\tau)=\Big[\frac{1}{2}\theta_2(\tau)\theta_3(\tau)\theta_4(\tau)\Big]^{1/3}$.
The crucial step is to transfer this exact information directly from the isotropic Ising model to the anisotropic Ising model by means of inverting the generalized shear transformation, without recourse to the  $\varphi^4$ model. This is achieved by defining the $(q^{\rm Is} ,\Omega^{\rm Is})$-dependent quantity
\begin{equation}
\label{tauqomega}
\tau(\rho,q^{\rm Is},\Omega^{\rm Is})= \tau\big(\alpha(q^{\rm Is},\Omega^{\rm Is}),\rho_{\rm{p}}(\rho,q^{\rm Is},\Omega^{\rm Is})\big)
\end{equation}
with $\alpha(q^{\rm Is},\Omega^{\rm Is})$ and $\rho_{\rm{p}}(\rho,q^{\rm Is},\Omega^{\rm Is})$ given by (\ref{alphaalt}) and (\ref{lengthratioPRL}) and by substituting (\ref{tauqomega}) into the isotropic formula (\ref{FcZCFT}). Using (\ref{FcIs}) we then obtain the exact result for the critical free energy ${\cal F}^{\rm Is}_c$ and the  Casimir amplitude $X^{\rm Is}_c$ of the anisotropic Ising model on the rectangle
as
\begin{eqnarray}
\label{FIS}
{\cal F}^{\rm Is}_c(\rho,q^{\rm Is},\Omega^{\rm Is})&=&-\ln Z^{\rm{CFT}}\big(\tau(\rho,q^{\rm Is},\Omega^{\rm Is})\big), \\
\label{FcasCFT}
X^{\rm Is}_c(\rho,q^{\rm Is},\Omega^{\rm Is})&=& -\rho^2 \;\partial {\cal F}^{\rm Is}_c(\rho,q^{\rm Is},\Omega^{\rm Is})/\partial\rho.
\end{eqnarray}
with two nonuniversal parameters $q^{\rm Is},\Omega^{\rm Is}$. Going from (\ref{FcZCFT}) to (\ref{FIS}) is equivalent to performing a nonuniversal inverse shear transformation from the isotropic to the anisotropic Ising model.
The difference between (\ref{FcZCFT}) and (\ref{FIS}) is that the former relation contains a universal function of the geometric variables $\alpha$ and $\rho_{\rm p}$ in agreement with two-scale factor universality whereas (\ref{FIS}) contains  a universal function  reflecting multiparameter universality with the two  nonuniversal anisotropy parameters $q^{\rm Is}$ and $\Omega^{\rm Is}$. They are introduced via the nonuniversal anisotropy parameters contained in the shear transformation formulae (\ref{ratioparazz})-(\ref{element21}).
The same structure was derived for the $\varphi^4$ model in \cite{DW2021}.
This proves the validity of the predictions of \cite{DW2021} not only for the triangular lattice Ising model but more generally for all other weakly anisotropic systems that belong to the $d=2,n=1$ universality class. In particular this establishes the validity of the self-similar structures of the critical free energy and the Casimir amplitude discovered in \cite{DW2021} for all weakly anisotropic systems with periodic boundary conditions in this universality class.

As noted in the context of the anisotropic {\it bulk} correlation function,  there exists a corresponding intrinsic diversity in {\it confined} anisotropic systems as compared to confined isotropic systems: It arises from the nonuniversal parameters $q,\Omega$ and $q^{\rm Is},\Omega^{\rm Is}$ that do not exist in isotropic systems. Furthermore there is a basic difference between $q(\{K_{i,j}\})$ and $\Omega(\{K_{i,j}\})$ of the $\varphi^4$ model (which are known exactly as functions of  $K_{i,j}$  \cite{dohm2019}) and $q^{\rm Is}(\{E_{i,j}\})$ and $\Omega^{\rm Is}(\{E_{i,j}\})$ of the Ising model which are generically unknown for the general Ising model (\ref{HspinIsing}). We conclude that weak anisotropy destroys the universality of the critical Casimir amplitude $X_c$ of the subclass of isotropic systems and makes $X_c$ to become an unknown quantity for general anisotropic systems whose principal angles and correlation lengths are unknown.
The triangular-lattice Ising model \cite{Vaidya1976,dohm2019} is a very rare example for a system other than the $\varphi^4$ model for  which these parameters are known exactly as a function of the microscopic couplings. It would be worthwhile to extend this knowledge to other anisotropic Ising models \cite{Perk1,Perk2,Perk3,Perk4} whose relevant anisotropy parameters are as yet unknown.

\subsection{Critical Casimir forces in anisotropic superconductors}
An  experimental verification of critical Casimir forces
has been achieved so far only in isotropic systems, most prominently in superfluid $^4$He films \cite{garcia}.
It has been pointed out by Williams \cite{wil-1} that a measurable critical Casimir force should occur also in superconducting films. Superconductors belong to the same universality class as superfluid  $^4$He and have the same (Dirichlet) boundary conditions but are anisotropic. Williams' scenario is the following: a superconducting film is connected to a bulk sample of the same material. He argues that below the bulk critical temperature the film-bulk-system can lower its free energy by a transfer of electrons (Cooper pairs) from the film to the bulk system which is analogous to helium atoms moving from the film to the superfluid bulk reservoir. While in the helium system this leads to a thinning of the film the corresponding effect in the superconducting film-bulk-system is a transfer of negative electrical charge from the film to the bulk system. Williams  \cite{wil-1} argues that this gives rise to an electrical potential difference which can be related to the free-energy difference (per unit area) between the film and the bulk and from which a Casimir force can be derived. He estimates the voltage difference to have a measurable magnitude.

The critical Casimir force is an observable only if the ordering degrees of freedom can enter and leave the system. Therefore it has been claimed in the literature \cite{toldin,diehl2009,diehl2010,DDreview} that this force can be active only in isotropic fluids and that the issue of spatial anisotropy is not relevant in the context of the critical Casimir force. We argue that, unlike the localized degrees of freedom (magnetic moments) of the order parameter of an anisotropic magnetic material, the ordering degrees of freedom (Cooper pairs) of a superconductor are not localized at lattice points but play the role of an electrical superfluid in an {\it anisotropic} environment that can leave and enter the film connected to the bulk of the same material, as anticipated by Williams \cite{wil-1}. So far no specific objection has been raised in the literature against this specific argumentation for superconductors, and in a comment \cite{comment} on \cite{wil-1} the measurability of the critical Casimir force in superconductors has not been questioned. Furthermore, we point to the largely unexplored area of thermodynamic Casimir forces in liquid crystals \cite{singh} which exhibit a wide variety of spatial anisotropy and whose ordering degrees of freedom can leave and enter the system.

In closing we note that experimental studies, Monte Carlo simulations, and further theoretical research are called for in view of the fact that at present no experimental or Monte Carlo data and no analytic predictions are available for the critical Casimir force in anisotropic systems with realistic boundary conditions. Also theoretical efforts based on the functional renormalization group \cite{metzner2021} applied to anisotropic confined systems could yield important contributions to this matter. Analytic results for the critical Casimir force in anisotropic films of finite thickness have been presented previously \cite{dohm2018} for the case of periodic boundary conditions which refute earlier results \cite{wil-1} where no anisotropy effect in anisotropic superconductors near $T_c$ was found. An analytic renormalization-group study in three dimensions with realistic Dirichlet boundary conditions  below $T_c$ without adjustable parameters was performed \cite{dohm2014} that explains the depth and position of the deep minimum of the Casimir force scaling function observed in isotropic $^4$He films \cite{garcia} in the temperature regime  $T_{\rm c,film}<T<T_c$ on a semiquantitative level. It is conceivable that this isotropic study can be extended to anisotropic film systems with Dirichlet boundary conditions which could lead to quantitative predictions of the critical Casimir force in real superconductors.
\renewcommand{\thesection}{\Roman{section}}
\renewcommand{\theequation}{10.\arabic{equation}}
\setcounter{equation}{0}

\section{Summary }
We have presented a general theory of bulk critical phenomena in weakly
$d$-dimensional $O(n)$-symmetric anisotropic systems in $2\leq d < 4$ dimensions
where our only assumptions are the existence of $d$ principal axes and correlation lengths
and the validity of two-scale-factor universality for isotropic systems.
Our general conclusions with regard to the validity of multiparameter universality
confirms and specifies the early
"belief in some form of universality not only for the two-dimensional Ising model but
also for a large class of two-dimensional models with short-range interactions" \cite{WuCoy}.
Our findings are supported by exact results for the $d=2,n=1$ universality class
and for the spherical and Gaussian universality classes in $d\geq2$ dimensions. On the other hand our theory reveals a high degree of intrinsic diversity in weakly anisotropic systems even in the asymptotic critical region. This limitation of universality was not anticipated in the traditional theory of critical phenomena and in the more recent development of the functional renormalization group.
Furthermore we have applied our theory to finite-size effects at $T_c$ of the
critical free energy and Casimir amplitude of anisotropic systems on a rectangle with periodic
boundary conditions of the $d=2,n=1$ universality class studied recently \cite{DW2021}.
Our main results are summarized in more detail as follows.

(i) After defining the anisotropic $O(n)$-symmetric $\varphi^4$ and $n$-vector models in Sec. II and summarizing several aspects of two-scale-factor universality in Sec. III we have introduced in Sec. IV  a generalized shear transformation that provides
exact relations between weakly anisotropic systems and isotropic
systems in the same universality class.
We have identified a temperature-independent universal structure of a reduced anisotropy matrix ${\bf \bar A}$ that depends on the ratios of principal correlation lengths and on the principal unit vectors describing the principal axes. The latter quantities depend on microscopic details such as coupling constants and the lattice structure, thus ${\bf \bar A}$  is a nonuniversal quantity that does not exist in isotropic systems.

(ii) In Sec. V a proof has been presented for the validity of multiparameter universality of the bulk order-parameter correlation function in  weakly anisotropic systems. It implies that the traditional notion of a universality class of critical phenomena must be revised in that it must be divided
into subclasses of isotropic and weakly anisotropic systems. The latter 
have up to $d(d+1)/2+1$ independent nonuniversal parameters. 
Only two of these parameters can be expressed in terms of thermodynamic amplitudes whereas the remaining $d(d+1)/2-1$ parameters enter the matrix ${\bf \bar A}$. We have also determined the exact structure of the anisotropic bulk order-parameter correlation function for $n>1$ in the Goldstone regime below $T_c$. Exact anisotropic results are presented in the large-$n$ limit and for the Gaussian model.
From the nonasymptotic nonuniversal result of the functional renormalization group \cite{hassel2007,sinner2008} we have identified the universal part of the isotropic correlation function of the $n=1$ Ising universality class in three dimensions. We have refuted the claim that an extended universality \cite{hassel2007,sinner2008} is valid in the nonasymptotic region. Our theory provides the opportunity of a quantitative comparison with the universal parts of numerical or experimental data in two-dimensional and three-dimensional anisotropic systems after the nonuniversal parameters have been determined.

(iii) In Sec. VI a bulk scaling variable $\widetilde x$, (\ref{scalinvariable}),
has been introduced for general $n$ in $2\leq d<4$ dimensions which is invariant under the shear transformation.
It permits us to represent the singular bulk part of the free energy of isotropic and anisotropic systems in a compact form.
We have also shown the validity of multiparameter universality of several critical bulk amplitude relations. In particular the amplitude of the bulk specific heat of anisotropic systems is shown to be universally related to the mean correlation length.

(iv) In Sec.  VII our theory has been applied to two dimensions.
The significance of our general results 
is that the universal validity of the structure of the correlation function
no longer rests upon exact calculations within special models on special lattices
or upon the hypothesis of multiparameter universality but is a proven fact that applies to all two-dimensional weakly anisotropic systems
of the $(d=2,n=1)$ universality class.
This
constitutes a fundamental simplification in the analytic theory
of two-dimensional anisotropic systems as well as in the analysis of numerical or experimental data.
 
(v) In Sec. VIII the anisotropic correlation function is written in terms of polar coordinates. An angular-dependent
correlation vector is introduced for all two-dimensional weakly anisotropic systems. Angular-dependent formulae 
of the shear transformation are derived that are applicable to any lattice vector in bulk and confined systems.

(vi) Application of these formulae to the boundaries  of  a finite anisotropic rectangle
in Sec. IX provides an analytic derivation of the aspect ratio and the angle of the transformed isotropic parallelogram.
Combining this result with the exact partition function of conformal field theory of the isotropic two-dimensional Ising model on a torus at $T_c$
proves the validity of the predictions of \cite{DW2021}
with regard to multiparameter universality and self-similar structures of the critical free energy and the Casimir amplitude
not only for the triangular lattice Ising model but more generally for all weakly anisotropic Ising models and other systems with periodic boundary conditions that belong to the $d=2,n=1$ universality class. In particular this identifies the previous  "effective shear transformation" \cite{DW2021} between anisotropic and isotropic two-dimensional Ising models.

(vii) We have not made progress in developing a systematic approach to determining the principal axes of weakly anisotropic systems other than $\varphi^4$ models. These axes are of fundamental physical importance in real systems.
So far the angles describing the principal directions, e.g., of the $n$-vector model, depend in an unknown way on the
microscopic anisotropic interactions
which must be determined for each special anisotropic system under consideration.
This lack of knowledge is not of a harmless kind and constitutes a
major challenge to future research.

Significant issues of weak anisotropy were not yet addressed in this paper and call for further research in several directions, e.g., extensions to

(a) bulk and finite-size properties of other lattice systems
\cite{Perk1,Perk2,Perk3,Perk4,CoyWu} and
other models \cite{Baxter1982}
with both isotropic and weakly anisotropic interactions,
to be analyzed in the framework of two-scale-factor universality and multiparameter universality,

(b) angular-dependent representations of correlation lengths and correlation functions in three dimensions for general $n$,

(c) other geometries beyond rectangles,

(d) effects of an ordering field,

(e) anisotropic effects near the Kosterlitz-Thouless transition of systems in the $d=2,n=2$
 universality class,

(f) finite-size effects in weakly anisotropic systems away from $T_c$,

(g) crossover from weak to strong anisotropy,

(h) finite-size effects in anisotropic systems in the presence of non-periodic
boundary conditions.

We consider item (h) to be most important as it is relevant for applications to real systems
such as magnetic materials and superconductors which require a description with free or Dirichlet boundary conditions.
\renewcommand{\thesection}{\Roman{section}}
\setcounter{equation}{0} \setcounter{section}{1}
\renewcommand{\theequation}{\Alph{section}.\arabic{equation}}

\section*{ Appendix A: 
Relation between $Q_3$ and $\widetilde Q_3$}

In the following the relation (\ref{tildeQ}) between the universal constants $Q_3$ and $\widetilde Q_3$ is derived \cite{Kastening}. The isotropic bulk correlation functions (\ref{3caltTc}) and (\ref{atTcxisoiso}) at $T_c$
\begin{eqnarray}
G_c(|{\bf x}|)&=&\frac{D_c}{|{\bf x}|^{d-2+\eta}}\;,\;\\
 \hat G_c(|{\bf k}|) &=&\frac{\hat D_c}{|{\bf k}|^{2-\eta}}\;,\\
\frac{D_c}{\hat D_c}&=& \frac{\widetilde Q_3}{ Q_3} \;,
\end{eqnarray}
are related by the Fourier transformation
\begin{eqnarray}
\label{Gc}
G_c(|{\bf x}|)= \int _{\bf k} \;e^{i{\bf k}\cdot{\bf x}}\hat G_c(|{\bf k}|)
\;,
\end{eqnarray}
where $\int _{\bf k}$ stands for $(2\pi)^{-d}\int d^d k$ with an infinite cutoff. We decompose ${\bf k}= {\bf k_0}+{\bf q}$ where ${\bf k_0}$ and ${\bf q}$ are parallel and perpendicular to ${\bf x}$, respectively. Then the ratio $D_c/\hat D_c$  is determined by
\begin{eqnarray}
\label{ralationQQ}
&&  \int _{\bf k}\frac{e^{i{\bf k}\cdot{\bf x}}}{|{\bf k}|^{2-\eta}}=\int^\infty_{-\infty}\frac{d k_0}{2\pi}e^{ik_0|{\bf x}|}\int \frac{d^{d-1}q}{(2\pi)^{d-1}}\frac{1}{(k_0^2+q^2)^{(2-\eta)/2}}\nonumber\\
  &&=\frac{1}{|{\bf x}|^{d-2+\eta}}\int^\infty_{-\infty}\frac{d y}{2\pi}\frac{e^{iy}}{|y|^{3-d-\eta}}\int\frac{d^{d-1}q}{(2\pi)^{d-1}}\frac{1}{(1+q^2)^{(2-\eta)/2}}
  \nonumber\\
  &&=\frac{1}{|{\bf x}|^{d-2+\eta}}\;\frac{-2\Gamma(d-2+\eta)\;\cos[(d+\eta)\pi/2]}{2\pi}\nonumber\\
  &&\times \frac{2}{(4\pi)^{(d-1)/2}\Gamma[(d-1)/2]}\;\frac{\Gamma[(d-1)/2]\;\Gamma[(3-d-\eta)/2]}{2\Gamma(1-\eta/2)}
 \nonumber \\
  &&=\frac{\sin[(3-d-\eta)\pi/2]\;\Gamma[(3-d-\eta)/2]\;\Gamma(d-2+\eta)}{|{\bf x}|^{d-2+\eta}\;\pi(4\pi)^{(d-1)/2}\;\Gamma(1-\eta/2)}
   \nonumber\\
   &&=\frac{1}{|{\bf x}|^{d-2+\eta}}\;\frac{\Gamma(d-2+\eta)}{(4\pi)^{(d-1)/2}\;\Gamma(1-\eta/2)\;\Gamma[(d-1+\eta)/2]}
\nonumber\\
   &&=\frac{1}{|{\bf x}|^{d-2+\eta}}\;\frac{2^{d-2+\eta}\;\Gamma[(d-2+\eta)/2]}{(4\pi)^{d/2}\;\Gamma[(2-\eta)/2]}\\
   &&=\frac{D_c}{\hat D_c}\;\frac{1}{|{\bf x}|^{d-2+\eta}}\;,
\end{eqnarray}
which yields (\ref{tildeQ}).

\section*{ACKNOWLEDGMENT}
I thank F. Kischel and S. Wessel for useful discussions
and J. H. H. Perk for calling attention to Refs. \cite{Perk1,Perk2,Perk3,Perk4}.

\end{document}